\documentclass[11pt]{article}

\usepackage[a4paper,hmarginratio=1:1,vmarginratio=2:3,
totalwidth=16cm,totalheight=23cm]{geometry}

\usepackage{dcolumn}
\usepackage{amsmath}
\usepackage{amssymb}
\usepackage{amsfonts}
\usepackage{epsfig}
\usepackage{graphics}
\usepackage{dcolumn}
\usepackage{longtable}

\newcounter{oldcounter}
\addtocounter{equation}{1}
\def\bm#1{\mbox{\boldmath$#1$\unboldmath}}

\begin{document}

\thispagestyle{empty} \vspace{3cm}

\begin{center}
{\Large \textbf{Charmless $B\to PP, PV, VV$ Decays Based on the six-quark Effective Hamiltonian with Strong Phase Effects I}}
\bigskip

\vspace{0.5cm}
Fang Su$^{\ast\dagger}$, Yue-Liang Wu$^{\ast}$,  Yi-Bo Yang$^{\ast\ddagger}$, Ci Zhuang$^{\ast}$\\
\vspace{0.5cm}
  $^\ast$ Kavli Institute for Theoretical Physics China \\ Key Laboratory of Frontiers in Theoretical Physics \\
Institute of Theoretical Physics, Chinese Academy of Science,
Beijing 100190, China \\
$^{\dagger}$ Institute of Particle Physics, Huazhong Normal University,
 Wuhan, Hubei, 430079, China\\
$^\ddagger$ Graduate School of the Chinese Academy of Sciences,
Beijing, 100039, China

\end{center}

\medskip
\begin{abstract}
\noindent Based on an approximate six-quark operator effective
Hamiltonian from perturbative QCD, we present a systematical study
of charmless $B \to PP, PV, VV$ decays~($P$ and $V$ denoting
pseudoscalar and vector mesons, respectively). The calculation of
the relevant hard-scattering kernels is completed, the resulting
transition form factors are consistent with the results of QCD sum
rule calculation. Important classes of power corrections include
``chirally-enhanced'' terms, vertex corrections and weak
annihilation contributions with non-trivial strong phase. With these
considerations, predictions are presented for the branching ratios
and CP asymmetries of B-meson decays into PP, PV and VV final
states, and also for the corresponding polarization observables in
VV final states. Several decay modes and observables, which are of
particular interest phenomenologically, are discussed in detail,
including the effect of annihilation amplitude with strong phase,
the $\pi\pi$, $\pi K$ and $\pi \rho$ systems, the longitudinal
polarization fraction $f_L$ in $\rho K^*$ and $\phi K^*$ systems and
so on. It is observed that predictions in our framework generally
agree with the current experimental data.

\end{abstract}

\newpage
\setcounter{page}{1}
\def\thefootnote{\arabic{footnote}}
\setcounter{footnote}{0}

\section{Introduction}

The study of hadronic charmless B-meson decays can provide not only an interesting avenue to understand the
CP violation and the flavor mixing of quark sector in the standard model~(SM), but also a powerful means to
probe different new physics scenarios beyond the SM~\cite{Antonelli:2009ws,Buchalla:2008jp}. With the operation
of dedicated B-factory experiments, a huge amount of experimental data on hadronic B-meson decays has been analyzed
with appreciative precision. To account for the experimental data, theorists are urged to gain deeper insight into
the mechanism of these decays.

Theoretically, in order to consistently predict hadronic $B$ decays, it needs to deal with the short-distance
contributions in a complete and systematic way from the high energy scale down to a proper low energy scale at
which the perturbative calculations still remain reliable, and to treat the long-distance contributions which
contain the non-perturbative strong interactions involved in these decays. The main task is to reliably compute
the hadronic matrix elements between the initial and final hadronic states. In the past years, much progress has
been made in our understandings of the hadronic charmless B-meson decays~\cite{Buchalla:2008jp}, and several novel
methods based on the ``naive'' factorization approach~(NF)~\cite{Wirbel}, such as the perturbative QCD method~(PQCD)~\cite{lihn},
the QCD factorization approach~(QCDF)~\cite{M}, and the soft-collinear effective theory (SCET)~\cite{SCET}, have been proposed.

All the above frameworks of weak decays are based on the four-fermion operator effective Hamiltonian derived via
operator product expansion and renormalization group evolution. In hadronic weak decays, the short-distance QCD
contributions are characterized by the Wilson coefficient functions of four-quark operators and the long-distance
ones are in principle obtained by evaluating the hadronic matrix elements of relevant four-quark operators.
The Wilson coefficient functions are in general calculated by perturbative QCD which is well developed,
while the evaluation of hadronic matrix elements remains a hard task as it involves non-perturbative effects
of QCD. In fact, for the mesonic two-body decays of B meson, it involves three quark-antiquark pairs once each
meson is regarded as a quark-antiquark bound state at the quark-level structure. This fact then naturally motivates
us to consider the six-quark~(rather than four-quark) operator effective Hamiltonian~\cite{Su:2008mc}.

For the infrared singularity caused by the gluon exchanging interaction during the  evaluation of the hadronic matrix
elements of effective six-quark operators, it is simply regulated by the introduction of a mass scale motivated from
the gauge invariant loop regularization method~\cite{LRC}, where the energy scale $\mu_g$ is introduced to play the
role of infrared cut-off energy scale without violating gauge invariance.

We have applied the QCD factorization based on six-quark operator
effective Hamiltonian to $B_{(s)}\to \pi\pi,\pi K, KK$
decays~\cite{Su:2008mc}, and the theoretical predictions for all the
branching ratios and $CP$ asymmetries in these decays are found to
be consistent with the current experimental data except for a few
decay modes. The strength of annihilation diagram contributions is a
widely discussed issue in hadronic B decays~\cite{Chay:2007ep}.
Their effects are assumed to be important both in pQCD and QCDF
framework, and could even be fine-tuned independently for different
decay modes~\cite{Beneke:2003zv,Cheng:2009cn}. Motivated by these
observations, in this work we try to find a simultaneous solution
for all $B \to PP, PV, VV$ decay modes in the framework of six-quark
operator effective Hamiltonian, by considering annihilation
contribution with non-trivial strong phase. However, the calculation
of strong phase from nonperturbative QCD effects is a hard task,
there exist no efficient approaches to evaluate reliably the strong
phases caused from nonperturbative QCD effects , and we has to set
the strong phase as an input parameter in our framework.

Our paper is organized as follows. In section~\ref{sec:sqeh}, after briefly reviewing the four-quark operator effective Hamiltonian,
we begin with the introduction of the primary six-quark diagrams with the exchanges of a single W-boson and a single gluon,
and the corresponding initial six-quark operator. It is shown that a complete six-quark operator effective Hamiltonian is in
general necessary to include all contributions from both perturbative and non-perturbative QCD corrections.
The treatments of the singularities caused by the gluon exchanging interactions and the on mass-shell fermion propagator
are presented in Section~\ref{sec:TOD}. Then in the next section, the vertex corrections and annihilation contributions are
presented. Section~\ref{Input} contains all the input parameters which will be used in our calculation. In Section~\ref{sec:fitting},
we will give our numerical predictions and discussions for $B\to PP, PV, VV$ decays. Our conclusions and the decay amplitudes are given
in the last section and in the Appendix, respectively.

\section{Effective Hamiltonian of Six-Quark Operators}\label{sec:sqeh}

\subsection{Four-Quark Operator Effective Hamiltonian}

Let us start from the four-quark effective operators in the effective weak Hamiltonian. The initial four-quark operator due to weak interaction via W-boson exchange is given as follows for B-meson decays
\begin{equation}
O_{1}=(\bar{q}^u_{i}b_{i})_{V-A}(\bar{q}^d_{j}u_{j})_{V-A}, \qquad
q^u=u,\ c, \quad  q^d = d,\ s.
\end{equation}

The complete set of four-quark operators are obtained from QCD and QED corrections which contain the gluon-exchange diagrams, strong penguin diagrams and electroweak penguin diagrams. The resulting effective Hamiltonian~(for $b\to s$ transition) with four-quark operators is given as follows~\cite{4qham}
\begin{eqnarray}
H_{\rm eff}\, =\, {G_F\over\sqrt{2}} \sum_{q=u,c}
\lambda_q^{s} \left[C_1(\mu)O_1^{(q)}(\mu) +C_2(\mu)O_2^{(q)}(\mu)+
\sum_{i=3}^{10}C_i(\mu)O_i(\mu)\right]+{\rm h.c.}\;,\label{eq:hpk}
\end{eqnarray}
where $\lambda_q^{s} = V_{qb}V^*_{qs}$ are products of the CKM matrix elements, $C_i(\mu)$ the Wilson coefficient functions~\cite{4qham}, and $O_i(\mu)$ the four-quark operators
\begin{eqnarray}
\begin{array}{ll}
\displaystyle O_1^{(q)}\, =\,
(\bar{q}_ib_i)_{V-A}(\bar{s}_jq_j)_{V-A}\;, & \displaystyle
O_2^{(q)}\, =\,(\bar{s}_ib_i)_{V-A}(\bar{q}_jq_j)_{V-A}\;,
\\
\displaystyle O_3\,
=\,(\bar{s}_ib_i)_{V-A}\sum_{q'}(\bar{q}'_jq'_j)_{V-A}\;,
&\displaystyle O_4\,
=\,\sum_{q'}(\bar{q}'_ib_i)_{V-A}(\bar{s}_jq'_j)_{V-A}\;,
\\
\displaystyle O_5\,
=\,(\bar{s}_ib_i)_{V-A}\sum_{q'}(\bar{q}'_jq'_j)_{V+A}\;,
&\displaystyle O_6\,
=\,-2\sum_{q'}(\bar{q}'_ib_i)_{S-P}(\bar{s}_jq'_j)_{S+P}\;,
\\
\displaystyle O_7\,
=\,\frac{3}{2}(\bar{s}_ib_i)_{V-A}\sum_{q'}e_{q'}(\bar{q}'_jq'_j)_{V+A}\;,&
\displaystyle O_8\, =\,
-3\sum_{q'}e_{q'}(\bar{q}'_ib_i)_{S-P}(\bar{s}_jq'_j)_{S+P}\;,
\\
\displaystyle O_9\,
=\,\frac{3}{2}(\bar{s}_ib_i)_{V-A}\sum_{q'}e_{q'}(\bar{q}'_jq'_j)_{V-A}\;,&
\displaystyle O_{10}\, =\,
\frac{3}{2}\sum_{q'}e_{q'}(\bar{q}'_ib_i)_{V-A}(\bar{s}_jq'_j)_{V-A}\;.\\
\end{array}
\label{eq:o}
\end{eqnarray}
Here the Fermi constant $G_F=1.16639\times 10^{-5}\;{\rm GeV}^{-2}$, $(\bar{q}'q')_{V\pm A} = \bar q' \gamma_\mu (1\pm \gamma_5)q'$, and $i, j$ are the color indices. The index $q'$ in the summation of the above operators runs through $u,\;d,\;s$, $c$, and $b$. The effective Hamiltonian for the $b\to d$ transition can be obtained by changing $s$ into $d$ in Eqs.~(\ref{eq:hpk}) and (\ref{eq:o}).

\subsection{\boldmath Six-quark Diagrams and Effective Operators}\label{sec:sqd}

\begin{figure}[h]
\begin{center}
\includegraphics[scale=0.60]{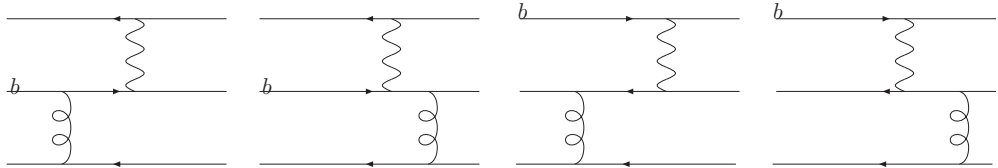}\\
  \caption{Four different six-quark diagrams with a single W-boson exchange and a single gluon exchange.}\label{pic:1insert}
  \end{center}
\end{figure}

As mesons are regarded as quark and anti-quark bound states, the
mesonic two body decays actually involve three quark-antiquark
pairs. It is then natural to consider the six-quark Feynman diagrams
which lead to three effective quark-antiquark currents. The initial
six-quark diagrams of weak decays contain one W-boson exchange and
one gluon exchange, thus there are four different diagrams as the
gluon exchange interaction can occur for each of four quarks in the
W-boson exchange diagram, see Fig.~\ref{pic:1insert}.

The resulting initial effective operators contain four terms corresponding to the four diagrams, respectively. In a good approximation, the four quarks via W-boson exchange can be regarded as a local four-quark interaction at the energy scale much below the W-boson mass, while the two QCD vertexes due to gluon exchange are at two independent space-time points, the resulting effective six-quark operators are hence in general non-local. The six-quark operators corresponding to the four diagrams in Fig.~\ref{pic:4insert} are found to be
\begin{eqnarray}
  O^{(6)}_{q_1}\, &=&\, 4\pi\alpha_s \int\!\!\int  \frac{\emph{d}^4k}{(2\pi)^4}\, \frac{\emph{d}^4p}{(2\pi)^4}\,e^{-i((x_1-x_2)p+(x_2-x_3)k)}
  (\bar{q'}(x_3)\gamma_{\nu}T^{a} q'(x_3))\frac{1}{k^2+i\epsilon}\nonumber\\
  &&\times(\bar{q}_{2}(x_1)\Gamma_{1}\frac{p\!\!\!/+m_b}{p^2-m_b^2+i\epsilon}\gamma^{\nu}T^{a} q_{1}(x_2))*
  (\bar{q}_{4}(x_1) \Gamma_{2} q_{3}(x_1)),\nonumber\\
  O^{(6)}_{q_2}\, &=&\, 4\pi\alpha_s \int\!\!\int  \frac{\emph{d}^4k}{(2\pi)^4}\, \frac{\emph{d}^4p}{(2\pi)^4}\,e^{-i((x_1-x_2)p+(x_2-x_3)k)}
  (\bar{q'}(x_3)\gamma_{\nu}T^{a} q'(x_3))\frac{1}{k^2+i\epsilon}\nonumber\\
  &&\times(\bar{q}_{2}(x_2)\frac{p\!\!\!/+m_{q_2}}{p^2-m_{q_2}^2+i\epsilon}\gamma^{\nu}T^{a}\Gamma_{1} q_{1}(x_1))*
  (\bar{q}_{4}(x_1) \Gamma_{2} q_{3}(x_1)),\nonumber\\
  O^{(6)}_{q_3}\, &=&\, 4\pi\alpha_s \int\!\!\int  \frac{\emph{d}^4k}{(2\pi)^4}\, \frac{\emph{d}^4p}{(2\pi)^4}\,e^{-i((x_1-x_2)p+(x_2-x_3)k)}
  (\bar{q'}(x_3)\gamma_{\nu}T^{a} q'(x_3))\frac{1}{k^2+i\epsilon}\nonumber\\
  &&\times(\bar{q}_{2}(x_1)\Gamma_{1} q_{1}(x_1))*
  (\bar{q}_{4}(x_1) \Gamma_{2} \frac{p\!\!\!/+m_{q_3}}{p^2-m_{q_3}^2+i\epsilon}\gamma^{\nu}T^{a} q_{3}(x_2)),\nonumber\\
  O^{(6)}_{q_4}\, &=&\, 4\pi\alpha_s \int\!\!\int  \frac{\emph{d}^4k}{(2\pi)^4}\, \frac{\emph{d}^4p}{(2\pi)^4}\,e^{-i((x_1-x_2)p+(x_2-x_3)k)}
  (\bar{q'}(x_3)\gamma_{\nu}T^{a} q'(x_3))\frac{1}{k^2+i\epsilon}\nonumber\\
  &&\times(\bar{q}_{2}(x_1)\Gamma_{1} q_{1}(x_1))*
  (\bar{q}_{4}(x_2)\frac{p\!\!\!/+m_{q_4}}{p^2-m_{q_4}^2+i\epsilon}\gamma^{\nu}T^{a} \Gamma_{2} q_{3}(x_1)),
\label{eq:six}
\end{eqnarray}
where $k$ and $p$ correspond to the momenta of gluon and quark in their propagators. $q_1$ is usually set to be the heavy b quark. $x_1$, $x_2$ and $x_3$ are space-time points corresponding to the three vertexes. The color index is summed between $q_1, q_2$ and $q_3, q_4$. Note that all the six-quark operators are proportional to the QCD coupling constant $\alpha_s$ due to gluon exchange. Thus the initial six-quark operator is given by summing over the above four operators
\begin{eqnarray}
  O^{(6)}=\sum_{j=1}^4 O^{(6)}_{q_j}.
\end{eqnarray}
Actually, the initial six-quark operators $O^{(6)}_{q_j}$~($j=1,2,3,4$) can be obtained from the following initial four-quark operator via a single gluon exchange
\begin{eqnarray}
  O \equiv (\bar{q}_{2} \Gamma_{1} q_{1})*(\bar{q}_{4} \Gamma_{2} q_{3}).
  \label{eq:any}
\end{eqnarray}

Unlike the classical four-quark effective operator, the six-quark operators used here are non-local with quark and gluon propagators inserted into them. Such operator is equal to the one used in SCET with $\frac{1}{D}$ and be found to be a more effective form to describe dynamics at low energy scale around 1.5~GeV.

With the above considerations, the QCD factorization approach with six-quark operator effective Hamiltonian enables us to evaluate all the hadronic matrix elements of nonleptonic two-body B-meson decays. The detailed calculations of the hadronic matrix elements in $B\to \pi^0\pi^0$, as an example, could be found in our previous paper~\cite{Su:2008mc}. As for the hadronic matrix elements in other $B\to PP, PV, VV$ decays, we shall list them in the Appendix.

\section{Treatment of Singularities}\label{sec:TOD}

Before proceeding, we would like to point out that there are two kinds of singularities in the evaluation of hadronic matrix elements. One singularity stems from the infrared divergence of gluon exchanging interaction, and the other one arises from the on mass-shell divergence of internal quark propagator.

In general, a Feynman diagram will yield an imaginary part for the decay amplitudes when the virtual particles in the diagram become on mass-shell, and the resulting diagram can be considered as a genuine physical process. It is well-known that when applying the Cutkosky rule~\cite{cutkosky} to deal with a physical-region singularity of all propagators, the following formula holds:
\begin{eqnarray}
\frac{1}{p^2-m_q^2+i\epsilon}=P\biggl[\frac{1}{p^2-m_q^2}
\biggl]-i\pi\delta[p^2-m_q^2],\label{quarkd}
\end{eqnarray}
which is known as the principal integration method, and the integration with the notation of capital letter $P$ is the so-called principal integration.

However, the Cutkosky rule is useless for singularities from infrared divergence of gluon propagator. Integration with those propagators is sensitive to the infrared cut-off for gluon and light-quark propagator, and diverge to infinity when the cut-off becomes to zero. A modified integration with different parameters for different channels is used in QCDF framework~\cite{Beneke:2003zv,Cheng:2009cn}, while the transverse momentum $k_T$ dominating in the zero momentum fraction is added to the propagator in pQCD framework~\cite{lihn}. In this work, we prefer to add the same dynamics mass for both gluon and light quark to investigate the infrared cut-off dependence of perturbative theory prediction:
\begin{eqnarray}
   \frac{1}{k^2}\frac{p\!\!/\,+m_q}{(p^2-m_q^2)} &\to&
   \frac{1}{(k^2-\mu_g^2+i\epsilon)}\frac{p\!\!/\,+\mu_q}{(p^2-\mu_q^2+i\epsilon)}~(\text{q is a light quark}).
\end{eqnarray}

\begin{figure}[htbp]
\begin{center}
  \includegraphics[scale=0.7]{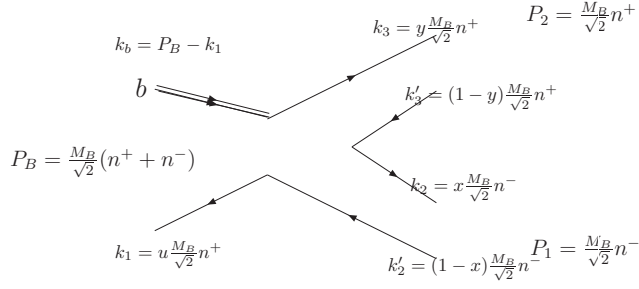}
  \caption{Definition of momentum in $B\rightarrow M_1M_2$ decay. The light-cone coordinate is adopted with $(n^+, n^-, \vec{k}_{\bot})$.} \label{pic:definition}
  \end{center}\vspace{-0.5cm}
\end{figure}

It is noted that, as the gauge dependent term $k_{\mu}k_{\nu}$ can always be written as linear combinations of the momenta $p_{\alpha}$ on the external lines of spectator quark, and they are all on mass-shell in our present consideration~(as defined in Fig.~\ref{pic:definition}, their contributions equal to zero once the equation of motion is used. Our results are therefore gauge independent.

\section{Vertex Corrections and Annihilation Contributions}\label{vetex corrections}

The short-distance contributions characterized by the Wilson coefficient functions for the effective four-quark operators were calculated by several groups at the leading order~(LO) and the next-to-leading order(NLO)~\cite{4qham,hep-ph/9806471}. Their values depend mainly on the choice for the running scale $\mu$. When considering the NLO Wilson coefficient functions and $\alpha_{s}$, one needs to include the magnetic penguin-like operator $O_{8g}$ which is defined as~\cite{4qham}
\begin{eqnarray}
O_{8g}\, =\,
\frac{g}{8\pi^2}m_b{\bar{q}}_i\sigma_{\mu\nu}(1+\gamma_5)T_{ij}^aG^{a\mu\nu}b_j\;,
\end{eqnarray}
where $i$, $j$ are the color indices. The magnetic-penguin contribution to a generic $B\to M_1 M_2$ decay leads to the modification for the Wilson coefficients corresponding to the QCD penguin operators,
\begin{eqnarray}
a_{4,6}(\mu)&\to& a_{4,6}(\mu) - \frac{\alpha_s(\mu)}{9\pi}
\frac{2m_B}{\sqrt{|l^2|}}C_{8g}^{\rm eff}(\mu),
\end{eqnarray}
with $C_{8g}^{\rm eff}=C_{8g}+C_5$, $|l^2| = m_B^2/2$, and $a_{4,6} = C_{4,6}+\frac{C_{3,5}}{N_c}$ which appear in the factorizable diagrams.

As shown in Ref.~\cite{0508041}, CP violating observables may be improved by adding vertex corrections. Furthermore, the vertex corrections were proposed to improve the scale dependence of Wilson coefficient functions of factorizable emission amplitudes in QCDF~\cite{NPB.606.245}. Those coefficients are always
combined as $C_{2n-1}+\frac{C_{2n}}{N_C}$ and $C_{2n}+\frac{C_{2n-1}}{N_C}$, which, after taken into account the vertex corrections, are modified to
\begin{eqnarray}
C_{2n-1}(\mu)+\frac{C_{2n}}{N_C}(\mu) \to
C_{2n-1}(\mu)+\frac{C_{2n}}{N_C}(\mu)
+\frac{\alpha_s(\mu)}{4\pi}C_F\frac{C_{2n}(\mu)}{N_c} V_{2n-1}(M_2)
\;,&&
\nonumber\\
C_{2n}(\mu)+\frac{C_{2n-1}}{N_C}(\mu) \to
C_{2n}(\mu)+\frac{C_{2n-1}}{N_C}(\mu)
+\frac{\alpha_s(\mu)}{4\pi}C_F\frac{C_{2n-1}(\mu)}{N_c} V_{2n}(M_2) \;,&&
\end{eqnarray}
with $n=1,...,5$, $M_2$ being the meson emitted from the weak vertex. In the naive dimensional regulation~(NDR) scheme, $V_i(M)$ are given by~\cite{Beneke:2003zv,NPB.606.245}
\begin{eqnarray}
V_i(M) &=&\left\{
\begin{array}{ll}
12\ln(\frac{m_b}{\mu})-18+\int_0^1 dx\, \phi_a(x)\, g(x)\;, &
\mbox{\rm for }i=1-4,9,10\;,
\\
-12\ln(\frac{m_b}{\mu})+6-\int_0^1dx\, \phi_a(1-x)\, g(1-x)\;, &
\mbox{\rm for }i=5,7\;,
\\
-6 +\int_0^1 dx\,\phi_b(x)\, h(x) \;, &
 \mbox{\rm for }i=6,8\;,
\end{array}
\right.\label{vim}
\end{eqnarray}
where $\phi_a(x)$ and $\phi_b(x)$ denote the leading-twist and twist-3 distribution amplitudes for a pseudoscalar meson or a longitudinally polarized vector meson, respectively. While for a perpendicularly polarized vector final state, $\phi_a(x)=\phi_{\pm}(x,\mu)$ and $\phi_b(x)=0$. The functions $g(x)$ and $h(x)$ used in the integration are given respectively as~\cite{Beneke:2003zv}

\begin{eqnarray}
g(x) &=& 3\left( \frac{1-2x}{1-x}\ln{x} -i\,\pi \right)\nonumber\\
& & +\left[ 2\,{\rm Li}_2(x)-\ln^2 x +\frac{2\ln
x}{1-x}-(3+2i\,\pi)\ln x - (x\leftrightarrow 1-x) \right] \;,\nonumber
\\
h(x) &=& 2\,{\rm Li}_2(x)-\ln^2 x -(1+2i\,\pi)\ln x -
(x\leftrightarrow 1-x) \;.
\end{eqnarray}

\begin{figure}[h]
\begin{center}
  \includegraphics[scale=0.6]{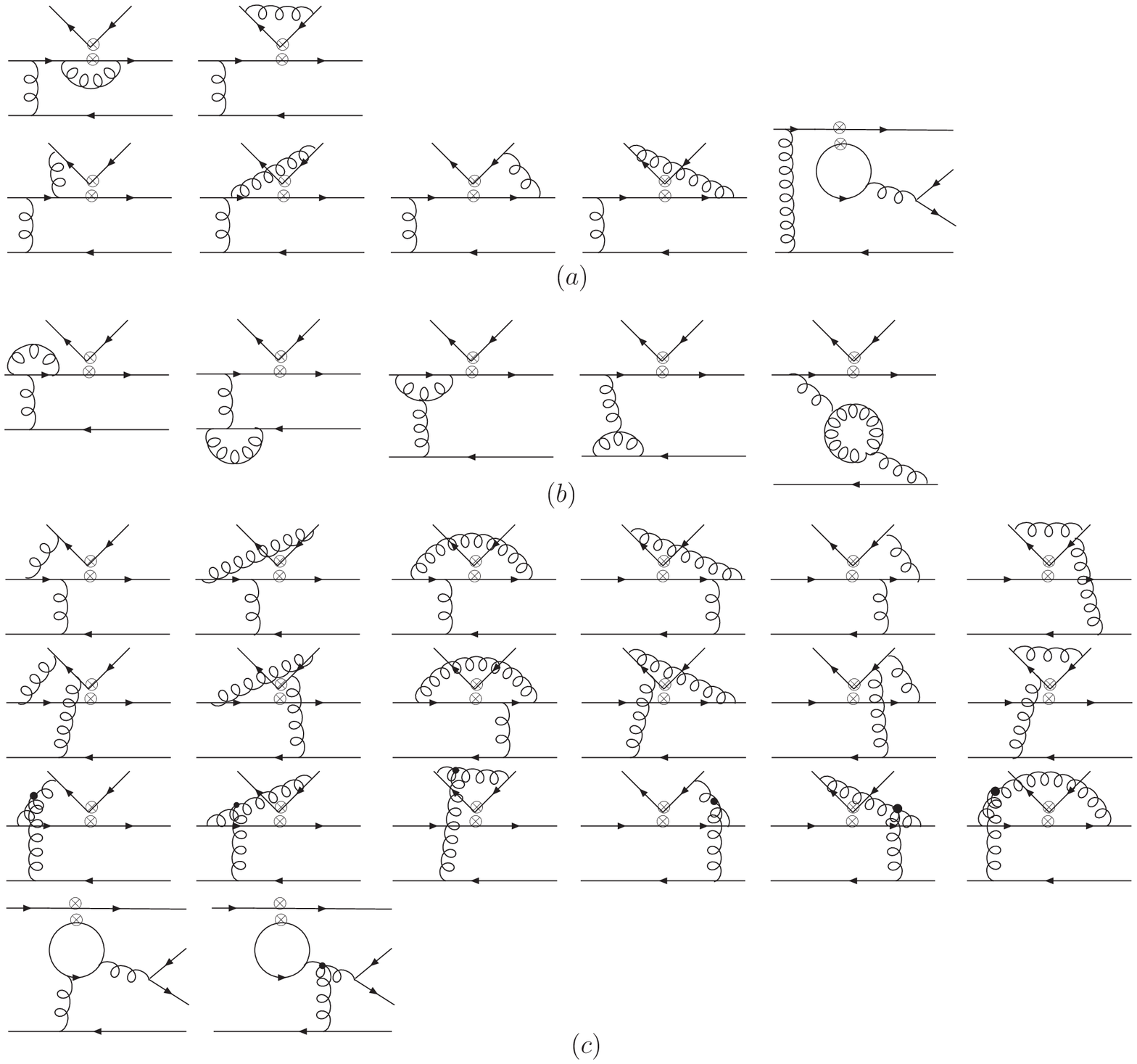}\\
  \caption{\small The diagrams in (a), (b) and (c) are loop contributions only to the effective weak vertex~(type I), only to the gluon vertex~(type II), and for both weak and strong vertexes~(type III), respectively.}\label{pic:4insert}
  \end{center}
\end{figure}

To further improve our predictions, we shall examine an interesting case that vertexes receive additional large non-perturbative contributions, namely the Wilson coefficients $a_i =C_i+\frac{C_{i\pm1}}{N_C}$ are modified to be the following effective ones:
\begin{eqnarray}\label{aeff}
a_i \to a_i^{eff} = C_i(\mu)+\frac{C_{i\pm1}}{N_C}(\mu)
+\frac{\alpha_s(\mu)}{4\pi}C_F\frac{C_{i\pm1}(\mu)}{N_c}( V_i(M_2)+\widetilde{V}_1(M_2))
\;, &&(i=1-4,9,10),
\nonumber\\
a_i \to a_i^{eff} = C_i(\mu)+\frac{C_{i\pm1}}{N_C}(\mu)
+\frac{\alpha_s(\mu)}{4\pi}C_F\frac{C_{i\pm1}(\mu)}{N_c}( V_i(M_2)+\widetilde{V}_2(M_2))
\;, &&(i=5-8).
\end{eqnarray}
The corrections $\widetilde{V}_1(M_2)$ and $\widetilde{V}_2(M_2)$ depend on whether the meson $M_2$ is a pseudoscalar or a vector. It could be caused from the higher order non-perturbative non-local effects, as shown in Fig.\ref{pic:4insert}. Adopting $\widetilde{V}_1(P) = 26 e^{-\frac{\pi}{3} i}$, $\widetilde{V}_2(P)= -26 $, $\widetilde{V}_1(V) = 15 e^{\frac{\pi}{8} i}$, and $\widetilde{V}_2(V)=-15 e^{\frac{\pi}{8} i}$, both the branching ratios and CP asymmetries of most $B\to PP, PV, VV$ decay modes are improved, which will be detailed later.

After involving the vertex corrections and the contributions of effective Wilson coefficients, we next
take account of the annihilation contributions with non-trivial strong phase. Both the vertex corrections and effective Wilson coefficients only affect emission diagrams while pall on exchange and annihilation diagrams. In our calculation, vertex correction for annihilation diagram might be sizable in PV and/or VV final state, and an extra strong phase for PV and/or VV channel would improve our theoretical predictions.

Most of the annihilation contributions are from factorizable annihilation diagrams with the $(S-P)\times (S+P)$ effective four-quark vertex:
\begin{eqnarray}\label{annihilation}
A_{SP}^{P_1 P_2}(M) \sim \int dxdy \frac{(\mu_{P_1}+\mu_{P_2})y(1-y)} {(x(1-y)m_B^2-\mu_g^2+i\epsilon)((1-y)m_B^2-m_{q}^2+i\epsilon)},\nonumber\\
A_{SP}^{P_1 V_2}(M) \sim \int dxdy \frac{(\mu_{P_1}-3(2x-1)m_{V_2})y(1-y)} {(x(1-y)m_B^2-\mu_g^2+i\epsilon)((1-y)m_B^2-m_{q}^2+i\epsilon)},\nonumber\\
A_{SP}^{V_1 P_2}(M) \sim \int dxdy \frac{(-3(1-2x)m_{V_1}-\mu_{P_2})y(1-y)} {(x(1-y)m_B^2-\mu_g^2+i\epsilon)((1-y)m_B^2-m_{q}^2+i\epsilon)},\nonumber\\
A_{SP}^{V_1 V_2}(M) \sim \int dxdy \frac{3(1-2x)(-m_{V_1}+3(2x-1)m_{V_2})y(1-y)} {(x(1-y)m_B^2-\mu_g^2+i\epsilon)((1-y)m_B^2-m_{q}^2+i\epsilon)}.
\end{eqnarray}
From Eq.~(\ref{annihilation}) and Eqs.~(\ref{eq:vvanf1})-(\ref{eq:vvanf2}), it can be seen that $A_{SPb}^{NM_1M_2}$ can be converted into $A_{SPa}^{NM_1M_2}$ by exchanging $x\leftrightarrow(1-y)$. Since the contributions of these amplitudes are dominated by the area $x\sim0$ or $y\sim1$, $A_{SP}^{P_1 P_2}(M)$ and $A_{SP}^{P_1 V_2}(M)$ have the same sign, while $A_{SP}^{V_1 P_2}(M)$ and $A_{SP}^{V_1 V_2}(M) $ have a different sign from $A_{SP}^{P_1 P_2}(M)$. As a result, we use the same strong phase for $A_{SP}^{P_1 P_2}(M)$ and $A_{SP}^{P_1 V_2}(M)$, and another one for $A_{SP}^{V_1 P_2}(M)$ and $A_{SP}^{V_1 V_2}(M)$.

\section{Theoretical Input Parameters} \label{Input}

The theoretical predictions in our calculations depend on many input parameters, such as the Wilson coefficients, the CKM matrix elements, the hadronic parameters, and so on. Here we present all
the relevant input parameters as follows.

\subsection{Light-Cone Distribution Amplitudes}

For the wave function of $B$ meson, we take the following form in our numerical calculations~\cite{h.y.cheng}:
\begin{equation}
\phi_B(x)=N_Bx^2(1-x)^2
\exp\left[-\frac{1}{2}\left(\frac{xm_B}{\omega_B}\right)^2 \right]
\;, \label{def:lcda1}
\end{equation}
where the shape parameter of B meson is $\omega_{B}$=0.30GeV, and $N_B$ is a normalization constant.

We next specify the light-cone distribution amplitudes~(LCDAs) for pseudoscalar and vector mesons, the definition of which could be found in Refs.~\cite{Beneke:2000wa,Braun:1989iv}. The general expressions of twist-2 LCDAs are
\begin{eqnarray}
\phi_{P}(x,\mu) &=& 6x(1-x) \big[1 + \sum^{\infty}_{n=1} a^P_n(\mu) C^{3/2}_n(2x-1)\big],\nonumber\\
\phi_{V}(x,\mu) &=& 6x(1-x) \big[1 + \sum^{\infty}_{n=1} a^V_n(\mu) C^{3/2}_n(2x-1)\big],\nonumber\\
\phi_{V}^T(x,\mu) &=& 6x(1-x) \big[1 + \sum^{\infty}_{n=1} a^{T,V}_n(\mu) C^{3/2}_n(2x-1)\big],
\end{eqnarray}
and those of twist-3 ones are
\begin{eqnarray}
\phi_{p}(x,\mu) &=&1,\  \hspace{0.5cm}
\phi_{\sigma}(x,\mu) = 6x(1-x),\nonumber\\
\phi_{\nu}(x,\mu) &=& 3 \big[2x - 1 + \sum^{\infty}_{n=1} a^{T,V}_n(\mu) P_{n+1}(2x-1)\big]\nonumber\\
\phi_{+}(x)& =& 3 (1-x)^2, \hspace{0.5cm} \phi_{-}(x) = 3 x^2,
\end{eqnarray}
where $C_n$(x) and $P_n$(x) are the Gegenbauer and Legendre polynomials, respectively. The shape parameters of light mesons are taken from~\cite{Ball:2007rt} and listed in Table~\ref{parameters}.

\begin{table}[t]
\vspace{-2em} \caption{\small{Values of Gegenbauer moments at the scale $\mu$=1GeV, which are taken from~\cite{Ball:2007rt}.}}\label{parameters}
\begin{tabular}{l|c|c|c|c|c|c}
\hline
        & $\pi$   &K  &$\rho$   &$K^*$    &$\phi$    &$\omega$  \\
\hline   \hline
$a_1$     &  --   &$0.06\pm0.03$ &--&$0.03\pm0.02$&--& -- \\
$a_2$&$0.25\pm0.15$&$0.25\pm0.15$&$0.15\pm0.07$&$0.11\pm0.09$&$0.15\pm0.07$&$0.18\pm0.08$ \\
$a_1^T$     &--& -- & --&$0.04\pm0.03$ &--& -- \\
$a_2^T$&--& -- &$0.14\pm0.06$&$0.10\pm0.08$&$0.16\pm0.06$&$0.14\pm0.07$ \\
\hline
\end{tabular}
\end{table}

The parameters in Table~\ref{parameters} are given at the scale
$\mu$=1.0GeV, and should run to our calculation scale $\mu =
\sqrt{2\Lambda_{QCD}m_b} \simeq 1.5\pm 0.1{\rm GeV}$, with
$\Lambda_{QCD}\simeq 288^{+21}_{-18}{\rm MeV}$. It is noted that
LCDAs of light mesons become more close to their asymptotic
forms~(all shape parameters become zero) when the scale runs to
higher values.

\subsection{Decay Constants and Other Input Parameters}

For decay constants of various mesons and other hadronic parameters, we list them in Table~\ref{input}. As for the CKM matrix elements, we shall use the Wolfenstein parametrization~\cite{Wolfenstein:1983yz} with the four parameters chosen as~\cite{Charles:2004jd}:
$A=0.798^{+0.023}_{-0.017}$, $\lambda=0.2252^{+0.00083}_{-0.00082}$, $\bar
\rho=0.141^{+0.035}_{-0.021}$, and $\bar \eta=0.340\pm0.016$.

\begin{table}[ht]
\begin{center}
\caption{The hadronic input parameters~\cite{Amsler:2008zzb} and the decay constants taken from the QCD sum rules~\cite{Ball:2006eu,ball3} and Lattice theory~\cite{lattice}.} \label{input}
\doublerulesep 0.8pt
\tabcolsep 0.08in \vspace{0.2cm}
\begin{tabular}{ccccccc} \hline\hline
$\tau_{B^\pm}$&$\tau_{B_d}$ &$m_B$  &$m_b$ &$m_t$&$m_u$ &$m_d$\\ \hline

$1.638$ps &$1.525$ps &$5.28$GeV  &$4.4$GeV &$173.3$GeV&$4.2$MeV &$7.6$MeV\\

$m_c$&$m_s$& $m_{\pi^\pm}$ & $m_{\pi^0}$ & $m_K $&$m_{\rho^0}$ &$m_{\rho^{\pm}}$ \\

$1.5$GeV&$0.122$GeV& $0.140$GeV & $0.135$GeV &$0.494$GeV &$0.775$GeV &$0.775$GeV\\

$m_{\omega}$& $m_{\phi}$&$m_{K^{*\pm}}$&$m_{K^{*0}}$& $\mu_\pi$ &$\mu_K$&$f_{\phi}$\\

1.7GeV &1.8GeV &$300$MeV &$0.78$GeV &$1.02$GeV&$0.892$GeV&$0.215$GeV \\

$f_B$ & $f_{\pi}$ &  $f_K$ &$f_{\rho}$&$f_{\omega}$&$f_{K^*}$&$f_{\omega}^T$\\

$0.210$GeV & $0.130$GeV & $0.16$GeV&$0.216$GeV&$0.187$GeV&$0.220$GeV&$0.151$GeV\\

$f_{K^*}^T$&$f_{\phi}^T$& $f_{\rho}^T$ \\

$0.185$GeV&$0.186$GeV &$0.165$GeV \\   \hline\hline
\end{tabular}
\end{center}
\end{table}

In our numerical calculations, the running scale is taken to be
\begin{eqnarray}
\mu = \sqrt{2\Lambda_{QCD}m_b} \simeq 1.5\pm0.1{\rm GeV}.
\end{eqnarray}
The scale of $\alpha_{s}$ in the six-quark operator effective
Hamiltonian is also taken at 1.5 GeV. Mass of b quark running to
$\mu=1.5\pm0.1$~GeV follows the framework in \cite{quarkmass} as:
\begin{eqnarray}
m^{}_q(\mu) &=& {{\cal R}(\alpha^{}_s(\mu))} \hat{m}^{}_q \
,\nonumber\\
 {\cal
R}(\alpha^{}_s) &=&
\left(\frac{\alpha^{}_s}{\pi}\right)^{\gamma^{}_0/\beta^{}_0}
\left[1 + \frac{\alpha^{}_s}{\pi}{\cal C}^{}_1  +
\frac{\alpha^2_s}{2\pi^2} \left({\cal C}^2_1 + {\cal C}^{}_2\right)
+ \frac{\alpha^3_s}{\pi^3} \left( \frac{1}{6} {\cal C}^3_1 +
\frac{1}{2} {\cal C}^{}_1 {\cal C}^{}_2 + \frac{1}{3}{\cal C}^{}_3
\right) \right] \; .
\end{eqnarray}
The definition of $C_i$ can be found in \cite{quarkmass}. After
calculation, we can get $m_{b}(\mu)\simeq$5.54 GeV.

In addition, the infrared cut-offs for gluon and light quarks are
the basic scale to determine annihilation diagram contributions~(the
smaller $\mu_g$, the larger contributions), and are set to be
$\mu_q=\mu_g$=0.37 GeV.

\subsection{Form Factors}

As is known, form factors have to be provided from outside in QCDF
and SCET frameworks~(such as by resorting to QCD sum rules or
lattice QCD), while can be calculated in pQCD approach. The method
developed here allows us to calculate the relevant transition form
factors as below:
 \begin{eqnarray}
&&F_0^{B\rightarrow M_1}=\frac{4\pi\alpha(\mu)C_F}{N_c m_B^2
 F_{M_2}}T_{LL}^{FM_1M_2}(B)(M_1,M_2 = P),\nonumber\\
&&V^{B\rightarrow M_1}=\frac{4\pi\alpha(\mu)C_F}{N_c m_B^2
F_{M_2}}T_{LL,\perp}^{FM_1M_2}(B)\frac{m_B^2(m_B+m_{M_1})}{m_{M_2}(m_B^2-m_{M_1}m_{M_2})}(M_1,M_2 = V),\nonumber\\
&&A_0^{B\rightarrow M_1}=\frac{4\pi\alpha(\mu)C_F}{N_c m_B^2
F_{M_2}}T_{LL}^{FM_1M_2}(B)(M_1 = V,M_2 = P),\nonumber\\
&&A_1^{B\rightarrow M_1}=\frac{4\pi\alpha(\mu)C_F}{N_c m_B^2
F_{M_2}}T_{LL,//}^{FM_1M_2}(B)\frac{m_B^2}{m_{M_2}(m_B+m_{M_1})}(M_1,M_2
= V),
\end{eqnarray}
with:
 \begin{eqnarray}
  T_{LL,\perp}=\frac{1}{2}(T_{LL,+}-T_{LL,-}),&& C_F=\frac{N_c^2-1}{2
  N_c}, \nonumber
\end{eqnarray}
Before giving predictions of the observables in $B \to PP, PV, VV$
decays, we first list our numerical results for the form factor at
$q^2=0$, and for comparison, we also list the results of QCD
sum-rules,light-cone sum rules and
PQCD~\cite{Wu:2006rd,pqcdformfactor,Ball:2004ye} in
Table~\ref{tab:formfactor}.

\begin{table}
\vspace{-2em}
\begin{center}
\caption{\small{The $B \to P,V$ form factors at $q^2 = 0$ in QCD Sum
Rules, Light Cone and our work,where the errors stem mainly from the
uncertainties in the global parameters $\mu_{scale}=1.5\pm0.1
\text{GeV}, \mu_g=0.37\pm0.037\text{GeV}$, and from the shape
parameters of light mesons. Within their respective uncertainties,
our predictions are consistent with the results of the other two
methods. }}\label{tab:formfactor}
\vspace{0.2cm}
\begin{tabular}{c|c|c|c|c|c}
\hline \hline
\ \ \ \ \ Mode          & F(0)  & QCDSR\cite{Ball:2004ye}& LCSR\cite{Wu:2006rd}&PQCD\cite{pqcdformfactor}& This work  \\
\hline
 $B \to K^*$ & V&0.411&0.339&0.406&$0.277^{+0.075+0.008}_{-0.043-0.006}$\\
  \cline{2-6}&$A_0$&0.374&0.280&0.455&$0.328^{+0.095+0.021}_{-0.048-0.014}$\\
  \cline{2-6}&$A_1$&0.292&0.274&0.30&$0.220^{+0.054+0.006}_{-0.031-0.005}$\\
\hline
 $B \to \rho$ &V&0.323&0.298&0.318&$0.233^{+0.063+0.006}_{-0.036-0.002}$\\
  \cline{2-6}&$A_0$&0.303&0.248&0.366&$0.280^{+0.077+0.012}_{-0.041-0.005}$\\
  \cline{2-6}&$A_1$&0.242&0.239&0.25&$0.193^{+0.048+0.004}_{-0.028-0.002}$\\
  \hline
 $B \to \omega$ &V&0.293&0.275&0.305&$0.206^{+0.056+0.005}_{-0.032-0.002}$\\
  \cline{2-6}&$A_0$&0.281&0.231&0.347&$0.251^{+0.070+0.010}_{-0.036-0.004}$\\
  \cline{2-6}&$A_1$&0.219&0.221&0.30&$0.170^{+0.043+0.004}_{-0.024-0.002}$\\
  \hline
 $B \to \pi$ &$F_0$&$0.258$&0.285&0.292&$0.269^{+0.053+0.008}_{-0.030-0.005}$\\
    \hline
 $B \to K$ &$F_0$&$0.331$&0.345&0.321&$0.349^{+0.070+0.012}_{-0.038-0.007}$\\
  \hline \hline
\end{tabular}
\end{center}
\end{table}


\section{Numerical Results and Discussions}\label{sec:fitting}

In this section, we shall classify the 43 channels of B decays into two light mesons according to the final states, and give our predictions for the branching ratios, the CP asymmetries, and the longitudinal polarization fractions. Comparisons with the current experiment data and other theoretical predictions, if
possible, are also made. According to different decay modes, we shall give our predictions for the observables one by one.

\subsection{$B\to PP$ decays}

The contributions of effective Wilson coefficients and effects of different strong phases for annihilation diagrams for $B \to PP$ are shown in Tables~\ref{tab:br010} and \ref{tab:br011}. For $B\to \pi \pi$ decay channel, the naive power-counting based on factorization theory predicts:
\begin{eqnarray}
Br(\pi^- \pi^+)> Br(\pi^- \pi^0)\gg Br(\pi^0 \pi^0).
\end{eqnarray}
However, from the current experimental data~\cite{HFAG}, we can see that Br($\pi^- \pi^+$)$\sim$Br($\pi^- \pi^0$) and Br($\pi^0 \pi^0$) is much larger than theory prophecy. From Table~\ref{tab:br011}, we can see that: with only the vertex and NLO corrections included, the branching ratio of $B\to \pi^+\pi^-$ and the ones of $B^+\to \pi^+\pi^0, \pi^0\pi^0$ modes are bigger and smaller than the current experiment data, respectively; after taking into account the contributions of effective Wilson coefficients, which can reduce the former but enhance the latter two, our predictions are well consistent with the data.

\begin{table}
\vspace{-2em} \caption{\small{The branching ratios~(in units of $10^{-6}$) and direct CP asymmetries in $B\to \pi K$ decays. The central values are obtained at $\mu_q=\mu_g$=0.37GeV, the first error stems from the uncertainties in the global parameters $\mu_{scale}=1.5\pm0.1 \text{GeV}, \mu_g=0.37\pm0.037\text{GeV}$, and the second from the shape parameters of light mesons. ${\rm NLO}^{eff}$ and ${\rm NLO}^{eff}(\theta^a)$ stand for results with ``NLO correction+effective Wilson coefficients" and ``NLO correction+effective Wilson coefficients+annihilation with strong phase", respectively.}}\label{tab:br010}
\scriptsize{
\vspace{0.2cm}
\begin{tabular}{l|c|c|c|ccc}
\hline \hline
\ \ \ \ \ Mode          & Data\cite{HFAG}  &\multicolumn{5}{c}{This work} \\
\cline{3-7}
                        &                  &NLO+Vertex & $\text{NLO}^{eff}$ &$\text{NLO}^{eff}$($-10^{\circ}$) &$\text{NLO}^{eff}$($5^{\circ}$)&$\text{NLO}^{eff}$($20^{\circ}$)   \\
\hline
$B^+ \to \pi^+ K^0$     & $ 23.1 \pm 1.0 $ &$22.5$ &21.4&19.0&22.6&25.9\\
$B^+\to \pi^0 K^+$      & $ 12.9 \pm 0.6 $ &$12.8$ &12.5 &11.2&13.1&14.9\\
$B^0 \to \pi^- K^+$     & $ 19.4 \pm 0.6 $ &$19.2$ &19.5 &17.4&20.5&23.3\\
$B^0 \to \pi^0 K^0 $    & $ 9.8 \pm 0.6 $ &$8.3$ &8.4&7.4&8.9&10.2\\
\hline
$A_{CP}(\pi^+ K^0)$     & $ 0.009 \pm0.025$     & $-0.006$&-0.006&-0.006&-0.007&-0.007\\
$A_{CP}(\pi^0 K^+)$     & $ 0.050\pm 0.025$     &-0.053& 0.012&0.003&0.018&0.034\\
$A_{CP}(\pi^- K^+)$     & $ -0.098\pm0.012$     & $-0.118$&-0.139&-0.158&-0.131&-0.105\\
$A_{CP}(\pi^0 K^0) $    & $ -0.01\pm 0.10 $     & $-0.052$&-0.139&-0.143&-0.138&-0.137\\
$S_{\pi^0 K_S}$         & $ 0.58\pm 0.17$    & $ 0.699$&0.760&0.768&0.756&0.745\\
\hline \hline
\end{tabular}
}
\end{table}

\begin{table}
\begin{center}
\vspace{-1em} \caption{\small{The same as Table~\ref{tab:br010} but for $B\to \pi\pi, KK$ decay modes.
}}\label{tab:br011} \scriptsize{
\vspace{0.2cm}
\begin{tabular}{l|c|c|c|ccc}
\hline \hline
\ \ \ \ \ Mode          & Data\cite{HFAG}  &\multicolumn{5}{c}{This work} \\
\cline{3-7}
                        &                  &NLO+Vertex & $\text{NLO}^{eff}$ &$\text{NLO}^{eff}$($-40^{\circ}$) &$\text{NLO}^{eff}$($5^{\circ}$)&$\text{NLO}^{eff}$($50^{\circ}$)   \\
\hline
$B^0\to\pi^- \pi^+  $   & $\,\,5.16\pm0.22$& $7.1$&6.5&6.00&$6.6$&7.6\\
$B^+\to\pi^+\pi^0$      & $\,\,5.59\pm 0.40$ & $4.1$&5.5&5.51&$5.5$&5.5\\
$B^0\to\pi^0\pi^0$      & $\,\,1.55\pm0.19$&$0.3$ &1.0&1.11&$1.0$ &1.0  \\
\hline
$B^+\to K^+ \bar{K}^0$  & $1.36\pm 0.28$  &$1.7$&1.6&1.0&$1.7$&2.2\\
$B^0\to K^0 \bar{K}^0$  & $0.96\pm0.20$  &$1.5$&1.4&0.7&$1.5$&2.2\\
$B^0\to K^+K^-  $ & $0.15\pm0.10$  &  $0.09$&0.09&0.09&$0.09$&0.09\\
\hline
$A_{CP}(\pi^- \pi^+)$     & $0.38\pm 0.06$   &0.206     &  $0.266$&0.239&$0.260$&0.141\\
$A_{CP}(\pi^+\pi^0)$      & $0.06\pm0.05$    &   $-0.000$&-0.001&-0.001&$-0.001$&-0.001\\
$A_{CP}(\pi^0\pi^0)$      & $0.43\pm0.25$    &  $0.382$&0.453&0.272&$0.485$&0.789\\
$S_{\pi\pi}$              & $-0.61\pm0.08$   &  $-0.504$&-0.506&-0.353&$-0.524$&-0.638\\
\hline
$A_{CP}(K^+ \bar{K}^0)$  & $0.12\pm 0.17$  &$0.101$&0.098&0.041& $0.101$&0.106\\
$A_{CP}(K^0 \bar{K}^0)$  & $-0.58\pm0.7$   &$0.000$&0.000&0.000&  $0.000$&0.000\\
$A_{CP}(K^+ K^-)$  &  \textbf{-}        & $-0.184$&-0.184&-0.184& $-0.184$&-0.184\\
\hline \hline
\end{tabular}
}
\end{center}
\end{table}

The decay $B\to \pi K$ are dominated by penguin contributions. For the ratios defined by~\cite{pikpuzzle}
\begin{eqnarray}
R_c=\frac{2 Br(B^+\to \pi^0 K^+)}{Br(B^+ \to \pi^+ K^0)},\ \qquad R_n=\frac{Br(B^0 \to \pi^- K^+)}{2Br( B^0 \to \pi^0 K^0)}.
\end{eqnarray}
There is a relation $R_c=R_n\approx1$ if other diagram amplitudes such as annihilation are negligible compared with penguin emission amplitude. The current experimental data are $R_c=1.12\pm0.07$ and $R_n=0.99\pm0.07$. In our framework, we have $R_c=1.15$ and $R_n=1.13$, which are in perfect agreement with the data.

If we only consider the NLO contribution and vertex corrections, the direct CP violations for $B^+\to \pi^0 K^+, \pi^+ K^0$ are wrong in signs compared to the experiments. To improve the predictions, we apply the effective Wilson coefficients to get more reasonable results, and add a strong phase $\theta^{a}_{\pi K}=5^{\circ}$ for annihilation diagrams. From Table~\ref{tab:br010}, it is noted that effective Wilson coefficients and annihilation contributions with strong phase could bring the signs of $A_{CP}(\pi^0 K^+)$ back to the right track, but do nothing with the sign of $A_{CP}(\pi^+ K^0)$.

As for tree diagram dominated channels $B \to \pi\pi, KK$, the contributions of strong phase are ignorable since they are doubly suppressed by CKM and small value of this strong phase. If we want to consider effect of strong phase, a large one such as $-40^{\circ}$ or $50^{\circ}$ are needed here. We list their effects in Table~\ref{tab:br011}.

In order to better test our theoretical framework, we list the most recent predictions based on QCDF with strong phase effects~\cite{Cheng:2009cn} and the predictions from pQCD~\cite{0508041} approach in Tables~\ref{tab:br10} and \ref{tab:br11}. The first theoretical error in our calculations is referred to the global parameters of running energy scale $\mu_{scale}$ and the infrared energy scale $\mu_g$, and the second one is from the shape parameters of light mesons.

\begin{table}[htbp]
\tabcolsep 0.03in\vspace{-0.5em} \caption{\small{Comparisons of predictions between our framework and QCDF, pQCD methods in $B \to \pi\pi, \pi K$ decays.}}\label{tab:br10}\scriptsize{
\begin{tabular}{l|c|c|cc|ccc}
\hline \hline
\ \ \ \ \ Mode          & Data\cite{HFAG}  &QCDF\cite{Cheng:2009cn}&\multicolumn{2}{c|}{pQCD\cite{0508041}}&\multicolumn{3}{c}{This work} \\
\cline{4-8}
                        &                  &           & LO       & NLO+Vertex & LO        & NLO+Vertex &NLO($a^{eff},\theta^{a}$)    \\
\hline
$B^+ \to \pi^+ K^0$     & $ 23.1 \pm 1.0 $ & $21.7^{+9.2+9.0}_{-6.0-6.9}$     & $17.0$    & $24.5^{+13.6\,(+12.9)}_{-\ 8.1\,(-\ 7.8)}$     & 17.5     &$22.5$ &$22.6^{+6.1+9.8}_{-3.5-2.8}$ \\
$B^+\to \pi^0 K^+$      & $ 12.9 \pm 0.6 $ & $12.5^{+4.7+4.9}_{-3.0-3.8}$     & $10.2$    & $13.9^{+10.0\,(+\ 7.0)}_{-\ 5.6\,(-\ 4.2)}$     &10.1     &$12.8$ &$13.1^{+3.7+6.0}_{-2.1-1.3}$ \\
$B^0 \to \pi^- K^+$     & $ 19.4 \pm 0.6 $ & $19.3^{+7.9+8.2}_{-4.8-6.2}$     & $14.2$    & $20.9^{+15.6\,(+11.0)}_{-\ 8.3\,(-\ 6.5)}$     & 14.8     &$19.2$ &$20.5^{+5.2+10.4}_{-3.0-3.0}$ \\
$B^0 \to \pi^0 K^0 $    & $ 9.8 \pm 0.6 $ & $\,\,8.6^{+3.8+3.8}_{-2.2-1.4}$  & $\,\,5.7$ & $\,\,9.1^{+\ 5.6\,(+\ 5.1)}_{-\ 3.3\,(-\ 2.9)}$  & 6.3  &$8.3$ &$8.9^{+2.1+5.0}_{-1.3-1.1}$\\
\hline
$B^0\to\pi^- \pi^+  $   & $\,\,5.16\pm0.22$& $\,\,7.0^{+0.4+0.7}_{-0.7-0.7}$  & $\,\,7.0$ & $\,\,6.5^{+\ 6.7\,(+\ 2.7)}_{-\ 3.8\,(-\ 1.8)}$  & 6.7  &$7.1$&$6.6^{+3.3+1.1}_{-1.3-0.3}$\\
$B^+\to\pi^+\pi^0$      & $\,\,5.59\pm 0.40$ & $\,\,5.9^{+.2+1.4}_{-1.1-.1.1}$  & $\,\,3.5$ & $\,\,4.0^{+\ 3.4\,(+\ 1.7)}_{-\ 1.9\,(-\ 1.2)}$  & 4.3  &$4.1$&$5.5^{+2.3+1.3}_{-1.1-0.4}$\\
$B^0\to\pi^0\pi^0$      & $\,\,1.55\pm0.19$& $\,\,1.1^{+1.0+0.7}_{-0.4-0.3}$ & $\,\,0.12$& $\,\,0.29^{+0.50\,(+0.13)}_{-0.20\,(-0.08)}$ & 0.1  &$0.3$ &$1.0^{+0.3+0.3}_{-0.1-0.1}$   \\
\hline
$A_{CP}(\pi^+ K^0)$     & $ 0.009 \pm0.025$& $0.0028^{+0.0003+0.0009}_{-0.0003-0.0010}$    & $-0.01$   & $-0.01\pm 0.00\,(\pm 0.00)$    &-0.008      & $-0.006$&$-0.007^{+0.002+0.003}_{-0.001-0.013}$\\
$A_{CP}(\pi^0 K^+)$     & $ 0.050\pm 0.025$& $0.049^{+0.039+0.044}_{-0.021-0.054}$    & $-0.08$   & $-0.01^{+0.03\,(+0.03)}_{-0.05\,(-0.05)}$    &-0.107      &-0.053&$0.018^{+0.014+0.022}_{-0.004-0.020}$ \\
$A_{CP}(\pi^- K^+)$     & $ -0.098\pm0.012$& $-0.074^{+0.017+0.043}_{-0.015-0.048}$    & $-0.12$   & $-0.09^{+0.06\,(+0.04)}_{-0.08\,(-0.06)}$    &-0.131     & $-0.118$&$-0.131^{+0.009+0.022}_{-0.003-0.004}$\\
$A_{CP}(\pi^0 K^0) $    & $ -0.01\pm 0.10 $& $-0.106^{+0.027+0.056}_{-0.038-0.043}$    & $-0.02$   & $-0.07^{+0.03\,(+0.01)}_{-0.03\,(-0.01)}$ &-0.001      & $-0.052$&$-0.138^{+0.003+0.004}_{-0.006-0.007}$\\
$S_{\pi^0 K_S}$             & $ 0.58\pm 0.17$  &   \textbf{-}    & $0.70$& $0.73^{+0.03\,(+0.01)}_{-0.02\,(-0.01)}$ & 0.703  & $ 0.699$& $ 0.756^{+0.002+0.002}_{-0.004-0.007}$\\
\hline
$A_{CP}(\pi^- \pi^+)$     & $0.38\pm 0.06$   & $0.170^{+0.013+0.043}_{-0.012-0.087}$   & $0.14$    & $0.18^{+0.20\,(+0.07)}_{-0.12\,(-0.06)}$     & 0.206     &  $0.225$&$0.260^{+0.043+0.059}_{-0.032-0.063}$\\
$A_{CP}(\pi^+\pi^0)$      & $0.06\pm0.05$    & $-0.0002$   & $0.00$    & $0.00\pm 0.00\,(\pm 0.00)$      &0.000          &  $-0.000$&$-0.001^{+0+0}_{-0-0}$\\
$A_{CP}(\pi^0\pi^0)$      & $0.43\pm0.25$    & $0.572^{+0.148+0.303}_{-0.208-0.346}$   & $-0.04$   & $0.63^{+0.35\,(+0.09)}_{-0.34\,(-0.15)}$    &-0.600      & $0.382$&$0.485^{+0.061+0.169}_{-0.025-0.070}$\\
$S_{\pi\pi}$              & $-0.61\pm0.08$   &  \textbf{-}  & $-0.34$ & $-0.43^{+1.00\,(+0.05)}_{-0.56\,(-0.05)}$  &-0.467   &  $-0.504$&$-0.524^{+0.017+0.003}_{-0.004-0.017}$\\
\hline \hline
\end{tabular}
}
\end{table}

\begin{table}[htbp]
\begin{center}
\vspace{-2em} \caption{\small{The same as Table~\ref{tab:br10} but for $B\to KK$ decay modes.}}\label{tab:br11} \scriptsize{
\vspace{0.2cm}
\begin{tabular}{l|c|c|c|ccc}
\hline \hline
\ \ \ \ \ Mode          & Data\cite{HFAG}  &QCDF\cite{Cheng:2009cn}&pQCD&\multicolumn{3}{c}{This work} \\
\cline{4-7}
                        &                  &           & LO       & LO        & NLO+Vertex &NLO($a^{eff},\theta^{a}$)    \\
\hline
$B^+\to K^+ \bar{K}^0$  & $1.36\pm 0.28$  &$1.8^{+0.9+0.7}_{-0.4-0.3}$&  1.65       &1.3     &$1.7$&$1.7^{+0.5+0.4}_{-0.2-0.1}$\\
$B^0\to K^0 \bar{K}^0$  & $0.96\pm0.20$  &$2.1^{+1.0+0.8}_{-0.5-0.5}$&  1.75        &1.2    &$1.5$&$1.5^{+0.4+0.7}_{-0.3-0.2}$\\
$B^0\to K^+K^-  $ & $0.15\pm0.10$  &$0.10^{+0.03+0.03}_{-0.02-0.02}$&  \textbf{-}  &0.08      &  $0.09$&$0.09^{+0.1+0.1}_{-0.0-0.1}$\\
\hline
$A_{CP}(K^+ \bar{K}^0)$  & $0.12\pm 0.17$  &$-0.064^{+0.008+0.018}_{-0.006-0.018}$&   \textbf{-}      &0.130    & $0.101$& $0.101^{+0.017+0.017}_{-0.013-0.069}$\\
$A_{CP}(K^0 \bar{K}^0)$  & $-0.58\pm0.7$   &$-0.100^{+0.007+0.010}_{-0.007-0.019}$&   \textbf{-}   &0.000          &  $0.000$&  $0.000^{+0+0}_{-0-0}$\\
$A_{CP}(K^+ K^-)$  &  \textbf{-}        &   \textbf{-}   &   \textbf{-}     &-0.182       & $-0.184$& $-0.184^{+0.050+0.051}_{-0.020-0.012}$\\
\hline \hline
\end{tabular}
}
\end{center}
\end{table}

\subsection{$B \to PV$ decays}

Power corrections to $a_i$ for $B \to PV, VV$ are not the same as that for $B \to PP$ as described by Eq.~(\ref{aeff}). From Tables~\ref{tab:br021}-\ref{tab:br22}, we can see that an enhancement of $a_i$ is needed to improve the rates of $B\to \rho K, \omega K, \rho^0\pi^0$. However, it is constrained by the measured rates of $\rho^+\pi^0$ and $\rho^+\pi^-$ modes. Because of that, with the requirement of PV final states, we use smaller Wilson coefficient if the emission meson in factorizable emission diagram is a vector.

\begin{table}[t]
\begin{center}
\caption{\small{The branching ratios~(in units of $10^{-6}$) and direct CP asymmetries in penguin dominated $B\to PV$ decays. The other captions are the same as Table~\ref{tab:br010}. }}\label{tab:br021}.
\scriptsize{
\vspace{0.2cm}
\begin{tabular}{l|c|ccccc}
\hline \hline
\ \ \ \ \ Mode          & Data\cite{HFAG}  &\multicolumn{5}{c}{This work} \\
\cline{3-7}
                        &                  &NLO+Vertex & $\text{NLO}^{eff}$ &$\text{NLO}^{eff}$($-10^{\circ}$) &$\text{NLO}^{eff}$($5^{\circ}$)&$\text{NLO}^{eff}$($20^{\circ}$)   \\
\hline
$B^+\to K^{*0} \pi^+$   & $9.9\pm0.8$           &$10.3$&9.0&7.6&9.8&11.8\\
$B^+\to K^{*+} \pi^0$   & $6.9\pm2.3$             &$6.2$&5.5&4.8&5.9&7.1\\
$B^0\to K^{*-} \pi^+$   & $8.6\pm0.9$            &$8.8$&8.3&7.2&8.9&10.6\\
$B^0\to K^{*0} \pi^0$   & $2.4\pm0.7$          &$3.5$&3.6&3.1&3.9&4.6\\
\hline
$B^+\to \phi K^+$       & $8.30\pm0.65$          &$9.3$&6.9&5.6&7.6&9.6\\
$B^0\to \phi K^0$       & $8.3\pm1.1$             &$8.9$&6.6&5.4&7.3&9.2\\
\hline
$A_{CP}(K^{*0} \pi^+)$   & $-0.038\pm0.042$     &$-0.017$&-0.018&-0.020&-0.017&-0.015\\
$A_{CP}(K^{*+} \pi^0)$   & $0.04\pm0.29$     &$-0.224$&-0.123&-0.164&-0.103&-0.045\\
$A_{CP}(K^{*-} \pi^+)$   & $-0.23\pm0.08$     &$-0.357$&-0.355&-0.415&-0.327&-0.251\\
$A_{CP}(K^{*0} \pi^0)$   & $-0.15\pm0.12$   &$-0.067$&-0.125&-.114&-0.129&-0.143\\
\hline
$A_{CP}(\phi K^+)$       & $0.23\pm0.15$      &$-0.022$&-0.025&-0.028&-0.023&-0.020\\
$A_{CP}(\phi K^0)$       & $-0.01\pm0.06$        &$0$&0&0&0&0\\
\hline\hline
\end{tabular}
}
\end{center}
\end{table}

\begin{table}[t]
\begin{center}
\caption{\small{The same as Table~\ref{tab:br021}. }} \label{tab:br022}. \scriptsize{
\vspace{0.2cm}
\begin{tabular}{l|c|ccccc}
\hline \hline
\ \ \ \ \ Mode          & Data\cite{HFAG}  &\multicolumn{5}{c}{This work} \\
\cline{3-7}
                        &                  &NLO+Vertex & $\text{NLO}^{eff}$ &$\text{NLO}^{eff}$($45^{\circ}$) &$\text{NLO}^{eff}$($60^{\circ}$)&$\text{NLO}^{eff}$($75^{\circ}$)   \\
\hline
$B^+\to \rho^+ K^0$     & $8.0\pm1.45$           &$5.2$&7.1&7.1&6.8&6.3\\
$B^+\to \rho^0 K^+$     & $3.81\pm0.48$          &$3.0$&3.3&2.8&2.6&2.4\\
$B^0\to \rho^- K^+$     & $8.6\pm1.0$            &$5.4$&6.2&7.3&7.3&7.2\\
$B^0\to \rho^0 K^0$     & $4.7\pm0.7$            &$2.8$&3.9&4.9&5.0&5.0\\
\hline
$B^+\to \omega K^+$     & $6.7\pm0.5$       &2.4       &$3.6$&4.3&5.3&5.2\\
$B^0\to \omega K^0$     & $5.0\pm0.6$      &1.9       &$3.2$&4.1&4.8&4.6\\
\hline
$A_{CP}(\rho^+ K^0)$     & $-0.12\pm0.17$      &$0.016$&0.013&0.014&0.014&0.014\\
$A_{CP}(\rho^0 K^+)$     & $0.37\pm0.11$       &$0.635$&0.727&0.594&0.463&0.285\\
$A_{CP}(\rho^- K^+)$     & $0.15\pm0.06$       &$0.605$&0.549&0.373&0.290&0.196\\
$A_{CP}(\rho^0 K^0)$     & $0.06\pm0.20$     &$0.056$&-0.136&-0.044&-0.015&0.015\\
\hline
$A_{CP}(\omega K^+)$     & $0.02\pm0.05$      &$0.453$&0.404&0.167&0.091&0.015\\
$A_{CP}(\omega K^0)$     & $0.32\pm0.17$    &$-0.011$&0.117&0.048&0.026&0.002\\
\hline\hline
\end{tabular}
}
\end{center}
\end{table}

\begin{table}
\begin{center}
\caption{\small{The same as Table~\ref{tab:br021} but for tree dominant $B\to PV$ decay modes.}} \label{tab:br023}. \scriptsize{
\vspace{0.2cm}
\begin{tabular}{l|c|cccccc}
\hline \hline
\ \ \ \ \ Mode          & Data\cite{HFAG}  &\multicolumn{6}{c}{This work} \\
\cline{3-8}
                        &                  &NLO+Vertex  & \multicolumn{5}{c}{$\text{NLO}^{eff}$}   \\
\cline{4-8}
                        &                  &            & default &($-45^{\circ}$,$0^{\circ}$) &($45^{\circ}$,$0^{\circ}$)&($0^{\circ}$,$-45^{\circ}$) &($0^{\circ}$,$45^{\circ}$)  \\
\hline
$B^+\to \rho^+ \pi^0$   & $10.9\pm1.5$      &$12.0$ &13.9&14.2& 13.5 &13.7&14.2\\
$B^+\to \rho^0 \pi^+$   & $8.3\pm1.3$       &$5.2$   &7.4&$7.0$&7.8&7.2&7.4\\
$B^0\to\rho^+\pi^-$&    $15.7\pm1.8$    &$19.6$  &17.4&$16.5$&19.1&17.5&17.4\\
$B^0\to\rho^-\pi^+$&    $7.3\pm1.2$    &$6.2$  &6.5&$6.5$&6.5&6.1&7.5\\
$B^0\to \rho^0 \pi^0  $ & $2.0\pm0.5$       &$0.2$ &1.3&$1.5$ &1.2 &1.5&1.1\\
\hline
$B^+\to \bar{K}^{*0} K^+$       & $0.68\pm0.19$     &$0.6$&0.3&$0.3$&0.3&0.3&0.3\\
$B^0\to K^{*0} \bar{K}^0$       & $<1.9$       &$0.6$&0.4&$0.2$&0.8&0.8&0.1\\
\hline
$A_{CP}(\rho^+ \pi^0)$   & $0.02\pm0.11$     & $0.255$  &0.199&$0.196$ &0.133&0.195 &0.131\\
$A_{CP}(\rho^0 \pi^+)$   & $0.18^{+0.09}_{-0.17}$    & $-0.308$  &-0.344&$-0.330$  &-0.269 &-0.309&-0.285\\
$A_{CP}(\rho^+\pi^-)$&    $0.11\pm0.06$     &$0.120$ &0.126&$0.108$&0.066 &0.121&0.127\\
$A_{CP}(\rho^-\pi^+)$&    $-0.18\pm0.12$     &$-0.281$  &-0.282&$-0.283$&-0.281&-0.217&-0.176\\
$A_{CP}(\rho^0 \pi^0  )$ & $-0.30\pm0.38$      &$0.058$ &0.187&$0.112$&0.381&0.258&-0.008\\
\hline
$A_{CP}(\bar{K}^{*0} K^+)$       & -        &$0.191$&0.257&$-0.342$&0.205&0.112&0.837\\
$A_{CP}(k^{*0} \bar{K}^0)$       & -         &$0.000$&0&$0$&0&0&0\\
\hline\hline
\end{tabular}
}
\end{center}
\end{table}

The decay modes $B\to \pi K^*, \rho K , \omega K, \phi K$ are all penguin dominant channels. Just as argued in Ref.~\cite{xnlPV}, contribution of annihilation diagram and NLO correction have been known to be able to remarkably improve theory prediction. The predictions in our framework also support such an argument, contributions from NLO are helpful for our theoretical predictions while still smaller than that in pQCD~\cite{xnlPV}. This may be due to the following fact: the system scale we use is higher than the stop scale used in pQCD~\cite{xnlPV}~(when running scale is lower than this scale, pQCD uses all running parameters in stop scale for present scale and then this stop scale is more or less the typical scale of pQCD calculation); as a result the NLO correction in our scale is smaller than that in pQCD framework.

For the case of $B\to K \phi$ or $B\to \pi K^*$ decays, contributions of annihilation diagrams are more important than that in PP modes. This is due to the fact that emission penguin amplitude does not receive contribution of $a_6$ but only from $a_4$. Our predictions without annihilation strong phase seem to  approach the data, while the CP violation is not so good. Therefore, a small strong phase $\theta^{a}=5^{\circ}$ is enough just like in $\pi K$ decays.

Predictions of $B\to\rho K$ decay modes are smaller than experimental data, because emission amplitudes of these channels are attributed to the destructive combination of the Wilson coefficients $a_4-2\mu_K a_6$,  and hence are almost canceled out; the decay amplitudes are therefore dominated by annihilation contributions. Adding a strong phase $\theta^{a}=60^{\circ}$ for annihilation amplitude will remarkably improve the branching ratio, especially the CP asymmetry. Moreover, the difference between Br($\rho^+ K^0$) and Br($\rho^- K^+$) is accommodated with this effect.

The case of $B\to\omega K$ is similar to $B\to\rho K$, where emission diagram is also almost canceled out. In pQCD framework~\cite{xnlPV}, vertex correction to $a_5$ might enhance branching ratio of this channel in lower scale diagram $P_c$~(color-suppressed penguin diagram), but this enhancement is not obviously in our framework since we use uniform scale for all diagrams. If considering correction of effective Wilson coefficient, we could note that the effect of larger $a_5$ is obvious. Just as in $B\to\rho K$ decay mode, we add a strong phase for annihilation amplitude to improve predictions of branching ratios and CP violations.

\begin{table}
\begin{center}
\tabcolsep 0.025in\vspace{-2em}
\caption{\small{The branching ratios~(in units of $10^{-6}$) and CP asymmetries for tree dominant $B\to PV$decays. Predictions of pQCD method correspond to NLO and vertex correction results~(LO results are listed in parenthesis). Here we use strong phase $\theta^{a}_{\pi \rho}$=$5^{\circ}$ and $\theta^{a}_{\rho \pi}$=$60^{\circ}$.}}\label{tab:br20} \scriptsize{
\begin{tabular}{l|c|c|c|ccc}
\hline \hline
\ \ \ \ \ Mode          & Data\cite{HFAG}   &QCDF\cite{Cheng:2009cn}&pQCD(LO)\cite{xnlPV,cdlu}&\multicolumn{3}{c}{this work}\\
\cline{5-7}
                        &                   &               &                       &LO                        & NLO+Vetex &NLO($a^{eff}$)\\
\hline
$B^+\to \rho^+ \pi^0$   & $10.9\pm1.5$      &$11.8^{+1.8+1.4}_{-1.1-1.4}$         &    $6\thicksim9$                  &12.0       &$12.0$ &$13.9^{+5.7+2.8}_{-3.2-0.9}$  \\
$B^+\to \rho^0 \pi^+$   & $8.3\pm1.3$       &$8.7^{+2.7+1.7}_{-1.3-.14}$         &     $5\thicksim6$                    &5.4       &$5.2$   &$7.4^{+3.7+1.0}_{-1.9-0.2}$\\
$B^0\to\rho^+\pi^-$&    $15.7\pm1.8$    &$15.9^{+1.1+0.9}_{-1.5-1.1}$         &                                   &18.6       &$19.6$  &$17.4^{+8.2+3.2}_{-4.2-0.9}$\\
$B^0\to\rho^-\pi^+$&    $7.3\pm1.2$    &$9.2^{+0.4+0.5}_{-0.7-0.7}$         &                                   &6.9       &$6.2$  &$6.5^{+4.3+0.1}_{-2.0-0.0}$\\
$B^0\to \rho^0 \pi^0  $ & $2.0\pm0.5$       &$1.3^{+1.7+1.2}_{-0.6-0.6}$          &        $0.07\thicksim0.11$            &0.2       &$0.2$ &$1.3^{+0.3+0.4}_{-0.2-0.2}$  \\
\hline
$B^+\to \bar{K}^{*0} K^+$       & $0.68\pm0.19$     &$0.80^{+0.20+0.31}_{-0.17-0.28}$   &$0.32^{+0.12}_{-0.07} $ &0.3       &$0.6$&$0.3^{+0.0+0.1}_{-0.0-0.2}$\\
$B^0\to K^{*0} \bar{K}^0$       & $<1.9$       &$0.47^{+0.36+0.43}_{-0.17-0.27}$          &$0.49^{+0.15}_{-0.09} $ &0.3       &$0.6$&$0.4^{+0.1+0.2}_{-0.1-0.2}$\\
\hline
$A_{CP}(\rho^+ \pi^0)$   & $0.02\pm0.11$     &$0.097^{+0.021+0.080}_{-0.031-0.103}$        &        $0\thicksim20$                 &0.251&  $0.255$  &$0.199^{+0.027+0.016}_{-0.044-0.053}$  \\
$A_{CP}(\rho^0 \pi^+)$   & $0.18^{+0.09}_{-0.17}$    &$-0.098^{+0.034+0.114}_{-0.026-0.104}$        &   $-20\thicksim0$         &-0.351& $-0.308$  &$-0.344^{+0.062+0.023}_{-0.41-0.086}$   \\
$A_{CP}(\rho^+\pi^-)$&    $0.11\pm0.06$    &$0.044^{+0.003+0.058}_{-0.003-0.068}$         &                                   &0.113       &$0.120$ &$0.126^{+0.015+0.004}_{-0.023-0.014}$ \\
$A_{CP}(\rho^-\pi^+)$&    $-0.18\pm0.12$    &$-0.227^{+0.009+0.082}_{-0.011-0.044}$         &                                   &-0.225       &$-0.281$  &$-0.284^{+0.064+0.045}_{-0.045-0.047}$\\
$A_{CP}(\rho^0 \pi^0  )$ & $-0.30\pm0.38$     &$0.110^{+0.050+0.235}_{-0.057-0.288}$        &      $-75\thicksim0$                  &0.048&  $0.058$ &$0.187^{+0.004+0.012}_{-0.001-0.007}$\\
\hline
$A_{CP}(\bar{K}^{*0} K^+)$       & -     &$-0.089^{+0.011+0.028}_{-0.011-0.024}$   &$-0.069^{+0.056+0.010+0.092+0.040}_{-0.053-0.003-0.065-0.0060} $ &  0.360     &$0.191$&$0.257^{+0.039+0.023}_{-0.042-0.019}$\\
$A_{CP}(k^{*0} \bar{K}^0)$       & -       &$-0.035^{+0.013+0.007}_{-0.017-0.020}$          &- &   0.000    &$0.000$&$0^{+0+0}_{-0-0}$\\
\hline \hline
\end{tabular}
}
\end{center}
\end{table}

\begin{table}
\begin{center}
\tabcolsep 0.03in\vspace{-2em}
\caption{\small{The branching ratios(in units of $10^{-6}$) and CP asymmetries for penguin dominant $B\to PV$decays.  Predictions of pQCD method correspond to NLO and vertex correction results~(LO results are listed in parenthesis ). Here we use strong phase $\theta^{a}_{\pi K^*}$=$\theta^{a}_{K \phi}$=$5^{\circ}$.}}\label{tab:br21}. \scriptsize{
\begin{tabular}{l|c|c|c|ccc}
\hline \hline
\ \ \ \ \ Mode          & Data\cite{HFAG}   &QCDF\cite{Cheng:2009cn}&pQCD(LO)\cite{xnlPV}&\multicolumn{3}{c}{this work}\\
\cline{5-7}
                         &                   &               &                &LO      & NLO+Vertex &NLO($a^{eff},\theta^{a}$)  \\
\hline
$B^+\to K^{*0} \pi^+$   & $9.9\pm0.8$      &$10.4^{+1.3+4.3}_{-1.5-3.9}$          &$6.0^{+2.8+2.7}_{-1.5-1.4}(5.5)$   &7.5       &$10.3$&$9.8^{+2.8+2.5}_{-1.7-0.9}$\\
$B^+\to K^{*+} \pi^0$   & $6.9\pm2.3$       &$6.7^{+0.7+2.4}_{-0.7-2.2}$          &$4.3^{+5.0+1.7}_{-2.2-1.0}(4.0)$   &4.7       &$6.2$&$5.9^{+2.0+1.4}_{-1.1-0.3}$\\
$B^0\to K^{*-} \pi^+$   & $8.6\pm0.9$      &$9.2^{+1.0+3.7}_{-1.0-3.3}$          &$6.0^{+6.8+2.4}_{-2.6-1.3}(5.1)$   &6.5       &$8.8$&$8.9^{+2.4+2.4}_{-1.4-1.0}$\\
$B^0\to K^{*0} \pi^0$   & $2.4\pm0.7$       &$3.5^{+0.4+1.6}_{-0.4-1.4}$          &$2.0^{+1.2+0.9}_{-0.6-0.4}(1.5)$   &2.5       &$3.5$&$3.9^{+0.8+1.1}_{-0.6-0.3}$\\
\hline
$B^+\to \phi K^+$       & $8.30\pm0.65$     &$8.8^{+2.8+4.4}_{-2.7-3.6}$          &$7.8^{+5.9+5.8}_{-1.8-1.7}(13.8) $ &10.8       &$9.3$&$7.6^{+1.8+0.6}_{-1.4-0.6}$\\
$B^0\to \phi K^0$       & $8.3\pm1.1$       &$8.1^{+2.6+4.4}_{-2.5-3.3}$          &$7.3^{+5.4+5.1}_{-1.8-1.5}(12.9) $ &10.4       &$8.9$&$7.3^{+1.6+0.5}_{-1.3-0.7}$\\
\hline
$A_{CP}(K^{*0} \pi^+)$   & $-0.038\pm0.042$   &$0.004^{+0.013+0.043}_{-0.016-0.039}$        &$-0.01^{+0.01+0.01}_{-0.00-0.00}(-0.03) $&-0.021    &$-0.017$&$-0.017^{+0.003+0.012}_{-0.002-0.003}$\\
$A_{CP}(K^{*+} \pi^0)$   & $0.04\pm0.29$     &$0.016^{+0.031+0.111}_{-0.017-0.144}$       &$-0.32^{+0.21+0.16}_{-0.28-0.19}(-0.38) $&-0.348   &$-0.224$&$-0.103^{+0.055+0.081}_{-0.027-0.016}$\\
$A_{CP}(K^{*+} \pi^-)$   & $-0.23\pm0.08$    &$-0.121^{+0.005+0.126}_{-0.005-0.160}$       &$-0.60^{+0.32+0.20}_{-0.19-0.15}(-0.56) $&-0.443   &$-0.357$&$-0.327^{+0.020+0.070}_{-0.008-0.016}$\\
$A_{CP}(K^{*0} \pi^0)$   & $-0.15\pm0.12$    &$-0.108^{+0.018+0.091}_{-0.028-0.063}$        &$-0.11^{+0.07+0.05}_{-0.05-0.02}(-0.60) $&0.012   &$-0.067$&$-0.130^{+0.016+0.004}_{-0.028-0.019}$\\
\hline
$A_{CP}(\phi K^+)$       & $0.23\pm0.15$    &$0.006^{+0.001+0.001}_{-0.001-0.001}$        &$0.01^{+0.00+0.00}_{-0.01-0.01}(-0.02)  $&-0.022    &$-0.022$&$-0.023^{+0.004+0.002}_{-0.002-0.014}$\\
$A_{CP}(\phi K^0)$       & $-0.01\pm0.06$    &$0.009^{+0.001+0.001}_{-0.001-0.001}$        &$0.03^{+0.01+0.00}_{-0.02-0.01}(0.00)   $&0        &$0$&$0^{+0+0}_{-0-0}$\\
\hline\hline
\end{tabular}
}
\end{center}
\end{table}

\begin{table}
\begin{center}
\tabcolsep 0.03in\vspace{-2em}
\caption{\small{The same as Table~\ref{tab:br21} but with $\theta^{a}_{\omega K}$=$\theta^{a}_{\rho K}$=$60^{\circ}$.}}\label{tab:br22}. \scriptsize{
\begin{tabular}{l|c|c|c|ccc}
\hline \hline
\ \ \ \ \ Mode          & Data\cite{HFAG}   &QCDF\cite{Cheng:2009cn}&pQCD(LO)\cite{xnlPV}&\multicolumn{3}{c}{this work}\\
\cline{5-7}
                         &                   &               &                &LO      & NLO+Vertex &NLO($a^{eff},\theta^{a}$)  \\
\hline
$B^+\to \rho^+ K^0$     & $8.0\pm1.45$      &$7.8^{+6.3+7.3}_{-2.9-4.4}$          &$8.7^{+6.8+6.4}_{-4.4-4.3}(3.6)$   &4.2        &$5.2$&$6.8^{+0.3+1.2}_{-0.2-1.1}$\\
$B^+\to \rho^0 K^+$     & $3.81\pm0.48$     &$3.5^{+2.9+2.9}_{-1.2-1.8}$          &$5.1^{+4.1+3.6}_{-2.8-2.6}(2.5)$   &2.3        &$3.0$&$2.6^{+0.3+0.4}_{-0.2-1.0}$\\
$B^0\to \rho^- K^+$     & $8.6\pm1.0$       &$8.6^{+5.7+7.4}_{-2.8-4.5}$         &$8.8^{+6.8+6.4}_{-4.4-4.3}(4.7)$   &4.9        &$5.4$&$7.3^{+0.8+2.0}_{-0.5-0.4}$\\
$B^0\to \rho^0 K^0$     & $4.7\pm0.7$       &$5.4^{+3.4+4.3}_{-1.7-2.8}$          &$4.8^{+4.3+3.2}_{-2.3-2.0}(2.5)$   &2.7        &$2.8$&$5.0^{+0.6+1.7}_{-0.4-0.3}$\\
\hline
$B^+\to \omega K^+$     & $6.7\pm0.5$       &$4.8^{+4.4+3.5}_{-1.9-2.3}$          &$10.6^{+10.4+7.2}_{-5.8-4.4}(2.1)$ &2.4       &$3.6$&$5.3^{+0.6+1.2}_{-0.4-0.5}$\\
$B^0\to \omega K^0$     & $5.0\pm0.6$       &$4.1^{+4.2+3.3}_{-1.7-2.2}$          &$9.8^{+8.8+8.7}_{-4.9-4.3}(1.9) $  &1.9       &$3.2$&$4.8^{+0.1+1.1}_{-0.3-0.5}$\\
\hline
$A_{CP}(\rho^+ K^0)$     & $-0.12\pm0.17$    &$0.003^{+0.002+0.005}_{-0.003-0.002}$        &$0.01^{+0.01+0.01}_{-0.01-0.01}(0.02)   $&0.019    &$0.016$&$0.014^{+0.001+0.017}_{-0.001-0.015}$\\
$A_{CP}(\rho^0 K^+)$     & $0.37\pm0.11$     &$0.454^{+0.178+0.314}_{-0.194-0.232}$        &$0.71^{+0.25+0.17}_{-0.35-0.14}(0.79)   $&0.726    &$0.635$&$0.463^{+0.041+0.025}_{-0.036-0.014}$\\
$A_{CP}(\rho^- K^+)$     & $0.15\pm0.06$     &$0.319^{+0.115+0.196}_{-0.110-0.127}$        &$0.64^{+0.24+0.07}_{-0.30-0.11}(0.83)   $&0.593    &$0.605$&$0.290^{+0.021+0.012}_{-0.020-0.007}$\\
$A_{CP}(\rho^0 K^0)$     & $0.06\pm0.20$     &$0.087^{+0.012+0.087}_{-0.012-0.068}$       &$0.07^{+0.08+0.07}_{-0.05-0.04}(0.07)   $&-0.040   &$0.056$&$-0.015^{+0.005+0.025}_{-0.002-0.021}$\\
\hline
$A_{CP}(\omega K^+)$     & $0.02\pm0.05$     &$0.221^{+0.137+0.140}_{-0.128-0.130}$        &$0.32^{+0.15+0.04}_{-0.17-0.05}(0.32)   $&0.688    &$0.453$&$0.091^{+0.031+0.020}_{-0.038-0.073}$\\
$A_{CP}(\omega K^0)$     & $0.32\pm0.17$   &$-0.047^{+0.018+0.055}_{-0.016-0.058}$        &$-0.03^{+0.02+0.02}_{-0.04-0.03}(-0.03) $&0.065    &$-0.011$&$0.026^{+0.001+0.005}_{-0.001-0.024}$\\
\hline\hline
\end{tabular}
}
\end{center}
\end{table}

\subsection{$B\to VV$ Decays}

Naive factorization without annihilation contribution predicts a longitudinal polarization fraction near 100\% for all $B\to VV $ decay modes, while the polarization anomaly~($f_L$ is about 50\%) is observed by the BarBar~\cite{babar}, Belle~\cite{belle} and CDF~\cite{cdf} experiments. Furthermore, experimental data on $B \to \rho\rho$ channels show that Br($\rho^+\rho^-$)$\sim$ Br($\rho^+\rho^0$) which breaks the factorization power-counting while the power-counting Br($\rho^+\rho^-$)$\gg$Br($\rho^0\rho^0$) still satisfies. Motivated by the anomaly and the unsatisfied power counting relations, we shall study in detail the polarization in $B\to VV$ decays, especially in $B\to \rho K^*$ and $\phi K^*$ decays, and the branching ratios of $B\to \rho \rho$ modes in this section. The relevant predictions are listed in Tables~\ref{tab:br030}-\ref{tab:br32}.

\begin{table}
\begin{center}
\caption{\small{Branching ratios for $B\to VV$ decay modes~(in unit of $10^{-6}$) which includes the contribution of effective Wilson coefficients and effect of different strong phase $\theta^{a}=60^{\circ}\pm15^{\circ}$) for annihilation diagram.}}\label{tab:br030}. \scriptsize{
\begin{tabular}{l|c|ccccc}
\hline \hline
\ \ \ \ \ Mode          & Data\cite{HFAG}  &\multicolumn{5}{c}{This work} \\
\cline{3-7}
                        &                  &NLO+Vertex & $\text{NLO}^{eff}$ &$\text{NLO}^{eff}$($45^{\circ}$) &$\text{NLO}^{eff}$($60^{\circ}$)&$\text{NLO}^{eff}$($75^{\circ}$)   \\
\hline
$B^+\to \rho^+ \rho^0$  & $24.0\pm2.0$             &$13.4$&16.8&16.8&16.8&16.8\\
$B^0\to \rho^+ \rho^-$  & $24.2\pm3.1$           &$22.3$ &19.8&21.7&22.3&22.7\\
$B^0\to \rho^0 \rho^0$  & $0.73\pm0.27$        &$0.4$ &0.92&0.67&0.61&0.57\\
\hline
$B^+\to K^{*0} \rho^+$  & $9.2\pm1.5$                        &$16.2$&14.0&9.6&8.3&7.2\\
$B^+\to K^{*+} \rho^0$  & $<6.1$                             &$9.9$&9.0&6.4&5.6&5.0\\
$B^0\to K^{*+} \rho^-$   & $<12$                 &$13.9$&13.0&9.1&7.9&6.9\\
$B^0\to K^{*0} \rho^0$  & $3.4\pm1.0$                &$5.6$&5.2&3.6&3.1&2.7\\
\hline
$B^+\to \bar{K}^{*0} K^{*+}$& $1.2\pm 0.5$                   &$0.9$  & 0.8&0.6&0.5&0.4\\
$B^0\to \bar{K}^{*0} K^{*0}$& $1.28\pm0.35$                 &$0.8$ &0.6&0.5&0.5&0.5\\
$B^0\to K^{*+} K^{*-}$& $<2$  &                 $0.07$&0.07&0.07&0.007&0.07\\
\hline
$B^+\to \phi K^{*+}$       & $10.0\pm1.1$          &$19.4$&15.2&10.9&9.5&8.4\\
$B^0\to \phi K^{*0}$       & $9.8\pm0.7$      &$18.7$&14.8&10.5&9.2&8.1\\
\hline
$B^+\to \omega K^{*+}$       & $<7.4$                 &$5.6$&4.2&3.3&3.0&2.8\\
$B^0\to \omega K^{*0}$       & $2.0\pm0.5$      &$6.2$&4.1&2.8&2.5&2.2\\
\hline \hline
\end{tabular}
}
\end{center}

\end{table}

\begin{table}
\begin{center}
\caption{\small{The same as Table~\ref{tab:br030} but for the longitudinal polarization fraction.}} \label{tab:br031}. \scriptsize{
\begin{tabular}{l|c|ccccc}
\hline \hline
\ \ \ \ \ Mode          & Data\cite{HFAG}  &\multicolumn{5}{c}{This work} \\
\cline{3-7}
                        &                  &NLO+Vertex & $\text{NLO}^{eff}$ &$\text{NLO}^{eff}$($45^{\circ}$) &$\text{NLO}^{eff}$($60^{\circ}$)&$\text{NLO}^{eff}$($75^{\circ}$)   \\
\hline
$f_L(\rho^+ \rho^0)$  & $0.950\pm0.016$           &$0.94$&0.92&0.95&0.95&0.95\\
$f_L(\rho^+ \rho^-)$  & $0.978\pm0.023$               &$0.95$&0.95&0.95&0.95&0.95\\
$f_L(\rho^0 \rho^0)$  & $0.75\pm0.15$                      &$0.84$ &0.86&0.77&0.74&0.71\\
\hline
$f_L(K^{*0} \rho^+)$  & $0.48\pm0.08$                      &$0.85$&0.82&0.57&0.45&0.32\\
$f_L(K^{*+} \rho^0)$  & $0.96^{+0.06}_{-0.16}$             &$0.86$&0.85&0.65&0.56&0.47\\
$f_L(K^{*-} \rho^+)$  & -                             &$0.81$&0.80&0.57&0.46&0.34\\
$f_L(K^{*0} \rho^0)$  & $0.57\pm0.12$                  &$0.78$&0.75&0.48&0.36&0.22\\
\hline
$f_L(\bar{K}^{*0} K^{*+})$& $0.75^{+0.16}_{-0.26}$            &$0.85$&0.81&0.60&0.49&0.37\\
$f_L(\bar{K}^{*0} K^{*0})$& $0.80\pm0.13$                     &$0.83$&0.63&0.60&0.53&0.46\\
$f_L(K^{*+} K^{*-})$&                       &$0.99$&0.99&0.99&0.99&0.99\\
\hline
$f_L(\phi K^{*+})$       & $0.50\pm0.05$                       &$0.87$&0.83&0.58&0.45&0.31\\
$f_L(\phi K^{*0})$       & $0.480\pm0.030$    &$0.87$&0.83&0.58&0.45&0.31\\
\hline
$f_L(\omega K^{*+})$       & $0.41\pm0.19$            &$0.90$ &0.86&0.68&0.58&0.48\\
$f_L(\omega K^{*0})$       & $0.70\pm0.13$         &$0.93$&0.89&0.67&0.53&0.37\\
\hline \hline
\end{tabular}
}
\end{center}

\end{table}
Before moving ahead, one point that should be noted is that we use a bigger gluon infrared cut-off in annihilation diagram, $\widetilde {\mu}_g$ =0.52GeV $\simeq\sqrt{2}\mu_g$, which is constrained by the huge penguin diagram contributions.

It is realized that the trick used in $\pi\pi$ puzzles is no longer useful here. If we add  effective $a_{1,2}$ just as we do in $\pi\pi$ mode, it is only helpful for branching ratios of $\rho^+\rho^-$ and $\rho^+\rho^0$ modes, but at the cost of overflowing that of $\rho^0\rho^0$. Branching ratios of $B\to \rho^0\rho^0$ can be decreased by introducing proper strong phase~(such as $\theta^{a} = 60^{\circ}$) in annihilation amplitudes, while be increased by considering effective Wilson coefficients $a^{eff'}_{1,2}$. Moreover, these two elements make predictions of branching ratios of $B \to \rho\rho, \rho K^*, \phi K^*$ and $\omega K^*$ to be better consistent with the experimental data. The effect of different strong phases to branching ratio, longitudinal polarization and direct CP asymmetry are listed in Tables~\ref{tab:br030}-\ref{tab:br032}. As for the channel $B\to K^*K^*$, predictions of the branching ratios are smaller than the experimental data. The strong phase effect does not increase the predictions, so the prediction without strong phase are better than the one with it.

Another important point should be noted that the predictions for the branching ratios of $B\to \rho K^*$, $\phi K^*$ and $\omega K^*$ modes are all bigger than the experimental data, which means that if we want to solve the observed polarization anomaly, we need to find some way to reduce the longitudinal amplitude and enhance the transverse ones simultaneously. Many studies have been made to try to provide possible resolutions to the anomaly both within the SM~\cite{Kagan:2004uw,Li:2004ti,Colangelo:2004rd,Beneke:2006hg} and in various new physics models~\cite{Bao:2008hd,Yang:2004pm,Chen:2005mka,Chang:2006dh}.

In our framework, sizable annihilation contributions appear in all the three parts of the amplitudes: longitudinal, parallel and transverse. In order to get a reconciled prediction for the polarization part, we shall add an extra strong phase~($\theta^{a} = 60^{\circ}$) in all annihilation diagrams. As a result, the longitudinal amplitude can be partly counteracted by annihilation contributions, while the parallel and transverse ones are not be influenced much. From Table~\ref{tab:br031}, we can see that our predictions for the longitudinal polarization fraction are generally consistent with the current data when considering complex annihilation contributions and effective Wilson coefficients.

In $B\to VV$ decays, there are another two interesting observables $\phi_{\parallel}$ and $\phi_{\perp}$, which are defined to be the relative phase between the parallel and the longitudinal amplitudes and that between the transverse and the longitudinal amplitudes, respectively. The corresponding definition can be found in Ref.~\cite{Beneke:2006hg}. To compare with the data, we only discuss $B\to \phi K^{*+}, \phi K^{*0}$ decay modes. Since $\phi_\parallel$ is the same as $\phi_\perp$, we simply write them as $\phi$ and the numerical results are as follows:
\begin{eqnarray}
\text{NLO+Vertex}:&\phi(\phi K^{*+})=99.0^{\circ},&\phi(\phi K^{*0})=100.7^{\circ},\nonumber\\
\text{with }\theta^{a}:&\phi(\phi K^{*+})=63.7^{\circ},&\phi(\phi K^{*0})=66.1^{\circ},\nonumber\\
\text{with }\theta^{a} {\rm and}~ a^{eff'}:&\phi(\phi K^{*+})=73.1^{\circ},&\phi(\phi K^{*0})=75.9^{\circ},\nonumber\\
\text{QCDF}\cite{Cheng:2009cn}:&\phi(\phi K^{*+})=(80^{+43}_{-83})^{\circ},&\phi(\phi K^{*0})=(78^{+54}_{-81})^{\circ},\nonumber\\
\text{expt}:&\phi_{\parallel}(\phi K^{*+})=(46\pm10)^{\circ},&\phi_{\parallel}(\phi K^{*0})=(44^{+8}_{-7})^{\circ},\nonumber\\
&\phi_{\perp}(\phi K^{*+})=(40\pm10)^{\circ},&\phi_{\perp}(\phi K^{*0})=(43\pm7)^{\circ}.
\end{eqnarray}
It is noted that, although our predictions are bigger than the data, they are roughly consistent with the results in QCDF~\cite{Cheng:2008gxa}.

\begin{table}
\begin{center}
\caption{The same as Table~\ref{tab:br030} but for direct CP asymmetries. }\label{tab:br032}. \scriptsize{
\begin{tabular}{l|c|ccccc}
\hline \hline
\ \ \ \ \ Mode          & Data\cite{HFAG}  &\multicolumn{5}{c}{This work} \\
\cline{3-7}
                        &                  &NLO+Vertex & $\text{NLO}^{eff}$ &$\text{NLO}^{eff}$($45^{\circ}$) &$\text{NLO}^{eff}$($60^{\circ}$)&$\text{NLO}^{eff}$($75^{\circ}$)   \\
\hline
$A_{CP}(\rho^+ \rho^0)$  & $-0.051\pm0.054$        &0.001&0.001&0.001&0.001&0.001\\
$A_{CP}(\rho^+ \rho^-)$  & $0.06\pm0.13$              &$-0.002$&-0.029&-0.041&-0.038&-0.033\\
$A_{CP}(\rho^0 \rho^0)$  & -                         &$0.702$ &0.177&0.350&0.417&0.475\\
\hline
$A_{CP}(K^{*0} \rho^+)$  & $-0.01\pm0.16$                &$-0.005$&-0.006&-0.009&-0.009&-0.008\\
$A_{CP}(K^{*+} \rho^0)$  & $0.20^{+0.32}_{-0.29}$      &$0.184$&0.182&0.266&0.273&0.257\\
$A_{CP}(K^{*-} \rho^+)$  & -                             &$0.148$&0.151&0.231&0.231&0.206\\
$A_{CP}(K^{*0} \rho^0)$  & $0.09\pm0.19$                  &$-0.090$&-0.101&-0.147&-0.155&-0.154\\
\hline
$A_{CP}(\bar{K}^{*0} K^{*+})$& -                    &$0.081$&0.085&0.141&0.143&0.128\\
$A_{CP}(\bar{K}^{*0} K^{*0})$& -                       &$0$&0&0&0&0\\
$A_{CP}(K^{*+} K^{*-})$& -                           &$-0.261$&-0.261&-0.261&-0.261&-0.261\\
\hline
$A_{CP}(\phi K^{*+})$       & $-0.01\pm0.08$              &$-0.003$&-0.003&-0.0006&-0.007&-0.007\\
$A_{CP}(\phi K^{*0})$       & $0.01\pm0.05$      &$0$&0&0&0&0\\
\hline
$A_{CP}(\omega K^{*+})$       & $0.29\pm0.35$          &$0.341$ &0.383&0.534&0.522&0.463\\
$A_{CP }(\omega K^{*0})$       & $0.45\pm0.25$    &$0.078$&0.116&0.170&0.182&0.185\\
\hline \hline
\end{tabular}
}
\end{center}

\end{table}

\begin{table}
\begin{center}
\tabcolsep 0.08in\vspace{-2em}
\caption{\small{Branching ratios for $B\to VV$ decay modes~(in unit of $10^{-6}$). The central values are obtained with $\tilde{\mu}_g$=0.52GeV and $\theta^{a} = 60^{\circ}$. The first error arises from the varying for $\mu_{scale}=1.4\sim 1.6$ GeV, the second one stems from the shape parameters of light mesons.}}\label{tab:br30}. \scriptsize{
\begin{tabular}{l|c|c|c|cccc}
\hline \hline
\ \ \ \ \ Mode          & Data\cite{HFAG}   &QCDF\cite{Cheng:2009cn}&pQCD~\cite{krophi,0602214}&\multicolumn{4}{c}{this work} \\
\cline{5-8}
                        &                   &               &                &LO      & NLO+Vertex &NLO($\theta^{a}$)&NLO($a^{eff},\theta^{a}$)  \\
\hline
$B^+\to \rho^+ \rho^0$  & $24.0\pm2.0$      &$20.0^{+4.0+2.0}_{-1.9-0.9}$           &$17\pm2\pm1$               &13.7       &$13.4$&13.4&$16.8^{+9.9+1.2}_{-4.4-0.7}$\\
$B^0\to \rho^+ \rho^-$  & $24.2\pm3.1$      &$25.5^{+1.5+2.4}_{-2.6-1.5}$           &$35\pm5\pm4$               &21.1       &$22.3$ &24.9&$22.2^{+14.0+1.3}_{-6.2-0.7}$\\
$B^0\to \rho^0 \rho^0$  & $0.73\pm0.27$     &$0.9^{+1.5+1.1}_{-0.4-0.2}$            &$0.9\pm0.1\pm0.1$          &0.3       &$0.4$ &0.4&$0.6^{+0.2+0.1}_{-0.1-0.0}$\\
\hline
$B^+\to K^{*0} \rho^+$  & $9.2\pm1.5$       &$9.2^{+1.2+3.6}_{-1.1-5.4}$            &17 (13)                       &11.4       &$16.2$&9.8&$8.8^{+3.4+3.5}_{-1.2-2.4}$\\
$B^+\to K^{*+} \rho^0$  & $<6.1$            &$5.5^{+0.6+1.3}_{-0.5-2.5}$            &9.0 (6.4)                       &7.3       &$9.9$&6.2&$5.9^{+2.1+1.9}_{-1.0-1.2}$\\
$B^0\to K^{*+} \rho^-$   & $<12$            &$8.9^{+1.1+4.8}_{-1.0-5.5}$            &13 (9.8)                   &10.2     &$13.9$&8.5&$8.3^{+2.1+3.2}_{-1.0-2.5}$\\
$B^0\to K^{*0} \rho^0$  & $3.4\pm1.0$       &$4.6^{+0.6+3.5}_{-0.5-3.5}$            &5.9 (4.7)                   &3.9       &$5.6$&3.4&$3.3^{+0.5+1.7}_{-0.2-1.1}$\\
\hline
$B^+\to \bar{K}^{*0} K^{*+}$& $1.2\pm 0.5$         &$0.6^{+0.1+0.3}_{-0.1-0.3}$            &0.48            &0.7       &$0.9$  &0.6&$0.5^{+0.2+0.2}_{-0.1-0.1}$ \\
$B^0\to \bar{K}^{*0} K^{*0}$& $1.28\pm0.35$ &$0.6^{+0.1+0.2}_{-0.1-0.3}$            &0.35                       &0.5     &$0.8$ &0.6&$0.5^{+0.2+0.2}_{-0.1-0.1}$\\
$B^0\to K^{*+} K^{*-}$& $<2$  &             &$0.1^{+0.0+0.1}_{-0.0-0.1}$                       &0.07     &$0.07$&0.07&$0.07^{+0.01+0.00}_{-0.01-0.01}$\\
\hline
$B^+\to \phi K^{*+}$       & $10.0\pm1.1$   &$10.0^{+1.4+12.3}_{-1.3-6.1}$          &$15.96$                    &15.9      &$19.4$&12.4&$9.6^{+2.5+2.4}_{-0.6-1.6}$\\
$B^0\to \phi K^{*0}$       & $9.8\pm0.7$    &$9.5^{+1.3+11.9}_{-1.2-5.9}$           &$14.86(10.2^{+2.5}_{-2.1})$&15.4   &$18.7$&11.8&$9.2^{+2.3+2.3}_{-0.5-1.6}$\\
\hline
$B^+\to \omega K^{*+}$       & $<7.4$   &$3.0^{+0.4+2.5}_{-0.3-1.5}$          &$7.9 (5.5)$                    &5.4       &$5.6$&3.7&$3.0^{+1.0+2.1}_{-0.4-1.0}$\\
$B^0\to \omega K^{*0}$       & $2.0\pm0.5$    &$2.5^{+0.4+2.5}_{-0.4-1.5}$           &$9.6 (6.6)$&5.8    &$6.2$&3.8&$2.5^{+0.7+1.0}_{-0.3-1.3}$\\
\hline \hline
\end{tabular}
}
\end{center}

\end{table}
\begin{table}
\begin{center}
\caption{The same as Table~\ref{tab:br30} but for longitudinal polarization fractions.}\label{tab:br31}. \scriptsize{
\begin{tabular}{l|c|c|c|cccc}
\hline \hline
\ \ \ \ \ Mode          & Data\cite{HFAG}   &QCDF~\cite{Cheng:2009cn}&pQCD~\cite{krophi,0602214}&\multicolumn{4}{c}{this work} \\
\cline{5-8}
                         &                   &               &                &LO      & NLO+Vertex &NLO($\theta^{a}$)&NLO($a^{eff},\theta^{a}$)  \\
\hline
$f_L(\rho^+ \rho^0)$  & $0.950\pm0.016$      &$0.96^{+0.01+0.02}_{-0.01-0.02}$ &$0.94$                     &0.95      &$0.94$&0.95&$0.95^{+0.00+0.00}_{-0.00-0.00}$\\
$f_L(\rho^+ \rho^-)$  & $0.978\pm0.023$    &$0.92^{+0.01+0.01}_{-0.02-0.02}$ &$0.94$                     &0.96      &$0.95$&0.95&$0.95^{+0.00+0.00}_{-0.00-0.00}$\\
$f_L(\rho^0 \rho^0)$  & $0.75\pm0.15$      &$0.92^{+0.03+0.06}_{-0.04-0.37}$  &$0.60$                     &0.79      &$0.84$ &0.56&$0.74^{+0.03+0.01}_{-0.03-0.01}$\\
\hline
$f_L(K^{*0} \rho^+)$  & $0.48\pm0.08$      &$0.48^{+0.03+0.52}_{-0.04-0.40}$      &0.82 (0.76)                       &0.79      &$0.85$&0.51&$0.45^{+0.05+0.02}_{-0.01-0.06}$\\
$f_L(K^{*+} \rho^0)$  & $0.96^{+0.06}_{-0.16}$&$0.67^{+0.02+0.31}_{-0.03-0.48}$     &0.85 (0.78)                     &0.83      &$0.86$&0.59&$0.57^{+0.05+0.02}_{-0.04-0.04}$\\
$f_L(K^{*-} \rho^+)$  & -                 &$0.53^{+0.02+0.45}_{-0.03-0.32}$      &0.78 (0.71)                    &0.77      &$0.81$&0.49&$0.46^{+0.04+0.03}_{-0.02-0.07}$\\
$f_L(K^{*0} \rho^0)$  & $0.57\pm0.12$      &$0.39^{+0.00+0.60}_{-0.00-0.31}$      &0.74 (0.68)                 &0.71      &$0.78$&0.38&$0.36^{+0.01+0.02}_{-0.01-0.05}$\\
\hline
$f_L(\bar{K}^{*0} K^{*+})$& $0.75^{+0.16}_{-0.26}$&$0.62^{+0.01+0.42}_{-0.02-0.33}$      &0.72                       &0.80     &$0.85$&0.53&$0.49^{+0.06+0.02}_{-0.03-0.05}$\\
$f_L(\bar{K}^{*0} K^{*0})$& $0.80\pm0.13$  &$0.69^{+0.01+0.34}_{-0.01-0.27}$      &0.67                       &0.77     &$0.83$&0.57&$0.53^{+0.07+0.01}_{-0.03-0.03}$\\
$f_L(K^{*+} K^{*-})$&    &       &0.99                       &0.99     &$0.99$&0.99&$0.99^{+0.00+0.00}_{-0.00-0.00}$\\
\hline
$f_L(\phi K^{*+})$       & $0.50\pm0.05$   &$0.49^{+0.04+0.51}_{-0.07-0.42}$            &$0.748$                    &0.80      &$0.87$&0.53&$0.45^{+0.03+0.02}_{-0.01-0.04}$\\
$f_L(\phi K^{*0})$       & $0.480\pm0.030$ &$0.50^{+0.04+0.51}_{-0.06-0.43}$            &$0.75(0.59^{+0.02}_{-0.02})$&0.80     &$0.87$&0.53&$0.45^{+0.03+0.03}_{-0.01-0.04}$\\
\hline
$f_L(\omega K^{*+})$       & $0.41\pm0.19$   &$0.67^{+0.03+0.32}_{-0.04-0.39}$            &$0.81 (0.73)$           & 0.84      &$0.90$ &0.63&$0.58^{+0.03+0.02}_{-0.01-0.02}$\\
$f_L(\omega K^{*0})$       & $0.70\pm0.13$ &$0.58^{+0.07+0.43}_{-0.10-0.14}$            &$0.82 (0.74)$&  0.86    &$0.93$&0.64&$0.53^{+0.03+0.04}_{-0.01-0.04}$\\
\hline \hline
\end{tabular}
}
\end{center}

\end{table}

\begin{table}
\begin{center}
\tabcolsep 0.07in\vspace{-2em}
\caption{The same as Table~\ref{tab:br30} but for direct CP asymmetries.}\label{tab:br32}. \scriptsize{
\begin{tabular}{l|c|c|c|cccc}
\hline \hline
\ \ \ \ \ Mode          & Data\cite{HFAG}   &QCDF\cite{Cheng:2009cn}&~pQCD\cite{pqcdvvcp}&\multicolumn{4}{c}{this work} \\
\cline{5-8}
                        &                   &               &                &LO      & NLO &NLO($\theta^{a}$)&NLO($a^{eff},\theta^{a}$)  \\
\hline
$A_{CP}(\rho^+ \rho^0)$  & $-0.051\pm0.054$      &$0.0006$ &$0$                     &0.001      &$0.001$&0.002&$0.001^{+0.000+0.000}_{-0.000-0.000}$\\
$A_{CP}(\rho^+ \rho^-)$  & $0.06\pm0.13$    &$-0.04^{+0.00+0.03}_{-0.00-0.03}$ &$-0.07$                     &-0.029      &$-0.002$&-0.036&$0.002^{+0.010+0.010}_{-0.006-0.003}$\\
$A_{CP}(\rho^0 \rho^0)$  & -      &$0.30^{+0.17+0.14}_{-0.16-0.26}$  &$0.80$                     &0.074      &$0.702$ &0.866&$0.417^{+0.004+0.005}_{-0.002-0.006}$\\
\hline
$A_{CP}(K^{*0} \rho^+)$  & $-0.01\pm0.16$      &$-0.003^{+0.00+0.02}_{-0.00-0.00}$      &+                       &-0.007      &$-0.005$&-0.009&$-0.009^{+0.001+0.001}_{-0.001-0.002}$\\
$A_{CP}(K^{*+} \rho^0)$  & $0.20^{+0.32}_{-0.29}$&$0.43^{+0.06+0.12}_{-0.03-0.28}$     &+                     &0.157      &$0.184$&0.267&$0.273^{+0.015+0.009}_{-0.019-0.004}$\\
$A_{CP}(K^{*-} \rho^+)$  & -                 &$0.32^{+0.01+0.02}_{-0.03-0.14}$      &+                    &0.189      &$0.148$&0.229&$0.231^{+0.00+0.010}_{-0.00-0.019}$\\
$A_{CP}(K^{*0} \rho^0)$  & $0.09\pm0.19$      &$-0.15^{+0.04+0.16}_{-0.08-0.14}$      &-                 &-0.041      &$-0.090$&-0.110&$-0.155^{+0.008+0.004}_{-0.008-0.005}$\\
\hline
$A_{CP}(\bar{K}^{*0} K^{*+})$& -&$0.16^{+0.01+0.17}_{-0.03-0.34}$      &-0.15                       &0.101     &$0.081$&0.139&$0.148^{+0.010+0.003}_{-0.023-0.002}$\\
$A_{CP}(\bar{K}^{*0} K^{*0})$& -  &$-0.14^{+0.01+0.06}_{-0.01-0.02}$      &-0.65                       &0.000     &$0.000$&0.000&$0.000^{+0+0}_{-0-0}$\\
$A_{CP}(K^{*+} K^{*-})$& - &       &0                       &-0.260     &$-0.261$&-0.261&$-0.261^{+0.023+0.077}_{-0.032-0.056}$\\
\hline
$A_{CP}(\phi K^{*+})$       & $-0.01\pm0.08$   &$0.0005$            &-                    &-0.003      &$-0.003$&-0.007&$-0.007^{+0.000+0.000}_{-0.000-0.000}$\\
$A_{CP}(\phi K^{*0})$       & $0.01\pm0.05$ &$0.008^{+0+0.004}_{-0-0.005}$            &-&0.000     &$0.000$&0.000&$0.000^{+0+0}_{-0-0}$\\
\hline
$A_{CP}(\omega K^{*+})$       & $0.29\pm0.35$   &$0.56^{+0.03+0.04}_{-0.04-0.43}$            &+           & 0.179      &$0.341$ &0.455&$0.522^{+0.002+0.005}_{-0.001-0.006}$\\
$A_{CP }(\omega K^{*0})$       & $0.45\pm0.25$ &$0.23^{+0.09+0.05}_{-0.05-0.18}$            &+&  0.0289   &$0.078$&0.102&$0.183^{+0.002+0.007}_{-0.007-0.007}$\\
\hline \hline
\end{tabular}
}
\end{center}

\end{table}

\section{Conclusions}\label{sec:conc}

Based on the approximate six-quark operator effective Hamiltonian derived from perturbative QCD, the QCD factorization approach has been naturally applied to evaluate the hadronic matrix elements for
charmless two body B-meson decays. It is shown that, with annihilation contribution and extra strong phase, our framework provides a simple way to evaluate the hadronic matrix elements of two body decays.

For $B\to PP$ final states, our predictions for branching ratios and CP asymmetries are generally consistent with the current experimental data within their respective uncertainties, once the effective Wilson coefficients and annihilation amplitude with small strong phase~($\theta^a=5^\circ$) are adopted. Especially for the branching ratio of $B\to \pi^0\pi^0$ mode, our result, although being still smaller than the data, is consistent with that in QCDF~\cite{Cheng:2009cn}.

As for $B\to PV$ decays, similar conclusions are found. The exceptions here are the branching ratio of $\rho^+ \pi^0$ and $B\to K^{*0}\pi^0$ modes, which are bigger than the data. An interesting point should be noted that our predictions are also consistent with the ones in QCDF~\cite{Cheng:2009cn}. Since the current data on CP asymmetries have large uncertainties in these modes, more precise experimental data are expected to further test our framework.

In $B\to VV$ decay modes, we have shed light on the polarization anomalies observed in $B\to \phi K^*, \rho K^*, \omega K^*$ decays. It is noted that these anomalies could be explained in our framework when considering annihilation contributions with a strong phase~($\theta^a=60^\circ$). Moreover, the annihilation contributions with a strong phase have remarkable effects on the branching ratios and CP asymmetries, especially on the observables of penguin dominated decay modes. As a result, with the effects, we also have good predictions for the branching ratios and direct CP asymmetries.

It is noted that the method developed in this paper allows us to calculate the relevant transition form factors. Our predictions(for B to light mesons form factors) are consistent with the results of light-core QCD sum-rules and pQCD.
We further apply the method to $B_s \to PP, PV, VV$ decays, and the paper~\cite{Bsdecays} is now in preparation.

\section*{\textbf{Acknowledgements: }}

This work was supported in part by the National Science Foundation
of China (NSFC) under the grant \# 10821504, 10975170 and the Project of
Knowledge Innovation Program (PKIP) of Chinese Academy of Science.

\section*{Appendix: Calculations of Hadronic Matrix Elements}
\def\theequation{A-\arabic{equation}}
\def\thesubsection{A}
\setcounter{equation}{0}
\label{sec:Amplitude}

With the considerations and analysis in the text, the QCD factorization approach with six-quark operator effective Hamiltonian enables us to evaluate all the hadronic matrix elements of nonleptonic two-body B-meson decays.

\begin{figure}[h]
\begin{center}
  \includegraphics[scale=0.6]{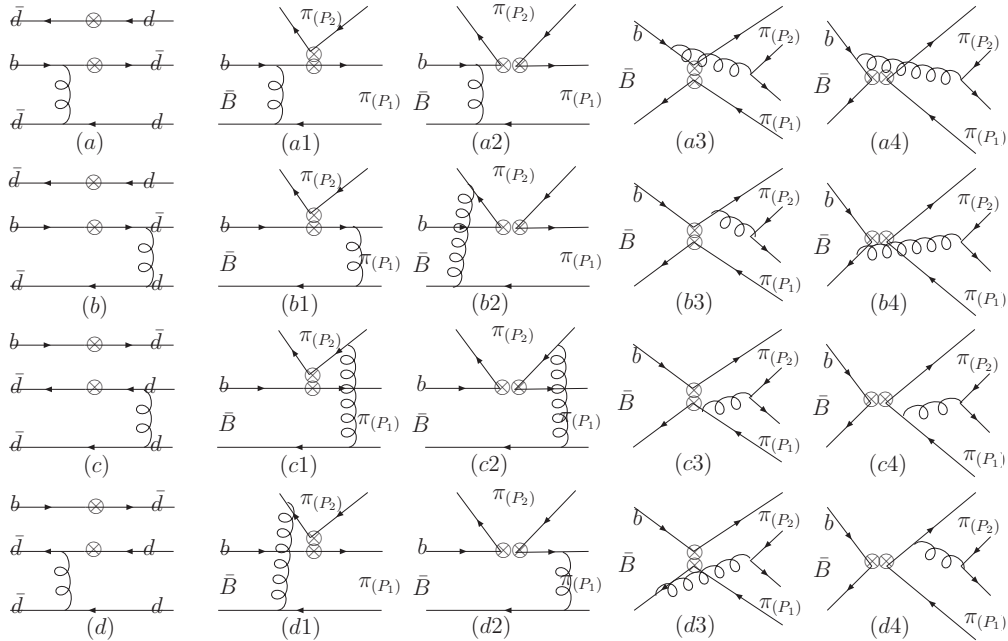}\\
  \caption{Four types of effective six quark diagrams lead to sixteen diagrams
  for hadronic two body decays of heavy meson via QCD factorization.}\label{pic:b-kpi}
  \end{center}
\end{figure}

When generalizing the above analysis to the present framework based on the approximate six-quark operator effective Hamiltonian, there are in general four types of six-quark diagrams, and each one corresponding to
four types of effective six-quark operators. So their hadronic matrix elements for two body mesonic decays lead to sixteen kinds of diagrams~(see Figs.~\ref{pic:b-kpi}(a1)-(d4)) as each of the effective six-quark operators leads to four kinds of amplitudes in the QCD factorization approach.
In order to get a global form of hadronic matrix elements, we should rewrite mesons with SU(3) symmetry into a vector/matric form. We use a column vector $\bm{\Lambda}_p$ containing CKM matrix elements, collect the three B-meson states into a row vector $\bm{B}$, and represent the final-state pseudoscalar and vector mesons by matrices $\bm{P}$ and $\bm{V}$
\begin{equation}
\begin{aligned}
  \bm{B} &= \left(B^+, B^0, B_s\right)
  \qquad
   \bm{P} = \left( \begin{array}{ccc}
    \frac{\pi^0}{\sqrt2} + \frac{\eta_q}{\sqrt 2}
     + \frac{\eta'_q}{\sqrt 2} & \pi^- & K^- \\
    \pi^+ & - \frac{\pi^0}{\sqrt2} + \frac{\eta_q}{\sqrt 2}
     + \frac{\eta'_q}{\sqrt 2} & \bar K^0 \\
    K^+ & K^0 & \eta_s + \eta'_s \end{array} \right) , \\
   \bm{V} &= \left( \begin{array}{ccc}
    \frac{\rho^0}{\sqrt2} + \frac{\omega}{\sqrt 2}
      & \rho^- & K^{*-} \\
    \rho^+ & - \frac{\rho^0}{\sqrt2} + \frac{\omega}{\sqrt 2}
      & \bar K^{*0} \\
    K^{*+} & K^{*0} & \phi \end{array} \right) ,
    \qquad
    \bm{\Lambda}_p = \left( \begin{array}{c}
    0 \\ \lambda_p^{(d)} \\ \lambda_p^{(s)}
   \end{array} \right).
\end{aligned}
\end{equation}

With the above definition, we can express hadronic matrix elements with the sixteen diagrams in Fig.\ref{pic:b-kpi} marked with (a1)-(d4):
\begin{eqnarray}\label{eq:matrix}
   &&\hspace{-1.0truecm}A(\bm{B}\to \bm{M_1}\bm{M_2})=\nonumber\\
   &&\hspace{-1.0truecm}\sum_{i=1,4,6,8,10}\sum_{p=u,c}\,\bigg\{
    C_i(M_i^{a1}+M_i^{b1}+M_i^{c1}+M_i^{d1})\bm{B}\bm{M_1} U_i\bm{M_2}\,
     \bm{\Lambda}_p \nonumber\\
   &&\qquad\mbox{}+ C_i(M_i^{a2}+M_i^{b2}+M_i^{c2}+M_i^{d2})\bm{B}\bm{M_1}\bm{\Lambda}_p\cdot
    \mbox{Tr}\left[U_i \bm{M_2}\right] \nonumber\\
   &&\qquad\mbox{}+ C_i(M_i^{a3}+M_i^{b3}+M_i^{c3}+M_i^{d3})\bm{B} U_i
    \bm{M_1}\bm{M_2}\bm{\Lambda}_p\nonumber\\
   &&\qquad\mbox{}+ C_i(M_i^{a4}+M_i^{b4}+M_i^{c4}+M_i^{d4})
    \bm{B}\bm{\Lambda}_p\cdot\mbox{Tr}\left[U_i \bm{M_1}\bm{M_2}\right]
    \bigg\} +\nonumber\\
    &&\hspace{-1.0truecm}\sum_{i=2,3,5,7,9}\sum_{p=u,c}\,\bigg\{
   C_i(M_i^{a1}+M_i^{b1}+M_i^{c1}+M_i^{d1})\bm{B}\bm{M_1}\bm{\Lambda}_p\cdot
    \mbox{Tr}\left[U_i \bm{M_2}\right]  \nonumber\\
   &&\qquad\mbox{}+ C_i(M_i^{a2}+M_i^{b2}+M_i^{c2}+M_i^{d2})\bm{B}\bm{M_1} U_i\bm{M_2}\,
     \bm{\Lambda}_p\nonumber\\
   &&\qquad\mbox{}+ C_i(M_i^{a3}+M_i^{b3}+M_i^{c3}+M_i^{d3})\bm{B}\bm{\Lambda}_p\cdot
    \mbox{Tr}\left[U_i \bm{M_1}\bm{M_2}\right] \nonumber\\
   &&\qquad\mbox{}+ C_i(M_i^{a4}+M_i^{b4}+M_i^{c4}+M_i^{d4})\bm{B} U_i
    \bm{M_1}\bm{M_2}\bm{\Lambda}_p
    \bigg\}\,,
\end{eqnarray}
with the matrices $\bm{U}_i$ defined as
\begin{equation}
\begin{aligned}
   \bm{U}_i&=\left\{
    \begin{array}{ll}
    \bm{U}_p\, &
        \mbox{\rm for }i=1,2\;,
    \\
    \bm{I}\, &
        \mbox{\rm for }i=3-6\;,
    \\
    \bm{Q}\, &
        \mbox{\rm for }i=7-10\;,
    \end{array}\right.
   \\
   \bm{U}_p = \left( \begin{array}{ccc}
    \delta_{pu} & 0 & ~0 \\
    0 & 0~ & ~0 \\
    0 & 0~ & ~0
   \end{array} \right),
   \qquad
   \bm{I} &= \left( \begin{array}{ccc}
    1 & 0& 0 \\
    0 & 1 & 0 \\
    0 & 0 & 1
   \end{array} \right) ,
   \qquad
   \bm{Q} = \left( \begin{array}{ccc}
    \frac23 & 0 & 0 \\
    0 & -\frac13 & 0 \\
    0 & 0 & -\frac13
   \end{array} \right) .
\end{aligned}
\end{equation}

Definitions of $M_i$ are attributed to three types of current-current four-quark vertexes:
\begin{equation}
\begin{aligned}
   M_i&=\left\{
    \begin{array}{lll}
    M_{LL}\, &
        \mbox{\rm for }i=1-4,9,10\;&(V-A)\times (V-A),
    \\
    M_{LR}\, &
        \mbox{\rm for }i=5,7\;&(V-A)\times (V+A),
    \\
    M_{SP}\, &
        \mbox{\rm for }i=6,8\;&(S-P)\times (S+P).
    \end{array}\right.
\end{aligned}
\end{equation}
Therefore, there are totally 48 kinds of hadronic matrix elements involved in the QCD factorization
approach, while it is easy to check that only half of them are independent with the following relations:
\begin{eqnarray}
\begin{array}{cccccccccccc}
  M^{a1}_{LL}&=&T^F_{LLa};&M^{a2}_{LL}&=&T^F_{LLa}/N_C;&M^{a3}_{LL}&=&A^N_{LLa}/N_C;&M^{a4}_{LL}&=&0;\\
  M^{a1}_{LR}&=&T^F_{LRa};&M^{a2}_{LR}&=&T^F_{SPa}/N_C;&M^{a3}_{LR}&=&A^N_{SPa}/N_C;&M^{a4}_{LR}&=&0;\\
  M^{a1}_{SP}&=&T^F_{SPa};&M^{a2}_{SP}&=&T^F_{LRa}/N_C;&M^{a3}_{SP}&=&A^N_{LRa}/N_C;&M^{a4}_{SP}&=&0;\\
  M^{b1}_{LL}&=&T^F_{LLb};&M^{b2}_{LL}&=&T^N_{LLb}/N_C;&M^{b3}_{LL}&=&A^F_{LLb}/N_C;&M^{b4}_{LL}&=&0;\\
  M^{b1}_{LR}&=&T^F_{LRb};&M^{b2}_{LR}&=&T^N_{SPb}/N_C;&M^{b3}_{LR}&=&A^F_{SPb}/N_C;&M^{b4}_{LR}&=&0;\\
  M^{b1}_{SP}&=&T^F_{LLb};&M^{b2}_{SP}&=&T^N_{LRb}/N_C;&M^{b3}_{SP}&=&A^F_{LRb}/N_C;&M^{b4}_{SP}&=&0;\\
  M^{c1}_{LL}&=&0;&M^{c2}_{LL}&=&T^N_{LLa}/N_C;&M^{c3}_{LL}&=&A^F_{LLa}/N_C;&M^{c4}_{LL}&=&A^F_{LLa};\\
  M^{c1}_{LR}&=&0;&M^{c2}_{LR}&=&T^N_{SPa}/N_C;&M^{c3}_{LR}&=&A^F_{SPa}/N_C;&M^{c4}_{LR}&=&A^F_{LRa};\\
  M^{c1}_{SP}&=&0;&M^{c2}_{SP}&=&T^N_{LRa}/N_C;&M^{c3}_{SP}&=&A^F_{LRa}/N_C;&M^{c4}_{SP}&=&A^F_{SPa};\\
  M^{d1}_{LL}&=&0;&M^{d2}_{LL}&=&T^F_{LLb}/N_C;&M^{d3}_{LL}&=&A^N_{LLb}/N_C;&M^{d4}_{LL}&=&A^F_{LLb};\\
  M^{d1}_{LR}&=&0;&M^{d2}_{LR}&=&T^F_{SPb}/N_C;&M^{d3}_{LL}&=&A^N_{SPb}/N_C;&M^{d4}_{LL}&=&A^F_{LRb};\\
  M^{d1}_{SP}&=&0;&M^{d2}_{SP}&=&T^F_{LRb}/N_C;&M^{d3}_{LL}&=&A^N_{LRb}/N_C;&M^{d4}_{LL}&=&A^F_{SPb},
\end{array}
\end{eqnarray}
where $T^F_{Xa}$ and  $T^F_{Xb}$ ($X=LL,LR,SP$) represent the factorizable emission diagram contributions, $T^N_{Xa}$ and $T^N_{Xb}$ ($X=LL,LR,SP$) are the non-factorizable emission diagram contributions. $A^F_{Xa}$, $A^F_{Xb}$ and $A^N_{Xa}$, $A^N_{Xb}$ ($X=LL,LR,SP$) denote the so-called factorizable and
non-factorizable annihilation diagram contributions, respectively. With this definition, we can rewrite Eq.~(\ref{eq:matrix}) into following form
\begin{eqnarray}
   &&\hspace{-1.0truecm}A(\bm{B}\to \bm{M_1}\bm{M_2})=\nonumber\\
   &&\hspace{-1.0truecm}
    \sum_{p=u,c}\,\bigg\{
    \bm{B}\bm{M_1} \left( T^{ {M_1} {M_2}}( \bm{B})\,\bm{U}_p + P^{ {M_1} {M_2}}( \bm{B})^p
    + P_{EW}^{C {M_1} {M_2}}(\bm{B})^p\,\bm{Q} \right)\bm{M_2}\,
     \bm{\Lambda}_p  \nonumber\\
   &&\qquad\mbox{}+ \bm{B}\bm{M_1}\bm{\Lambda}_p\cdot
    \mbox{Tr}\left[\left( C^{M_1 M_2}(\bm{B})\,\bm{U}_p + P_C^{ {M_1} {M_2}}( {\bm B})^p
     + P_{EW}^{ {M_1} {M_2}}( \bm{B})\,\bm{Q} \right) \bm{M_2}\right]\nonumber\\
   &&\qquad\mbox{}+ \bm{B} \left( A^{ {M_1} {M_2}}(\bm{B})\,\bm{U}_p +P_A^{ {M_1} {M_2}}(\bm{B})^p
    + P_{EW}^{E {M_1} {M_2}}(\bm{B})^p\,\bm{Q} \right)
    \bm{M_1}\bm{M_2}\bm{\Lambda}_p \nonumber\\
   &&\qquad\mbox{}+ \bm{B}\bm{\Lambda}_p\cdot
    \mbox{Tr}\left[ \left( E^{ {M_1} {M_2}}(\bm{B})\,\bm{U}_p + P_E^{ {M_1} {M_2}}(\bm{B})^p
    + P_{EW}^{A {M_1} {M_2}}(\bm{B})^p\,\bm{Q} \right)\bm{M_1}\bm{M_2}\right]
    \bigg\}\,,
\end{eqnarray}
where ${M_1}$ and ${M_2}$ are pseudoscalar (${P}$) or vector(${V}$) mesons. The twelve types of amplitudes  $T^{M_1M_2}(M)$, $C^{M_1M_2}(M)$,
$P^{M_1M_2}(M)$, $P_C^{M_1M_2}(M)$, $P_{EW}^{M_1M_2}(M)$, $A^{M_1M_2}(M)$,
$E^{M_1M_2}(M)$, $P_E^{M_1M_2}(M)$, $P_A^{M_1M_2}(M)$,
$P_{EW}^{CM_1M_2}(M)$, $P_{EW}^{EM_1M_2}(M)$, $P_{EW}^{AM_1M_2}(M)$,
 are defined
as follows
\begin{eqnarray}
  \label{eq:pptopologyT}
  T^{M_1M_2}(M)&=&4\pi\alpha_s(\mu)\frac{G_F}{\sqrt{2}}\big\{[C_1(\mu)+\frac{1}{N_C}C_2(\mu)]T_{LL}^{FM_1M_2}(M)+\frac{1}{N_C}C_2(\mu)T_{LL}^{NM_1M_2}(M)\big\},\nonumber\\
  C^{M_1M_2}(M)&=&4\pi\alpha_s(\mu)\frac{G_F}{\sqrt{2}}\big\{[C_2(\mu)+\frac{1}{N_C}C_1(\mu)]T_{LL}^{FM_1M_2}(M)+\frac{1}{N_C}C_1(\mu)T_{LL}^{NM_1M_2}(M)\big\},\nonumber\\
  P^{M_1M_2}(M)&=&4\pi\alpha_s(\mu)\frac{G_F}{\sqrt{2}}\big\{[C_4(\mu)+\frac{1}{N_C}C_3(\mu)]T_{LL}^{FM_1M_2}(M)+\frac{1}{N_C}C_3(\mu)T_{LL}^{NM_1M_2}(M)
  \nonumber\\&&+[C_6(\mu)+\frac{1}{N_C}C_5(\mu)]T_{SP}^{FM_1M_2}(M)+\frac{1}{N_C}C_5(\mu)T_{LR}^{NM_1M_2}(M)\big\},\nonumber\\
  P_C^{M_1M_2}(M)&=&4\pi\alpha_s(\mu)\frac{G_F}{\sqrt{2}}\big\{[C_3(\mu)+\frac{1}{N_C}C_4(\mu)]T_{LL}^{FM_1M_2}(M)+\frac{1}{N_C}C_4(\mu)T_{LL}^{NM_1M_2}(M)
  \nonumber\\&&+[C_5(\mu)+\frac{1}{N_C}C_6(\mu)]T_{SP}^{FM_1M_2}(M)+\frac{1}{N_C}C_6(\mu)T_{LR}^{NM_1M_2}(M)\big\},\nonumber\\
  P_{EW}^{M_1M_2}(M)&=&4\pi\alpha_s(\mu)\frac{G_F}{\sqrt{2}}\frac{3}{2}\big\{[C_9(\mu)+\frac{1}{N_C}C_{10}(\mu)]T_{LL}^{FM_1M_2}(M)+\frac{1}{N_C}C_{10}(\mu)T_{LL}^{NM_1M_2}(M)
  \nonumber\\&&+[C_7(\mu)+\frac{1}{N_C}C_8(\mu)]T_{LR}^{FM_1M_2}(M)+\frac{1}{N_C}C_8(\mu)T_{SP}^{NM_1M_2}(M)\},
  \nonumber\\
  P_{EW}^{CM_1M_2}(M)&=&4\pi\alpha_s(\mu)\frac{G_F}{\sqrt{2}}\frac{3}{2}\big\{[C_{10}(\mu)+\frac{1}{N_C}C_9(\mu)]T_{LL}^{FM_1M_2}(M)+\frac{1}{N_C}C_9(\mu)T_{LL}^{NM_1M_2}(M)
  \nonumber\\&&+[C_8(\mu)+\frac{1}{N_C}C_7(\mu)]T_{SP}^{FM_1M_2}(M)+\frac{1}{N_C}C_7(\mu)T_{LR}^{NM_1M_2}(M)\big\},
\end{eqnarray}
for the so-called emission diagrams, and
\begin{eqnarray}
  \label{eq:pptopologyA}
  A^{M_1M_2}(M)&=&4\pi\alpha_s(\mu)\frac{G_F}{\sqrt{2}}\big\{[C_1(\mu)+\frac{1}{N_C}C_2(\mu)]A_{LL}^{FM_1M_2}(M)+\frac{1}{N_C}C_2(\mu)A_{LL}^{NM_1M_2}(M)\}
  .\nonumber\\
  E^{M_1M_2}(M)&=&4\pi\alpha_s(\mu)\frac{G_F}{\sqrt{2}}\big\{[C_2(\mu)+\frac{1}{N_C}C_1(\mu)]A_{LL}^{FM_1M_2}(M)+\frac{1}{N_C}C_1(\mu)A_{LL}^{NM_1M_2}(M)\big\},\nonumber\\
  P_{E}^{M_1M_2}(M)&=&4\pi\alpha_s(\mu)\frac{G_F}{\sqrt{2}}\big\{[C_4(\mu)+\frac{1}{N_C}C_3(\mu)]A_{LL}^{FM_1M_2}(M)+\frac{1}{N_C}C_3(\mu)A_{LL}^{NM_1M_2}(M)
  \nonumber\\&&+[C_6(\mu)+\frac{1}{N_C}C_5(\mu)]A_{SP}^{FM_1M_2}(M)+\frac{1}{N_C}C_5(\mu)A_{LR}^{NM_1M_2}(M)\}
  ,\nonumber\\
  P_{A}^{M_1M_2}(M)&=&4\pi\alpha_s(\mu)\frac{G_F}{\sqrt{2}}\big\{[C_3(\mu)+\frac{1}{N_C}C_4(\mu)]A_{LL}^{FM_1M_2}(M)+\frac{1}{N_C}C_4(\mu)A_{LL}^{NM_1M_2}(M)
  \nonumber\\&&+[C_5(\mu)+\frac{1}{N_C}C_6(\mu)]A_{LR}^{FM_1M_2}(M)+\frac{1}{N_C}C_6(\mu)A_{SP}^{NM_1M_2}(M)\}
  ,\nonumber\\
  P_{EW}^{AM_1M_2}(M)&=&4\pi\alpha_s(\mu)\frac{G_F}{\sqrt{2}}\frac{3}{2}\big\{[C_9(\mu)+\frac{1}{N_C}C_{10}(\mu)]A_{LL}^{FM_1M_2}(M)+\frac{1}{N_C}C_{10}(\mu)A_{LL}^{NM_1M_2}(M)
  \nonumber\\&&+[C_7(\mu)+\frac{1}{N_C}C_8(\mu)]A_{LR}^{FM_1M_2}(M)+\frac{1}{N_C}C_8(\mu)A_{SP}^{NM_1M_2}(M)\}
  ,\nonumber\\
  P_{EW}^{EM_1M_2}(M)&=&4\pi\alpha_s(\mu)\frac{G_F}{\sqrt{2}}\frac{3}{2}\big\{[C_{10}(\mu)+\frac{1}{N_C}C_9(\mu)]A_{LL}^{FM_1M_2}(M)+\frac{1}{N_C}C_9(\mu)A_{LL}^{NM_1M_2}(M)
  \nonumber\\&&+[C_8(\mu)+\frac{1}{N_C}C_7(\mu)]A_{SP}^{FM_1M_2}(M)+\frac{1}{N_C}C_7(\mu)A_{LR}^{NM_1M_2}(M)\},
\end{eqnarray}
for the so-called annihilation diagrams. Where $T^F_{XA}$,
$T^N_{XA}$, $A^F_{XA}$, $A^N_{XA}$ ($X=LL,LR,SP$, $A=a,b$) arise
from the hadronic matrix elements and their detailed expressions are
given below.

We now list our results for the various decay amplitudes expressed in
terms of above topological amplitudes.
The detailed calculations of the hadronic matrix elements for $B\to PP$ decays could be found in our previous paper~\cite{Su:2008mc}. As for decay amplitudes and the hadronic matrix elements for $B\to PV, VV$ decays, we shall list them one by one as follows. Firstly, for $B\to \pi\rho$ decay channels, we have
\begin{eqnarray}
  A(B^0\to \pi^-\rho^+)&=&V_{td}V^*_{tb}[P^{\pi\rho}(B)+\frac{2}{3}P_{EW}^{C\pi\rho}(B)+P_E^{\pi\rho}(B)+P_A^{\pi\rho}(B)+P_A^{\rho\pi}(B)+
  \frac{2}{3}P_{EW}^{A\rho\pi}(B)\nonumber\\&&-\frac{1}{3}P_{EW}^{A\pi\rho}(B)-\frac{1}{3}A_{EW}^{E\pi\rho}(B)]
  \nonumber\\&&-V_{ud}V^*_{ub}[T^{\pi\rho}(B)+E^{\pi\rho}(B)],\nonumber\\
  A(B^0\to \rho^-\pi^+)&=&V_{td}V^*_{tb}[P^{\rho\pi}(B)+\frac{2}{3}P_{EW}^{C\rho\pi}(B)+P_E^{\rho\pi}(B)+P_A^{\rho\pi}(B)+P_A^{\pi\rho}(B)+
  \frac{1}{3}P_{EW}^{A\rho\pi}(B)\nonumber\\&&-\frac{2}{3}P_{EW}^{A\pi\rho}(B)-\frac{1}{3}A_{EW}^{E\rho\pi}(B)]
  \nonumber\\&&-V_{ud}V^*_{ub}[T^{\rho\pi}(B)+E^{\rho\pi}(B)],\nonumber\\
  A(B^+\to \pi^+\rho^0)&=&\frac{1}{\sqrt{2}}\{V_{td}V^*_{tb}[-P^{\pi\rho}(B)+P^{\rho\pi}(B)+P_{EW}^{\pi\rho}(B)+\frac{1}{3}P_{EW}^{C\pi\rho}(B)
  +\frac{2}{3}P_{EW}^{C\rho\pi}(B)\nonumber\\&&-P^{A\pi\rho}(B)+P_A^{\rho\pi}(B)-P_{EW}^{A\pi\rho}(B)+P_{EW}^{A\rho\pi}(B)]-V_{ud}V^*_{ub}
  [T^{\rho\pi}(B)\nonumber\\&&+C^{\rho\pi}(B)+A^{\rho\pi}(B)-A^{\pi\rho}(B)]\},\nonumber\\
  A(B^+\to \rho^+\pi^0)&=&\frac{1}{\sqrt{2}}\{V_{td}V^*_{tb}[-P^{\rho\pi}(B)+P^{\pi\rho}(B)+P_{EW}^{\rho\pi}(B)+\frac{1}{3}P_{EW}^{C\rho\pi}(B)
  +\frac{2}{3}P_{EW}^{C\pi\rho}(B)\nonumber\\&&-P^{A\rho\pi}(B)+P_A^{\pi\rho}(B)-P_{EW}^{A\rho\pi}(B)+P_{EW}^{A\pi\rho}(B)]-V_{ud}V^*_{ub}
  [T^{\pi\rho}(B)\nonumber\\&&+C^{\pi\rho}(B)+A^{\pi\rho}(B)-A^{\rho\pi}(B)]\},\nonumber\\
  A(B^0\to \pi^0\pi^0)&=&\frac{1}{2}\{A(B^0\to \rho^+\pi^-)+A(B^0\to \rho^-\pi^+)\nonumber\\&&-\sqrt{2}[A(B^+\to \pi^+\rho^0)+A(B^+\to \rho^+\pi^0)]\},
\end{eqnarray}

For $B\to \pi K^*$ decay channels, the decay amplitudes are
\begin{eqnarray}
  A(B^+\to \pi^+ K^{*0})&=&-V_{ts}V^*_{tb}[P^{\pi K^*}(B)-\frac{1}{3} P_{EW}^{C\pi K^*}(B)+P_E^{\pi K^*}(B)+
  \frac{2}{3}P_{EW}^{E\pi K^*}(B)]+V_{us}V^*_{ub}A^{\pi K^*}(B),\nonumber\\
  A(B^+\to \pi^0 K^{*+})&=&\frac{1}{\sqrt{2}}\{V_{td}V^*_{tb}[P^{\pi K^*}(B)+P_{EW}^{K^*\pi}(B)+\frac{2}{3}P_{EW}^{C\pi K^*}(B)+
  P_E^{\pi K^*}(B)+\frac{2}{3}P_{EW}^{E\pi K^*}(B)]
  \nonumber\\&&-V_{us}V^*_{ub}[T^{\pi K^*}(B)+C^{K^*\pi}(B)+A^{\pi K^*}(B)]\},\nonumber\\
  A(B^0\to \pi^- K^{*+})&=&V_{td}V^*_{tb}[P^{\pi K^*}(B)+\frac{2}{3}P_{EW}^{C\pi K^*}(B)+P_E^{\pi K^*}(B)-\frac{1}{3}P_{EW}^{E\pi K^*}(B)]-
  V_{us}V^*_{ub}T^{\pi K^*}(B),\nonumber\\
  A(B^0\to \pi^0 K^0)&=&-\frac{1}{\sqrt{2}}\{V_{td}V^*_{tb}[P^{\pi K^*}(B)-P_{EW}^{K^*\pi}(B)-\frac{1}{3}P_{EW}^{C\pi K^*}(B)
  +P_E^{\pi K^*}(B)\nonumber\\&&
  -\frac{1}{3}P_{EW}^{E\pi K^*}(B)]+V_{us}V^*_{ub}C^{K^*\pi}(B)\},
\end{eqnarray}

For $B\to \rho K $ decay channels, we have:
\begin{eqnarray}
  A(B^+\to \rho^+ K^0)&=&-V_{ts}V^*_{tb}[P^{\rho K}(B)-\frac{1}{3} P_{EW}^{C\rho K}(B)+P_E^{\rho K}(B)+\frac{2}{3}P_{EW}^{E\rho K}(B)]+V_{us}V^*_{ub}A^{\rho K}(B),\nonumber\\
  A(B^+\to \rho^0 K^+)&=&\frac{1}{\sqrt{2}}\{V_{td}V^*_{tb}[P^{\rho K}(B)+P_{EW}^{K\rho}(B)+\frac{2}{3}P_{EW}^{C\rho K}(B)+P_E^{\rho K}(B)+\frac{2}{3}P_{EW}^{E\rho K}(B)]
  \nonumber\\&&-V_{us}V^*_{ub}[T^{\rho K}(B)+C^{K\rho}(B)+A^{\rho K}(B)]\},\nonumber\\
  A(B^0\to \rho^- K^+)&=&V_{td}V^*_{tb}[P^{\rho K}(B)+\frac{2}{3}P_{EW}^{C\rho K}(B)+P_E^{\rho K}(B)-\frac{1}{3}P_{EW}^{E\rho K}(B)]-V_{us}V^*_{ub}T^{\rho K}(B),\nonumber\\
  A(B^0\to \rho^0 K^0)&=&-\frac{1}{\sqrt{2}}\{V_{td}V^*_{tb}[P^{\rho K}(B)-P_{EW}^{K\rho}(B)-\frac{1}{3}P_{EW}^{C\rho K}(B)+P_E^{\rho K}(B)\nonumber\\&&
  -\frac{1}{3}P_{EW}^{E\rho K}(B)]+V_{us}V^*_{ub}C^{K\rho}(B)\},
\end{eqnarray}

For $B\to \omega K $ decay channels, we have:
\begin{eqnarray}
  A(B^+\to \omega K^+)&=&\frac{1}{\sqrt{2}}\{V_{td}V^*_{tb}[P^{\omega K}(B)+\frac{2}{3}P_{EW}^{C\omega K}(B)+\frac{1}{3}P_{EW}^{K\omega}(B)
  +2 P_C^{K\omega}(B)+P_E^{\omega K}(B)\nonumber\\&&
  +\frac{2}{3}P_{EW}^{E\omega K}(B)]
  -V_{us}V^*_{ub}[T^{\omega K}(B)+C^{K\omega}(B)+A^{\omega K}(B)]\},\nonumber\\
  A(B^0\to \omega K^0)&=&-\frac{1}{\sqrt{2}}\{V_{td}V^*_{tb}[P^{\omega K}(B)-\frac{1}{3}P_{EW}^{C\omega K}(B)+\frac{1}{3}P_{EW}^{K\omega}(B)
  +2 P_C^{K\omega}(B)+P_E^{\omega K}(B)\nonumber\\&&
  -\frac{1}{3}P_{EW}^{E\omega K}(B)]+V_{us}V^*_{ub}C^{K\rho}(B)\},
\end{eqnarray}

For $B\to \phi K $ decay channels, we have:
\begin{eqnarray}
  A(B^+\to \phi K^+)&=&V_{td}V^*_{tb}[P^{K\phi}(B)-\frac{1}{3}P_{EW}^{C\ K\phi}(B)-\frac{2}{3}P_{EW}^{K\phi}(B)
  +2 P_C^{K\phi}(B)+P_E^{\phi K}(B)\nonumber\\&&
  +\frac{2}{3}P_{EW}^{E\phi K}(B)]
  -V_{us}V^*_{ub}A^{\phi K}(B),\nonumber\\
  A(B^0\to \phi K^0)&=&-V_{td}V^*_{tb}[P^{K\phi}(B)-\frac{1}{3}P_{EW}^{C\ K\phi}(B)-\frac{2}{3}P_{EW}^{K\phi}(B)
  +2 P_C^{K\phi}(B)+P_E^{\phi K}(B)\nonumber\\&&
  -\frac{1}{3}P_{EW}^{E\phi K}(B)],
\end{eqnarray}

For $B\to K^* K $ decay channels, we have:
\begin{eqnarray}
  A(B^+\to \bar{K}^{*0} K^+)&=&-V_{td}V^*_{tb}[P^{K\bar{K}^*}(B)-\frac{1}{3}P_{EW}^{C\ K\bar{K}^*}(B)+
  P_E^{K\bar{K}^*}(B)+\frac{2}{3}P_{EW}^{E\bar{K}K}(B)]\nonumber\\&&+V_{ud}V^*_{ub}A^{K\bar{K}}(B),
\nonumber\\
  A(B^0\to K^{*0} \bar{K}^0)&=&-V_{td}V^*_{tb}[P^{K^*\bar{K}}(B)-\frac{1}{3}P_{EW}^{C\ K^*\bar{K}}(B)+
  P_E^{K^*\bar{K}}(B)+P_A^{K^*\bar{K}}(B)\nonumber\\&&+P_A^{\bar{K}K^*}(B)
  -\frac{1}{3}P_{EW}^{A\ K^*\bar{K}}(B)-\frac{1}{3}P_{EW}^{A\ \bar{K}K^*}(B)-\frac{1}{3}P_{EW}^{E\ K^*\bar{K}}(B)],\nonumber\\
\end{eqnarray}

For $B\to \rho \rho$ decay channels, we have(three part of amplitude are similar):
\begin{eqnarray}
  A(B^0\to \rho^+\rho^-)&=&V_{td}V^*_{tb}[P^{\rho\rho}(B)+\frac{2}{3}P_{EW}^{C\rho\rho}(B)+P_E^{\rho\rho}(B)+
  2P_A^{\rho\rho}(B)+\frac{1}{3}P_{EW}^{A\rho\rho}(B)\nonumber\\&&-\frac{1}{3}A_{EW}^{E\rho\rho}(B)]
  \nonumber\\&&-V_{ud}V^*_{ub}[T^{\rho\rho}(B)+E^{\rho\rho}(B)],\nonumber\\
  A(B^+\to \rho^+\rho^0)&=&\frac{1}{\sqrt{2}}\{V_{td}V^*_{tb}[P_{EW}^{\rho\rho}(B)+P_{EW}^{C\rho\rho}(B)]
  -V_{ud}V^*_{ub}[T^{\rho\rho}(B)+C^{\rho\rho}(B)]\},\nonumber\\
  A(B^0\to \rho^0\rho^0)&=&\frac{1}{\sqrt{2}}\{-V_{td}V^*_{tb}[P^{\rho\rho}(B)-P_{EW}^{\rho\rho}(B)-\frac{1}{3}P_{EW}^{C\rho\rho}(B)
  +P_E^{\rho\rho}(B)+2P_A^{\rho\rho}(B)
  \nonumber\\&&+\frac{1}{3}P_{EW}^{A\rho\rho}(B)-\frac{1}{3}P_{EW}^{E\rho\rho}(B)]+V_{ud}V^*_{ub}[-C^{\rho\rho}(B)+E^{\rho\rho}(B)]\},
\end{eqnarray}

For $B\to \rho K^* $ decay channels, we have(three part of amplitude are similar):
\begin{eqnarray}
  A(B^+\to \rho^+ K^{*0})&=&-V_{ts}V^*_{tb}[P^{\rho K^*}(B)-\frac{1}{3} P_{EW}^{C\rho K^*}(B)+P_E^{\rho K^*}(B)+
  \frac{2}{3}P_{EW}^{E\rho K^*}(B)]\nonumber\\&&+V_{us}V^*_{ub}A^{\rho K^*}(B),\nonumber\\
  A(B^+\to \rho^0 K^{*+})&=&\frac{1}{\sqrt{2}}\{V_{td}V^*_{tb}[P^{\rho K^*}(B)+P_{EW}^{K^*\rho}(B)+\frac{2}{3}P_{EW}^{C\rho K^*}(B)+
  P_E^{\rho K^*}(B)\nonumber\\&&+\frac{2}{3}P_{EW}^{E\rho K^*}(B)]
  -V_{us}V^*_{ub}[T^{\rho K^*}(B)+C^{K^*\rho}(B)+A^{\rho K^*}(B)]\},\nonumber\\
  A(B^0\to \rho^- K^{*+})&=&V_{td}V^*_{tb}[P^{\rho K^*}(B)+\frac{2}{3}P_{EW}^{C\rho K^*}(B)+P_E^{\rho K^*}(B)-\frac{1}{3}P_{EW}^{E\rho K^*}(B)]
  \nonumber\\&&-V_{us}V^*_{ub}T^{\rho K^*}(B),\nonumber\\
  A(B^0\to \rho^0 K^0)&=&-\frac{1}{\sqrt{2}}\{V_{td}V^*_{tb}[P^{\rho K^*}(B)-P_{EW}^{K^*\rho}(B)-\frac{1}{3}P_{EW}^{C\rho K^*}(B)
  +P_E^{\rho K^*}(B)\nonumber\\&&
  -\frac{1}{3}P_{EW}^{E\rho K^*}(B)]+V_{us}V^*_{ub}C^{K^*\rho}(B)\},
\end{eqnarray}

For $B\to \omega K^* $ decay channels, we have(three part of amplitude are similar):
\begin{eqnarray}
  A(B^+\to \omega K^{*+})&=&\frac{1}{\sqrt{2}}\{V_{td}V^*_{tb}[P^{\omega K^*}(B)+\frac{2}{3}P_{EW}^{C\omega K^*}(B)+\frac{1}{3}P_{EW}^{K^*\omega}(B)
  +2 P_C^{K^*\omega}(B)+P_E^{\omega K^*}(B)\nonumber\\&&
  +\frac{2}{3}P_{EW}^{E\omega K^*}(B)]
  -V_{us}V^*_{ub}[T^{\omega K^*}(B)+C^{K^*\omega}(B)+A^{\omega K^*}(B)]\},\nonumber\\
  A(B^0\to \omega K^{*0})&=&-\frac{1}{\sqrt{2}}\{V_{td}V^*_{tb}[P^{\omega K^*}(B)-\frac{1}{3}P_{EW}^{C\omega K^*}(B)+\frac{1}{3}P_{EW}^{K^*\omega}(B)
  +2 P_C^{K^*\omega}(B)+P_E^{\omega K^*}(B)\nonumber\\&&
  -\frac{1}{3}P_{EW}^{E\omega K^*}(B)]+V_{us}V^*_{ub}C^{K^*\rho}(B)\},
\end{eqnarray}

For $B\to \phi K^* $ decay channels, we have(three part of amplitude are similar):
\begin{eqnarray}
  A(B^+\to \phi K^{*+})&=&V_{td}V^*_{tb}[P^{K^*\phi}(B)-\frac{1}{3}P_{EW}^{C\ K^*\phi}(B)-\frac{2}{3}P_{EW}^{K^*\phi}(B)
  +2 P_C^{K^*\phi}(B)+P_E^{\phi K^*}(B)\nonumber\\&&
  +\frac{2}{3}P_{EW}^{E\phi K^*}(B)]
  -V_{us}V^*_{ub}A^{\phi K^*}(B),\nonumber\\
  A(B^0\to \phi K^{*0})&=&-V_{td}V^*_{tb}[P^{K^*\phi}(B)-\frac{1}{3}P_{EW}^{C\ K^*\phi}(B)-\frac{2}{3}P_{EW}^{K^*\phi}(B)
  +2 P_C^{K^*\phi}(B)+P_E^{\phi K^*}(B)\nonumber\\&&
  -\frac{1}{3}P_{EW}^{E\phi K^*}(B)],
\end{eqnarray}

For $B\to K^* K^* $ decay channels, we have(three part of amplitude are similar):
\begin{eqnarray}
  A(B^0\to K^{*+}K^{*-})&=&-V_{td}V^*_{tb}*[P_A^{K^*\bar{K}^*}(B)+P_A^{\bar{K}^*K^*}(B)+\frac{2}{3}P_{EW}^{A\ K^*\bar{K}^*}(B)
  \nonumber\\&&-\frac{1}{3}P_{EW}^{A\bar{K}^*K^*}(B)]+V_{ud}V^*_{ub}E^{K^*\bar{K}^*}(B),
\nonumber\\
  A(B^+\to K^{*+}\bar{K}^{*0})&=&-V_{td}V^*_{tb}[P^{K^*\bar{K}^*}(B)-\frac{1}{3}P_{EW}^{C\ K^*\bar{K}^*}(B)
  +P_E^{K^*\bar{K}^*}(B)+\frac{2}{3}P_{EW}^{E\ \bar{K}^*K^*}(B)]\nonumber\\&&+V_{ud}V^*_{ub}A^{K^*\bar{K}^*}(B),
\nonumber\\
  A(B^0\to K^{*0}\bar{K}^{*0})&=&-V_{td}V^*_{tb}[P^{K^*\bar{K}^*}(B)-\frac{1}{3}P_{EW}^{C\ K^*\bar{K}^*}(B)
  +P_E^{K^*\bar{K}^*}(B)+P_A^{K^*\bar{K}^*}(B)\nonumber\\&&+P_A^{\bar{K}^*K^*}(B)
  -\frac{1}{3}P_{EW}^{A\ K^*\bar{K}^*}(B)-\frac{1}{3}P_{EW}^{A\ \bar{K}^*K^*}(B)-\frac{1}{3}P_{EW}^{E\ K^*\bar{K}^*}(B)],\nonumber\\
\end{eqnarray}

Let us first give the factorizable emission contributions for the
$(V-A)\times (V-A)$ and $(V-A)\times (V+A)$ effective four-quark
vertexes, they are simply denoted by $LL$ and $LR$
\begin{eqnarray}
  \label{eq:ppf1}
  T_{LL}^{FM_1M_2}(M)&=&T_{LLa}^{FM_1M_2}(M)+T_{LLb}^{FM_1M_2}(M),\nonumber\\
  T_{LLa}^{FM_1M_2}(M)&=&i \frac{1}{4}\frac{C_{F}}{N_C}\ F_M\ F_{M_1}\
  F_{M_2} \int_0^1\int_0^1\int_0^1  \emph{d}u\,\emph{d}x\,\emph{d}y\,m_B^2\phi_{M}(u)\nonumber\\&&
  \big\{m_B (2 m_b-m_B x) \phi_{M_1}(x)+\mu_{M_1}(2 m_B x-m_b)[\phi^{p}_{M_1}(x)-\phi^{T}_{M_1}(x)]\big\}
  \phi_{M_2}(y)h_{Ta}^F(u,x),\nonumber\\
  T_{LLb}^{FM_1M_2}(M)&=&i \frac{1}{2}  \frac{C_{F}}{N_C}\ F_M\ F_{M_1}\
  F_{M_2} \int_0^1\int_0^1\int_0^1  \emph{d}u\,\emph{d}x\,\emph{d}y\, m_B^3 \mu_{M_1}  \phi_{M}(u)\phi_{M_2}(y) \phi^{p}_{M_1}(x)h_{Tb}^F(u,x),\nonumber\\
  T_{LR}^{FM_1M_2}(M)&=&T_{LLa}^{FM_1M_2}(M)+T_{LLb}^{FM_1M_2}(M),\nonumber\\
  T_{LRa}^{FM_1M_2}(M)&=&- T_{LLa}^{FM_1M_2}(M),\nonumber\\
  T_{LRb}^{FM_1M_2}(M)&=&- T_{LLb}^{FM_1M_2}(M).
\end{eqnarray}
The factorizable emission contributions for the $(S-P)\times (S+P)$
effective four-quark vertex are found to be
\begin{eqnarray}
  \label{eq:ppf2}
  T_{SP}^{FM_1M_2}(M)&=&T_{SPa}^{FM_1M_2}(M)+T_{SPb}^{FM_1M_2}(M),\nonumber\\
  T_{SPa}^{FM_1M_2}(M)&=&- \frac{1}{2} \frac{C_{F}}{N_C}\ F_M\ F_{M_1}\
  F_{M_2} \int_0^1\int_0^1\int_0^1  \emph{d}u\,\emph{d}x\,\emph{d}y\, \ m_B\ \mu_{M_2}
  \phi_{M}(u)\nonumber\\&&\big\{m_B(2 m_B-m_b)\phi_{M_1}(x)+\mu_{M_1}[4 m_b- (x+1)m_B] \phi^{p}_{M_1}(x)
  +\mu_{M_1}m_B(1-x)\phi^{T}_{M_1}(x)\big\}\nonumber\\&& \phi^{p}_{M_2}(y)
   h_{Ta}^F(u,x),\nonumber\\
  T_{SPb}^{FM_1M_2}(M)&=&- \frac{1}{2} \frac{C_{F}}{N_C}\ F_M\ F_{M_1}\
  F_{M_2} \int_0^1\int_0^1\int_0^1  \emph{d}u\,\emph{d}x\,\emph{d}y\, \ m_B^2\ \mu_{M_2}\phi_{M}(u)
  \nonumber\\&&[m_B u \phi_{M_1}(x)+2 (1-u)
  \mu_{M_1}\phi^{p}_{M_1}(x)]\phi^{p}_{M_2}(y)h_{Tb}^F(u,x).
\end{eqnarray}
Similarly, we obtain
\begin{eqnarray}
  \label{eq:ppnf1}
  T_{LL}^{NM_1M_2}(M)&=&T_{LLa}^{NM_1M_2}(M)+T_{LLb}^{NM_1M_2}(M),\nonumber\\
  T_{LLa}^{NM_1M_2}(M)&=&\frac{1}{4}  \frac{C_{F}}{N_C}\ F_M\ F_{M_1}\ F_{M_2} \int_0^1\int_0^1\int_0^1  \emph{d}u\,\emph{d}x\,\emph{d}y\,
  m_B^3\phi_{M}(u)\nonumber\\&&\big\{ (u-y) m_B\phi_{M_1}(x)
  + (1-x)\mu_{M_1} [\phi^{p}_{M_1}(x)+\phi^{T}_{M_1}(x)]\big\}\phi_{M_2}(y)
  h_{Ta}^N(u,x,y),\nonumber\\
  T_{LLb}^{NM_1M_2}(M)&=&-\frac{1}{4}  \frac{C_{F}}{N_C}\ F_M\ F_1
  \ F_{M_2} \int_0^1\int_0^1\int_0^1  \emph{d}u\,\emph{d}x\,\emph{d}y\,\
  m_B^3\phi_{M}(u)\nonumber\\&&\big\{(u+x+y-2) m_B \phi_{M_1}(x)
  +(1-x)\mu_{M_1}  [\phi^{p}_{M_1}(x)-\phi^{T}_{M_1}(x)]\big\}\phi_{M_2}(y)
  h_{Tb}^N(u,x,y)\nonumber\\
\end{eqnarray}
for non-factorizable emission contributions with the $(V-A)\times
(V-A)$ effective four-quark vertex, and
\begin{eqnarray}
  \label{eq:ppnf3}
  T_{LR}^{NM_1M_2}(M)&=&T_{LRa}^{NM_1M_2}(M)+T_{LRb}^{NM_1M_2}(M)\nonumber\\
  T_{LRa}^{NM_1M_2}(M)&=&\frac{1}{4} \frac{C_{F}}{N_C}\ F_M\ F_{M_1}\
  F_{M_2} \int_0^1\int_0^1\int_0^1  \emph{d}u\,\emph{d}x\,\emph{d}y\,\ m_B^2\phi_{M}(u)
  \nonumber\\&&\Big\{\mu_{M_2}\ \mu_{M_1}\big\{[(u-x-y+1) \phi^{T}_{M_1}(x)+(u+x-y-1) \phi^{p}_{M_1}(x)]\phi^{p}_{M_2}(y)
  \nonumber\\&&-[(u-x-y+1) \phi^{p}_{M_1}(x)+(u+x-y-1) \phi^{T}_{M_1}(x)]\phi^{T}_{M_2}(y)\}
  \nonumber\\&&+(u-y) m_B\ \mu_{M_2} [\phi^{p}_{M_2}(y)-\phi^{T}_{M_2}(y)]\phi_{M_1}(x)\Big\}h_{Ta}^N(u,x,y)\nonumber\\
  T_{LRb}^{NM_1M_2}(M)&=&-\frac{1}{4} \frac{C_{F}}{N_C}\ F_M\ F_{M_1}\
  F_{M_2} \int_0^1\int_0^1\int_0^1  \emph{d}u\,\emph{d}x\,\emph{d}y\,\ m_B^2\phi_{M}(u)
  \nonumber\\&&\Big\{\mu_{M_2}\ \mu_{M_1}\big\{[(u-x+y)\phi^{T}_{M_1}(x)
  +(u+x+y-2) \phi^{p}_{M_1}(x)] \phi^{p}_{M_2}(y)
  \nonumber\\&&+[(u-x+y)\phi^{p}_{M_1}(x)+(u+x+y-2)\phi^{T}_{M_1}(x)]\phi^{T}_{M_2}(y)\big\}
  \nonumber\\&&+(u+y-1) m_B\ \mu_{M_2}[\phi^{p}_{M_2}(y)+\phi^{T}_{M_2}(y)]\phi_{M_1}(x)\Big\}h_{Tb}^N(u,x,y)
\end{eqnarray}
for non-factorizable emission contributions with the $(V-A)\times
(V+A)$ effective four-quark vertex, and
\begin{eqnarray}
  \label{eq:ppnf2}
  T_{SP}^{NM_1M_2}(M)&=&T_{SPa}^{NM_1M_2}(M)+T_{SPb}^{NM_1M_2}(M)\nonumber\\
  T_{SPa}^{NM_1M_2}(M)&=&- \frac{1}{4}\frac{C_{F}}{N_C}\ F_M\ F_{M_1}\
  F_{M_2} \int_0^1\int_0^1\int_0^1  \emph{d}u\,\emph{d}x\,\emph{d}y\,\ m_B^3\phi_{M}(u)\nonumber\\&&
  \big\{(u+x-y-1)m_B  \phi_{M_1}(x)+ (1-x)\mu_{M_1} [\phi^{p}_{M_1}(x)-\phi^{T}_{M_1}(x)]\big\}
  \phi_{M_2}(y)h_{Ta}^N(u,x,y)\nonumber\\
  T_{SPb}^{NM_1M_2}(M)&=& \frac{1}{4}\frac{C_{F}}{N_C}\ F_M\ F_{M_1}\
  F_{M_2} \int_0^1\int_0^1\int_0^1  \emph{d}u\,\emph{d}x\,\emph{d}y\,\ m_B^3\phi_{M}(u)\nonumber\\&&
  \big\{(u+y-1)m_B \phi_{M_1}(x)+(1-x)\mu_{M_1} [\phi^{p}_{M_1}(x)+\phi^{T}_{M_1}(x)]\big\}\phi_{M_2}(y)
  h_{Tb}^N(u,x,y)
\end{eqnarray}
for non-factorizable emission contributions with the $(S-P)\times
(S+P)$ effective four-quark vertex.

We now present the results from annihilation diagram contributions,
\begin{eqnarray}
  \label{eq:af1}
A_{LL}^{FM_1M_2}(M)&=&A_{LLa}^{FM_1M_2}(M)+A_{LLb}^{FM_1M_2}(M),\nonumber\\
A_{LLa}^{FM_1M_2}(M)&=&- \frac{1}{4} \frac{C_{F}}{N_C}\ F_M\
F_{M_1}\ F_{M_2} \int_0^1\int_0^1\int_0^1
\emph{d}u\,\emph{d}x\,\emph{d}y\,\ m_B^2 \phi_{M}(u)
\nonumber\\&&\big\{(1-y)m_B^2 \phi_{M_2}(y) \phi_{M_1}(x) +2
\mu_{M_2}\ \mu_{M_1} [(2-y) \phi^{p}_{M_2}(y)+y\phi^{T}_{M_2}(y)]
 \phi^{p}_{M_1}(x)\big\}h_{Aa}^F(x,y),\nonumber\\
A_{LLb}^{FM_1M_2}(M)&=& \frac{1}{4} \frac{C_{F}}{N_C}\ F_M\
F_{M_1} \ F_{M_2} \int_0^1\int_0^1\int_0^1
\emph{d}u\,\emph{d}x\,\emph{d}y\,\ m_B^2 \phi_{M}(u)
\nonumber\\&&\big\{x m_B^2 \phi_{M_2}(y) \phi_{M_1}(x)+2 \mu_{M_2}\
\mu_{M_1} [(1+x)\phi^{p}_{M_1}(x)-
(1-x)\phi^{T}_{M_1}(x)]\phi^{p}_{M_2}(y)\}
h_{Ab}^F(x,y),\nonumber\\
A_{LR}^{FM_1M_2}(M)&=&A_{LRa}^{FM_1M_2}(M)+A_{LRb}^{FM_1M_2}(M),\nonumber\\
A_{LRa}^{FM_1M_2}(M)&=&A_{LLa}^{FM_1M_2}(M),\nonumber\\
A_{LRb}^{FM_1M_2}(M)&=&A_{LLb}^{FM_1M_2}(M)
\end{eqnarray}
for the factorizable annihilation contributions with the
$(V-A)\times (V-A)$ and $(V-A)\times (V+A)$ effective four-quark
vertexes, and
\begin{eqnarray}
  \label{eq:af3}
A_{SP}^{FM_1M_2}(M)&=&A_{SPa}^{FM_1M_2}(M)+A_{SPb}^{FM_1M_2}(M),\nonumber\\
A_{SPa}^{FM_1M_2}(M)&=&-\frac{1}{2} \frac{C_{F}}{N_C}\ F_M\ F_{M_1}\
F_{M_2} \int_0^1\int_0^1\int_0^1  \emph{d}u\,\emph{d}x\,\emph{d}y\,\
m_B^3\phi_{M}(u)
 \nonumber\\&&[(1-y)\mu_{M_2}
[\phi^{p}_{M_2}(y)+\phi^{T}_{M_2}(y)]\phi_{M_1}(x) +2 \mu_{M_1}
\phi_{M_2}(y) \phi^{p}_{M_1}(x)]h_{Aa}^F(x,y),\nonumber\\
A_{SPb}^{FM_1M_2}(M)&=&-\frac{1}{2} \frac{C_{F}}{N_C}\ F_M\ F_{M_1}\
F_{M_2} \int_0^1\int_0^1\int_0^1  \emph{d}u\,\emph{d}x\,\emph{d}y\,\
m_B^3\phi_{M}(u)
 \big\{2 \mu_{M_2}
\phi^{p}_{M_2}(y) \phi_{M_1}(x)\nonumber\\&&+ x\ \mu_{M_1}
\phi_{M_2}(y)[\phi^{p}_{M_1}(x)-\phi^{T}_{M_1}(x)]\big\}h_{Ab}^F(x,y)
\end{eqnarray}
for the factorizable annihilation contributions with the
$(S-P)\times (S+P)$ effective four-quark vertex, and
\begin{eqnarray}
  \label{eq:anf1}
A_{LL}^{NM_1M_2}(M)&=&A_{LLa}^{NM_1M_2}(M)+A_{LLb}^{NM_1M_2}(M),\nonumber\\
A_{LLa}^{NM_1M_2}(M)&=&\frac{1}{4} \frac{C_{F}}{N_C}\ F_M\
F_{M_1}\ F_{M_2} \int_0^1\int_0^1\int_0^1
\emph{d}u\,\emph{d}x\,\emph{d}y\,\ m_B^2 \phi_{M}(u) \Big\{[m_b+m_B
(u-y)] m_B^2 \phi_{M_2}(y)
\phi_{M_1}(x)\nonumber\\&&+\mu_{M_1}\mu_{M_2} \big\{[-(u-x-y+1)m_B
\phi^{p}_{M_1}(x)+ (-u-x+y+1)m_B\phi^{T}_{M_1}(x)]
 \phi^{T}_{M_2}(y)\big\}\nonumber\\&&+[\big(4 m_b+(u+x-y-1)m_B \big)\phi^{p}_{M_1}(x)+(u-x-y+1)m_B \phi^{T}_{M_1}(x)]
 \phi^{p}_{M_2}(y)\Big\}
 h_{Aa}^N(u,x,y),\nonumber\\
A_{LLb}^{NM_1M_2}(M)&=&-\frac{1}{4} \frac{C_{F}}{N_C}\ F_M\ F_{M_1}\
F_{M_2} \int_0^1\int_0^1\int_0^1  \emph{d}u\,\emph{d}x\,\emph{d}y\,\
m_B^2 \phi_{M}(u) \Big\{x m_B^2 \phi_{M_2}(y)
\phi_{M_1}(x)\nonumber\\&&+\mu_{M_1}\mu_{M_2}\big\{
-[(u+x+y-1)\phi^{p}_{M_1}(x)+(-u+x-y+1)\phi^{T}_{M_1}(x)]
 \phi^{T}_{M_2}(y)\nonumber\\&&+[(-u+x-y+1)\phi^{p}_{M_1}(x)
+(u+x+y-1)\phi^{T}_{M_1}(x)]
 \phi^{p}_{M_2}(y)\big\}\Big\}h_{Ab}^N(u,x,y)
\end{eqnarray}
for the non-factorizable annihilation contributions with the
$(V-A)\times (V-A)$ effective four-quark vertex, and
\begin{eqnarray}
  \label{eq:anf3}
A_{LR}^{NM_1M_2}(M)&=&A_{LRa}^{NM_1M_2}(M)+A_{LRb}^{NM_1M_2}(M),\nonumber\\
A_{LRa}^{NM_1M_2}(M)&=&- \frac{1}{4} \frac{C_{F}}{N_C}\ F_M\
F_{M_1}\ F_{M_2} \int_0^1\int_0^1\int_0^1
\emph{d}u\,\emph{d}x\,\emph{d}y\,\ m_B^2\phi_{M}(u)
\nonumber\\&&\big\{\mu_{M_2} [m_b+(y-u)m_B
][\phi^{p}_{M_2}(y)-\phi^{T}_{M_2}(y)]
\phi_{M_1}(x)\nonumber\\&&-\mu_{M_1} [
(1-x)m_B+m_b][\phi^{p}_{M_1}(x)+\phi^{T}_{M_1}(x)]\phi^{p}_{M_2}(y)\big\}
h_{Aa}^N(u,x,y),\nonumber\\
A_{LRa}^{NM_1M_2}(M)&=& \frac{1}{4} \frac{C_{F}}{N_C}\ F_M\
F_{M_1}\ F_{M_2} \int_0^1\int_0^1\int_0^1
\emph{d}u\,\emph{d}x\,\emph{d}y\, \ m_B^3\phi_{M}(u)\big\{x\
\mu_{M_1} [\phi^{p}_{M_1}(x)+\phi^{T}_{M_1}(x)]
\phi^{p}_{M_2}(y)\nonumber\\&&-(1-u-y)\mu_{M_2}
[\phi^{p}_{M_2}(y)-\phi^{T}_{M_2}(y)] \phi_{M_1}(x)\}
h_{Ab}^N(u,x,y)
\end{eqnarray}
for the non-factorizable annihilation contributions with the
$(V-A)\times (V-A)$ and $(V-A)\times (V+A)$ effective four-quark
vertexes, and
\begin{eqnarray}
  \label{eq:anf2}
A_{SP}^{NM_1M_2}(M)&=&A_{SPa}^{NM_1M_2}(M)+A_{SPb}^{NM_1M_2}(M),\nonumber\\
A_{SPa}^{NM_1M_2}(M)&=& \frac{1}{4} \frac{C_{F}}{N_C}\ F_M\
F_{M_1}\  F_{M_2} \int_0^1\int_0^1\int_0^1
\emph{d}u\,\emph{d}x\,\emph{d}y\,\ m_B \phi_{M}(u)
\Big\{[m_b+(x-1)m_B ] m_B^2 \phi^{A}
 _{M_2}(y)\nonumber\\&&+\mu_{M_1}\mu_{M_2} \big\{ \phi_{M_1}(x)[(u-x-y+1)m_B \phi^{p}_{M_1}(x)+
(-u-x+y+1)m_B\phi^{T}_{M_1}(x)]
 \phi^{T}_{M_2}(y)\big\}\nonumber\\&&+\big\{[4 m_b-(-u-x+y+1)m_B ]\phi^{p}_{M_1}(x)-(u-x-y+1)m_B \phi^{T}_{M_1}(x)\big\}
 \phi^{p}_{M_2}(y)\Big\}
 h_{Aa}^N(u,x,y),\nonumber\\
A_{SPb}^{NM_1M_2}(M)&=&- \frac{1}{4} \frac{C_{F}}{N_C}\ F_M\
F_{M_1}\  F_{M_2} \int_0^1\int_0^1\int_0^1
\emph{d}u\,\emph{d}x\,\emph{d}y\,\ m_B^2 \phi_{M}(u) \Big\{(-u-y+1)
m_B^2 \phi_{M_2}(y) \phi_{M_1}(x)\nonumber\\&&+\mu_{M_1}\mu_{M_2}
\big\{[(u+x+y-1)\phi^{p}_{M_1}(x)-(-u+x-y+1)\phi^{T}_{M_1}(x)]
 \phi^{T}_{M_2}(y)\big\}\nonumber\\&&+[(-u+x-y+1)\phi^{p}_{M_1}(x)-(u+x+y-1)\phi^{T}_{M_1}(x)]
 \phi^{p}_{M_2}(y) \Big\} h_{Ab}^N(u,x,y)
\end{eqnarray}
for the non-factorizable annihilation contributions with the
$(S-P)\times (S+P)$ effective four-quark vertex.

The functions $h^{Y}_{XA}$ with $(A=a,b)$ from Eqs.~(\ref{eq:ppf1})
to (\ref{eq:anf3}) arise from propagators of gluon and quark, here
$Y=F,N$ denote the factorizable and non-factorizable contributions
respectively, and $X=T,A$ the emission and annihilation diagrams
respectively. They have the following explicit forms:
\begin{eqnarray}
  &&h_{Ta}^F(u,x)=\frac{1}{(-u(1-x)m_B^2-\mu_g^2+i\epsilon)(x m_B^2-m_b^2+i\epsilon)},\nonumber\\
  &&h_{Tb}^F(u,x)=\frac{1}{(-u(1-x)m_B^2-\mu_g^2+i\epsilon)(-u m_B^2-\mu_{q}^2+i\epsilon)},\nonumber\\
  &&h_{Ta}^N(u,x,y)=\frac{1}{(-u(1-x)m_B^2-\mu_g^2+i\epsilon)((1-x)(1-u-y)m_B^2-\mu_{q}^2+i\epsilon)},\nonumber\\
  &&h_{Tb}^N(u,x,y)=\frac{1}{(-u(1-x)m_B^2-\mu_g^2+i\epsilon)((1-x)(y-u)m_B^2-\mu_{q}^2+i\epsilon)},\nonumber\\
  &&h_{Aa}^F(x,y)=\frac{1}{(x(1-y)m_B^2-\mu_g^2+i\epsilon)((1-y)m_B^2-\mu_{q}^2+i\epsilon)},\nonumber\\
  &&h_{Ab}^F(x,y)=\frac{1}{(x(1-y)m_B^2-\mu_g^2+i\epsilon)(x\,m_B^2-\mu_{q}^2+i\epsilon)},\nonumber\\
  &&h_{Aa}^N(u,x,y)=\frac{1}{(x(1-y)m_B^2-\mu_g^2+i\epsilon)((y-u)(1-x)m_B^2-m_b^2+i\epsilon)},\nonumber\\
  &&h_{Ab}^N(u,x,y)=\frac{1}{(x(1-y)m_B^2-\mu_g^2+i\epsilon)((1-u-y)x\;m_B^2-\mu_{q}^2+i\epsilon)}.\label{eq:propN}
\end{eqnarray}

For the $P_1V_2$ final states, we can replace wave function in Eqs.~(\ref{eq:ppf1}) to (\ref{eq:anf3})with£º
\begin{eqnarray}
&&\phi_{P_2}(x)\to \phi_{V_2}(x),\ \phi^p_{P_2}(x)\to -\phi^{\nu}_{V_2}(x),\ \phi^T_{P_2}(x)\to \phi^{\nu}_{V_2}(x)(2\ x-1), \mu_{P_2}\to -m_{V_2}
  \nonumber\\
\end{eqnarray}
in which $\phi^T_{P_1}(x)=\frac{\phi_{\sigma}'}{6}$ with $\phi_{\sigma}'=\frac{\partial \phi_{\sigma}}{\partial x}$.

For $V_1P_2$ final states, we can replace wave function in Eqs.~(\ref{eq:ppf1})to(\ref{eq:anf3})with£º
\begin{eqnarray}
  &&\phi_{P_1}(x)\to \phi_{V_1}(x),\ \phi^p_{P_1}(x)\to -\phi^{\nu}_{V_1}(x),\ \phi^T_{P_1}(x)\to \phi^{\nu}_{V_1}(x)(2\ x-1), \mu_{P_1}\to -m_{V_1}\,,\nonumber\\
  &&\mu_{P_2}\to -\mu_{P_2}\,.
 \end{eqnarray}

For the $V_1V_2$ final states, we can replace wave function in Eqs.~(\ref{eq:ppf1})to(\ref{eq:anf3})with£º
\begin{eqnarray}
  &&\phi_{P_1}(x)\to \phi_{V_1}(x),\ \phi^p_{P_1}(x)\to -\phi^{\nu}_{V_1}(x),\ \phi^T_{P_1}(x)\to \phi^{\nu}_{V_1}(x)(2\ x-1), \mu_{P_1}\to -m_{V_1}\,,\nonumber\\
&&\phi_{P_2}(x)\to \phi_{V_2}(x),\ \phi^p_{P_2}(x)\to -\phi^{\nu}_{V_2}(x),\ \phi^T_{P_2}(x)\to \phi^{\nu}_{V_2}(x)(2\ x-1), \mu_{P_2}\to m_{V_2}\,.
 \end{eqnarray}

For the transverse part of two vector meson final states, we list all hadronic matrix elements based on six-quark approach(separate to plus/minus parts, which is valuable when considering vertex correction)£º
we first get the factorizable emission contributions for the $(V-A)\times (V-A)$ ºÍ $(V-A)\times (V+A)$effective four-quark
vertexes, they are simply denoted by $LL$ and $LR$
\begin{eqnarray}
  \label{eq:vvf1}
  T_{LL,+/-}^{FM_1M_2}(M)&=&T_{LLa,+/-}^{FM_1M_2}(M)+T_{LLb,+/-}^{FM_1M_2}(M),\nonumber\\
  T_{LLa,+}^{FM_1M_2}(M)&=& \frac{1}{2}\frac{C_{F}}{N_C}\ F_M\ F_{M_1}\
  F_{M_2} \int_0^1\int_0^1\int_0^1  \emph{d}u\,\emph{d}x\,\emph{d}y\,m_B m_{M_1} m_{M_2} \phi_{M}(u)\nonumber\\&&
  (2m_b-m_B x)\phi^+_{M_1}(x)\phi^+_{M_2}(y)h_{Ta}^F(u,x),\nonumber\\
  T_{LLa,-}^{FM_1M_2}(M)&=& \frac{1}{2}  \frac{C_{F}}{N_C}\ F_M\ F_{M_1}\
  F_{M_2} \int_0^1\int_0^1\int_0^1  \emph{d}u\,\emph{d}x\,\emph{d}y\, m_B m_{M_1} m_{M_2} \phi_{M}(u)\nonumber\\&&
  \big\{(2 m_b-m_B)m_{M_1} \phi^-_{M_1}(x)+m_B(2 m_B-m_b)\phi^T_{M_1}(x)\big\}\phi^-_{M_2}(y)h_{Tb}^F(u,x),\nonumber\\
  T_{LLb,+}^{FM_1M_2}(M)&=& -\frac{1}{2}\frac{C_{F}}{N_C}\ F_M\ F_{M_1}\
  F_{M_2} \int_0^1\int_0^1\int_0^1  \emph{d}u\,\emph{d}x\,\emph{d}y\,m_B m_{M_1} m_{M_2} \phi_{M}(u)\nonumber\\&&
  \big\{m_b u \phi^+_{M_1}(x)-m_q\phi^-_{M_1}(x)\big\}
  \phi^+_{M_2}(y)h_{Ta}^F(u,x),\nonumber\\
  T_{LLb,-}^{FM_1M_2}(M)&=& \frac{1}{2}  \frac{C_{F}}{N_C}\ F_M\ F_{M_1}\
  F_{M_2} \int_0^1\int_0^1\int_0^1  \emph{d}u\,\emph{d}x\,\emph{d}y\, m_B m_{M_1} m_{M_2} \phi_{M}(u)\nonumber\\&&
  \big\{m_q \phi^+_{M_1}(x)+ m_B\phi^-_{M_1}(x)\big\}\phi^{-}_{M_2}(y)h_{Tb}^F(u,x),\nonumber\\
  T_{LR,+/-}^{FM_1M_2}(M)&=&T_{LLa,+/-}^{FM_1M_2}(M)+T_{LLb,+/-}^{FM_1M_2}(M),\nonumber\\
  T_{LRa,+/-}^{FM_1M_2}(M)&=&T_{LLa,+/-}^{FM_1M_2}(M),\nonumber\\
  T_{LRb,+/-}^{FM_1M_2}(M)&=&T_{LLb,+/-}^{FM_1M_2}(M).
\end{eqnarray}

The factorizable emission contributions for the $(S-P)\times (S+P)$
effective four-quark vertex are found to be
\begin{eqnarray}
  \label{eq:vvf2}
  T_{SP,+/-}^{FM_1M_2}(M)&=&T_{SPa,+/-}^{FM_1M_2}(M)+T_{SPb,+/-}^{FM_1M_2}(M),\nonumber\\
  T_{SPa,+/-}^{FM_1M_2}(M)&=&0\,,\nonumber\\
  T_{SPb,+/-}^{FM_1M_2}(M)&=&0\,.
\end{eqnarray}
Similarly, we obtain:
\begin{eqnarray}
  \label{eq:vvnf1}
  T_{LL,+/-}^{NM_1M_2}(M)&=&T_{LLa,+/-}^{NM_1M_2}(M)+T_{LLb,+/-}^{NM_1M_2}(M),\nonumber\\
  T_{LLa,+}^{NM_1M_2}(M)&=&\frac{1}{2}  \frac{C_{F}}{N_C}\ F_M\ F_{M_1}\ F_{M_2} \int_0^1\int_0^1\int_0^1  \emph{d}u\,\emph{d}x\,\emph{d}y\,
  m_B^2\phi_{M}(u)m_{M_1}m_q \phi^+_{M_1}(x)\phi^T_{M_2}(y)
  h_{Ta}^N(u,x,y),\nonumber\\
  T_{LLa,-}^{NM_1M_2}(M)&=&\frac{1}{2}  \frac{C_{F}}{N_C}\ F_M\ F_{M_1}\ F_{M_2} \int_0^1\int_0^1\int_0^1  \emph{d}u\,\emph{d}x\,\emph{d}y\,
  m_B^3\phi_{M}(u)m_{M_2}(u-y)\phi^T_{M_1}(x)\phi^-_{M_2}(y)
  \nonumber\\&&h_{Ta}^N(u,x,y),\nonumber\\
  T_{LLb,+}^{NM_1M_2}(M)&=&\frac{1}{2}  \frac{C_{F}}{N_C}\ F_M\ F_1
  \ F_{M_2} \int_0^1\int_0^1\int_0^1  \emph{d}u\,\emph{d}x\,\emph{d}y\,\
  m_B^2\phi_{M}(u)m_{M_1}m_{M_2}(-2+u+x+y)\nonumber\\&&\phi^+_{M_1}(x)\phi^+_{M_2}(y)
  h_{Tb}^N(u,x,y)\nonumber\\
  T_{LLb,-}^{NM_1M_2}(M)&=&\frac{1}{2}  \frac{C_{F}}{N_C}\ F_M\ F_1
  \ F_{M_2} \int_0^1\int_0^1\int_0^1  \emph{d}u\,\emph{d}x\,\emph{d}y\,\
  m_B^2\phi_{M}(u)\big\{m_q \phi^T_{M_2}(y)[m_{M_1}\phi^-_{M_1}(x)\nonumber\\&&-2m_B \phi^T_{M_1}(x)]-m_{M_2}\phi^-_{M_2}(y)[-m_{M_1}\phi^-_{M_1}(x)
  (-2+u+x+y)\nonumber\\&&+m_B(-1+u+y)\phi^T_{M_1}(x)] \big\}
  h_{Tb}^N(u,x,y)
\end{eqnarray}
for non-factorizable emission contributions with the $(V-A)\times
(V-A)$ effective four-quark vertex, and
\begin{eqnarray}
  \label{eq:vvnf3}
  T_{LR,+/-}^{NM_1M_2}(M)&=&T_{LRa,+/-}^{NM_1M_2}(M)+T_{LRb,+/-}^{NM_1M_2}(M)\nonumber\\
  T_{LRa,+}^{NM_1M_2}(M)&=&\frac{1}{2} \frac{C_{F}}{N_C}\ F_M\ F_{M_1}\
  F_{M_2} \int_0^1\int_0^1\int_0^1  \emph{d}u\,\emph{d}x\,\emph{d}y\,\ m_B m_{M_2} m_q\phi_{M}(u)
  \nonumber\\&&\Big\{\phi^-{M_2}(y)[m_{M_1}\phi^-_{M_1}(x)-m_B\phi^T_{M_1}(x)]\Big\}h_{Ta}^N(u,x,y)\nonumber\\
  T_{LRa,-}^{NM_1M_2}(M)&=&\frac{1}{2} \frac{C_{F}}{N_C}\ F_M\ F_{M_1}\
  F_{M_2} \int_0^1\int_0^1\int_0^1  \emph{d}u\,\emph{d}x\,\emph{d}y\,\ m_B m_{M_1}\phi_{M}(u)
  \Big\{m_B^2(x-1)\phi^T_{M_2}(y)\nonumber\\&&+m_{M_2}m_q\phi^+_{M_2}(y)\Big\}\phi^+_{M_1}(x)h_{Ta}^N(u,x,y)\nonumber\\
  T_{LRb,+}^{NM_1M_2}(M)&=&-\frac{1}{2} \frac{C_{F}}{N_C}\ F_M\ F_{M_1}\
  F_{M_2} \int_0^1\int_0^1\int_0^1  \emph{d}u\,\emph{d}x\,\emph{d}y\,\ m_B m_{M_2} m_q\phi_{M}(u)
  \nonumber\\&&\Big\{\phi^-{M_2}(y)[m_{M_1}\phi^-_{M_1}(x)-m_B\phi^T_{M_1}(x)]\Big\}h_{Tb}^N(u,x,y)\nonumber\\
  T_{LRb,-}^{NM_1M_2}(M)&=&\frac{1}{2} \frac{C_{F}}{N_C}\ F_M\ F_{M_1}\
  F_{M_2} \int_0^1\int_0^1\int_0^1  \emph{d}u\,\emph{d}x\,\emph{d}y\,\ m_B m_{M_1}\phi_{M}(u)
  \Big\{m_B^2(x-1)\phi^T_{M_2}(y)\nonumber\\&&+m_{M_2}m_q\phi^-_{M_2}(y)\Big\}\phi^+_{M_1}(x)h_{Tb}^N(u,x,y)
\end{eqnarray}
for non-factorizable emission contributions with the $(V-A)\times
(V+A)$ effective four-quark vertex, and
\begin{eqnarray}
  \label{eq:vvnf2}
  T_{SP,+/-}^{NM_1M_2}(M)&=&T_{SPa,+/-}^{NM_1M_2}(M)+T_{SPb,+/-}^{NM_1M_2}(M)\nonumber\\
  T_{SPa,+}^{NM_1M_2}(M)&=& -\frac{1}{2}\frac{C_{F}}{N_C}\ F_M\ F_{M_1}\
  F_{M_2} \int_0^1\int_0^1\int_0^1  \emph{d}u\,\emph{d}x\,\emph{d}y\,\
  m_B^2\phi_{M}(u) m_{M_1} m_{M_2}\nonumber\\&&(-1+u+x-y)\phi^+_{M_1}(x)\phi^-_{M_2}(y)h_{Ta}^N(u,x,y)\nonumber\\
  T_{SPa,-}^{NM_1M_2}(M)&=& \frac{1}{2}\frac{C_{F}}{N_C}\ F_M\ F_{M_1}\
  F_{M_2} \int_0^1\int_0^1\int_0^1  \emph{d}u\,\emph{d}x\,\emph{d}y\,\
  m_B^2\phi_{M}(u)\big\{m_q \phi^T_{M_2}(y)[-m_{M_1}\phi^-_{M_1}(x)\nonumber\\&&+2m_B \phi^T_{M_1}(x)]+m_{M_2}\phi^+_{M_2}(y)[-m_{M_1}\phi^-_{M_1}(x)
  (-1+u+x-y)\nonumber\\&&+m_B(u-y)\phi^T_{M_1}(x)] \big\}
  h_{Ta}^N(u,x,y)\nonumber\\
  T_{SPb,+}^{NM_1M_2}(M)&=&\frac{1}{2}\frac{C_{F}}{N_C}\ F_M\ F_{M_1}\
  F_{M_2} \int_0^1\int_0^1\int_0^1  \emph{d}u\,\emph{d}x\,\emph{d}y\,\ m_B\phi_{M}(u)
  m_B^2\phi_{M}(u)m_{M_1}m_q \nonumber\\&&\phi^+_{M_1}(x)\phi^T_{M_2}(y)h_{Tb}^N(u,x,y)\nonumber\\
  T_{SPb,-}^{NM_1M_2}(M)&=&\frac{1}{2}\frac{C_{F}}{N_C}\ F_M\ F_{M_1}\
  F_{M_2} \int_0^1\int_0^1\int_0^1  \emph{d}u\,\emph{d}x\,\emph{d}y\,\ m_B\phi_{M}(u)
  m_B^3\phi_{M}(u)m_{M_2}\nonumber\\&&(-1+u+y)\phi^T_{M_1}(x)\phi^+_{M_2}(y)h_{Tb}^N(u,x,y)
\end{eqnarray}
for non-factorizable emission contributions with the $(S-P)\times
(S+P)$ effective four-quark vertex.

We now present the results from annihilation diagram contributions,
\begin{eqnarray}
  \label{eq:vvaf1}
A_{LL,+/-}^{FM_1M_2}(M)&=&A_{LLa,+/-}^{FM_1M_2}(M)+A_{LLb,+/-}^{FM_1M_2}(M),\nonumber\\
A_{LLa,+}^{FM_1M_2}(M)&=&-\frac{1}{2} \frac{C_{F}}{N_C}\ F_M\
F_{M_1}\ F_{M_2} \int_0^1\int_0^1\int_0^1
\emph{d}u\,\emph{d}x\,\emph{d}y\,\ m_B^2 m_{M_2} \phi_{M}(u)
\nonumber\\&&[\phi^+_{M_2}(y)\phi^+_{M_1}(x)m_{M_1}x-m_q\phi^-_{M_2}(y)\phi^T_{M_1}(x)]h_{Aa}^F(x,y),\nonumber\\
A_{LLa,-}^{FM_1M_2}(M)&=&-\frac{1}{2} \frac{C_{F}}{N_C}\ F_M\
F_{M_1}\ F_{M_2} \int_0^1\int_0^1\int_0^1
\emph{d}u\,\emph{d}x\,\emph{d}y\,\ m_B^2 m_{M_1} m_{M_2} \phi_{M}(u)
\nonumber\\&&\phi^-_{M_2}(y)\phi^-_{M_1}(x)h_{Aa}^F(x,y),\nonumber\\
A_{LLb,+}^{FM_1M_2}(M)&=&\frac{1}{2} \frac{C_{F}}{N_C}\ F_M\
F_{M_1} \ F_{M_2} \int_0^1\int_0^1\int_0^1
\emph{d}u\,\emph{d}x\,\emph{d}y\,\ m_B^2 m_{M_1} m_{M_2} \phi_{M}(u)
\nonumber\\&&\phi^+_{M_2}(y)\phi^+_{M_1}(x)
h_{Ab}^F(x,y),\nonumber\\
A_{LLb,-}^{FM_1M_2}(M)&=& -\frac{1}{2} \frac{C_{F}}{N_C}\ F_M\
F_{M_1} \ F_{M_2} \int_0^1\int_0^1\int_0^1
\emph{d}u\,\emph{d}x\,\emph{d}y\,\ m_B^2 m_{M_1} \phi_{M}(u)
\nonumber\\&&[\phi^-_{M_2}(y)\phi^-_{M_1}(x)m_{M_2}(-1+y)+m_q\phi^+_{M_1}(x)\phi^T_{M_2}(y)]
h_{Ab}^F(x,y),\nonumber\\
A_{LR,+/-}^{FM_1M_2}(M)&=&A_{LRa,+/-}^{FM_1M_2}(M)+A_{LRb,+/-}^{FM_1M_2}(M),\nonumber\\
A_{LRa,+/-}^{FM_1M_2}(M)&=&A_{LLa,-/+}^{FM_1M_2}(M),\
A_{LRb,+/-}^{FM_1M_2}(M)=A_{LLb,-/+}^{FM_1M_2}(M)
\end{eqnarray}
for the factorizable annihilation contributions with the
$(V-A)\times (V-A)$ and $(V-A)\times (V+A)$ effective four-quark
vertexes, and
\begin{eqnarray}
  \label{eq:vvaf3}
A_{SP,+/-}^{FM_1M_2}(M)&=&A_{SPa,+/-}^{FM_1M_2}(M)+A_{SPb,+/-}^{FM_1M_2}(M),\nonumber\\
A_{SPa£¬+}^{FM_1M_2}(M)&=&-\frac{C_{F}}{N_C}\ F_M\ F_{M_1}\
F_{M_2} \int_0^1\int_0^1\int_0^1  \emph{d}u\,\emph{d}x\,\emph{d}y\,\
m_B m_q m_{M_1} m_{M_2}\phi_{M}(u)\nonumber\\&&
\phi^-_{M_1}(x)\phi^-_{M_y}(x)h_{Aa}^F(x,y),\nonumber\\
A_{SPa£¬-}^{FM_1M_2}(M)&=&-\frac{C_{F}}{N_C}\ F_M\ F_{M_1}\
F_{M_2} \int_0^1\int_0^1\int_0^1  \emph{d}u\,\emph{d}x\,\emph{d}y\,\
m_B m_{M_2}\phi_{M}(u)[-\phi^-_{M_2}(y)m_B^2\phi^T_{M_1}(x)\nonumber\\&&+\phi^+_{M_2}(y)m_q m_{M_1}\phi^+_{M_1}(x)]
h_{Aa}^F(x,y),\nonumber\\
A_{SPb£¬+}^{FM_1M_2}(M)&=&-\frac{C_{F}}{N_C}\ F_M\ F_{M_1}\
F_{M_2} \int_0^1\int_0^1\int_0^1  \emph{d}u\,\emph{d}x\,\emph{d}y\,\
m_B m_q m_{M_1} m_{M_2}\phi_{M}(u)\nonumber\\&&
\phi^+_{M_1}(x)\phi^+_{M_y}(x)
h_{Ab}^F(x,y)\nonumber\\
A_{SPb£¬-}^{FM_1M_2}(M)&=&\frac{C_{F}}{N_C}\ F_M\ F_{M_1}\
F_{M_2} \int_0^1\int_0^1\int_0^1  \emph{d}u\,\emph{d}x\,\emph{d}y\,\
m_B m_{M_1}\phi_{M}(u)[\phi^+_{M_1}(x)m_B^2\phi^T_{M_2}(y)\nonumber\\&&-\phi^-_{M_2}(y)m_q m_{M_2}\phi^-_{M_1}(x)]
h_{Ab}^F(x,y)
\end{eqnarray}
for the factorizable annihilation contributions with the
$(S-P)\times (S+P)$ effective four-quark vertex, and
\begin{eqnarray}
  \label{eq:vvanf1}
A_{LL,+/-}^{NM_1M_2}(M)&=&A_{LLa,+/-}^{NM_1M_2}(M)+A_{LLb,+/-}^{NM_1M_2}(M),\nonumber\\
A_{LLa,+}^{NM_1M_2}(M)&=&-\frac{1}{2} \frac{C_{F}}{N_C}\ F_M\
F_{M_1}\ F_{M_2} \int_0^1\int_0^1\int_0^1
\emph{d}u\,\emph{d}x\,\emph{d}y\,\ m_B  m_b m_{M_1} m_{M_2}\phi_{M}(u)
\phi^+_{M_1}(x) \nonumber\\&&\phi^+_{M_2}(y)h_{Aa}^N(u,x,y),\nonumber\\
A_{LLa,-}^{NM_1M_2}(M)&=&-\frac{1}{2} \frac{C_{F}}{N_C}\ F_M\
F_{M_1}\ F_{M_2} \int_0^1\int_0^1\int_0^1
\emph{d}u\,\emph{d}x\,\emph{d}y\,\ m_B  m_b m_{M_1} m_{M_2} \phi_{M}(u)
\phi^-_{M_1}(x) \nonumber\\&&\phi^-_{M_2}(y)h_{Aa}^N(u,x,y),\nonumber\\
A_{LLb,+}^{NM_1M_2}(M)&=&\frac{1}{2} \frac{C_{F}}{N_C}\ F_M\ F_{M_1}\
F_{M_2} \int_0^1\int_0^1\int_0^1  \emph{d}u\,\emph{d}x\,\emph{d}y\,\
m_B  m_q m_{M_1} m_{M_2}  \phi_{M}(u) \nonumber\\&&\phi^+_{M_1}(x) \phi^+_{M_2}(y)h_{Ab}^N(u,x,y)\nonumber\\
A_{LLb,-}^{NM_1M_2}(M)&=&\frac{1}{2} \frac{C_{F}}{N_C}\ F_M\ F_{M_1}\
F_{M_2} \int_0^1\int_0^1\int_0^1  \emph{d}u\,\emph{d}x\,\emph{d}y\,\
m_B  m_q m_{M_1} m_{M_2}  \phi_{M}(u) \nonumber\\&&\phi^-_{M_1}(x) \phi^-_{M_2}(y)h_{Ab}^N(u,x,y)
\end{eqnarray}
for the non-factorizable annihilation contributions with the
$(V-A)\times (V-A)$ effective four-quark vertex, and
\begin{eqnarray}
  \label{eq:vvanf3}
A_{LR,+/-}^{NM_1M_2}(M)&=&A_{LRa,+/-}^{NM_1M_2}(M)+A_{LRb,+/-}^{NM_1M_2}(M),\nonumber\\
A_{LRa,-}^{NM_1M_2}(M)&=&0,\nonumber\\
A_{LRa,-}^{NM_1M_2}(M)&=&\frac{1}{2} \frac{C_{F}}{N_C}\ F_M\
F_{M_1}\ F_{M_2} \int_0^1\int_0^1\int_0^1
\emph{d}u\,\emph{d}x\,\emph{d}y\,\ m_B^2\phi_{M}(u)\nonumber\\&&
\big\{m_{M_2}[m_b+m_B(-u+y)]\phi^-_{M_2}(y)\phi^T_{M_1}(x)\nonumber\\&&+m_{M_1}(-m_B-m_b+m_B x)\phi^T_{M_2}(y)\phi^+_{M_1}(x)\big\}
h_{Aa}^N(u,x,y),\nonumber\\
A_{LRa,-}^{NM_1M_2}(M)&=&0,\nonumber\\
A_{LRa,-}^{NM_1M_2}(M)&=&-\frac{1}{2} \frac{C_{F}}{N_C}\ F_M\
F_{M_1}\ F_{M_2} \int_0^1\int_0^1\int_0^1
\emph{d}u\,\emph{d}x\,\emph{d}y\, \ m_B^2\phi_{M}(u)\nonumber\\&&
\big\{m_{M_2}[m_q+m_B(-1+u+y)]\phi^-_{M_2}(y)\phi^T_{M_1}(x)\nonumber\\&&-m_{M_1}(m_q-m_B x)\phi^T_{M_2}(y)\phi^+_{M_1}(x)\big\}
h_{Ab}^N(u,x,y)
\end{eqnarray}
for the non-factorizable annihilation contributions with the
$(V-A)\times (V-A)$ and $(V-A)\times (V+A)$ effective four-quark
vertexes, and
\begin{eqnarray}
  \label{eq:vvanf2}
A_{SP,+/-}^{NM_1M_2}(M)&=&A_{SPa,+/-}^{NM_1M_2}(M)+A_{SPb,+/-}^{NM_1M_2}(M),\nonumber\\
A_{SPa,+}^{NM_1M_2}(M)&=& -\frac{1}{2} \frac{C_{F}}{N_C}\ F_M\
F_{M_1}\  F_{M_2} \int_0^1\int_0^1\int_0^1
\emph{d}u\,\emph{d}x\,\emph{d}y\,\ m_B  m_b m_{M_1} m_{M_2}\phi_{M}(u)
\nonumber\\&&\phi^-_{M_1}(x) \phi^-_{M_2}(y)
 h_{Aa}^N(u,x,y),\nonumber\\
A_{SPa,-}^{NM_1M_2}(M)&=& -\frac{1}{2} \frac{C_{F}}{N_C}\ F_M\
F_{M_1}\  F_{M_2} \int_0^1\int_0^1\int_0^1
\emph{d}u\,\emph{d}x\,\emph{d}y\,\ m_B  m_b m_{M_1} m_{M_2}\phi_{M}(u)
\nonumber\\&&\phi^+_{M_1}(x) \phi^+_{M_2}(y)
 h_{Aa}^N(u,x,y),\nonumber\\
A_{SPb,+}^{NM_1M_2}(M)&=& \frac{1}{2} \frac{C_{F}}{N_C}\ F_M\
F_{M_1}\  F_{M_2} \int_0^1\int_0^1\int_0^1
\emph{d}u\,\emph{d}x\,\emph{d}y\,\ m_B  m_q m_{M_1} m_{M_2}\phi_{M}(u)
\nonumber\\&&\phi^-_{M_1}(x) \phi^-_{M_2}(y)  h_{Ab}^N(u,x,y)\nonumber\\
A_{SPb,-}^{NM_1M_2}(M)&=&\frac{1}{2} \frac{C_{F}}{N_C}\ F_M\
F_{M_1}\  F_{M_2} \int_0^1\int_0^1\int_0^1
\emph{d}u\,\emph{d}x\,\emph{d}y\,\ m_B  m_q m_{M_1} m_{M_2}\phi_{M}(u)
\nonumber\\&&\phi^+_{M_1}(x) \phi^+_{M_2}(y)  h_{Ab}^N(u,x,y)
\end{eqnarray}
for the non-factorizable annihilation contributions with the
$(S-P)\times (S+P)$ effective four-quark vertex.

The functions $h^{Y}_{XA}$ with $(A=a,b)$ from Eqs.~(\ref{eq:ppf1})to(\ref{eq:anf3})Àïarise from propagators of gluon and quark, here
$Y=F,N$ denote the factorizable and non-factorizable contributions
respectively, and $X=T,A$ the emission and annihilation diagrams
respectively. Their definitions refer to Eqs.~(\ref{eq:propN}).


\end{document}